\documentclass[article]{svmult}

\usepackage{mathptmx}       
\usepackage{helvet}         
\usepackage{courier}        
\usepackage{type1cm}        
\usepackage{graphicx}        
\usepackage[square,numbers]{natbib}
\makeindex             

\usepackage{amsmath,amssymb}
\usepackage{comment}

\newcommand{\rmd}{\mathrm{d}}
\newcommand{\rmi}{\mathrm{i}}
\def\rme{\mathrm{e}}
\def\rmR{\mathrm{R}}
\usepackage{upgreek}
\usepackage{dsfont}
\usepackage{slashed}

\usepackage{multicol,color,xcolor,longtable}

\begin{document}
\title*{Non-Lorentzian Supergravity}
\author{Eric A. Bergshoeff \thanks{corresponding author} and Jan Rosseel}
\institute{Eric A Bergshoeff \at Van Swinderen Institute, Nijenborgh 4, 9747AG Groningen, The Netherlands,\\ \email{E.A.Bergshoeff@rug.nl}
\and Jan Rosseel \at Division of Theoretical Physics, Rudjer Bo\v skovi\'c Institute, Bijeni\v cka 54, 10000 Zagreb, Croatia,
 \email{rosseelj@gmail.com}}
%
%
\maketitle
\abstract{We give an overview of the different non-Lorentzian supergravity theories in diverse dimensions that have been constructed in recent years. After giving a detailed discussion of non-Lorentzian geometries as compared to Lorentzian geometries, we outline some of the construction methods that have been applied to obtain non-Lorentzian supergravity. Explicit results are given for non-Lorentzian supergravity theories in three and ten dimensions.
}

\section*{Keywords}
Non-Lorentzian geometry; supersymmetry, gravity, higher dimensions

\section{Introduction}

General relativity as the theory of classical relativistic gravity is  a well-established framework already for more than a hundred years. It has confronted many experimental tests with flying colours. The combination of quantum mechanics and gravity is expected to lead to a new theory called quantum gravity that is valid up to Planck scales. Despite tremendous effort, no one has succeeded in constructing such a theory to date.

Newtonian gravity being the precursor of general relativity, is only able to describe gravitational physics for velocities that are low compared to the speed of light and that involve weak gravitational fields. Moreover, Newtonian gravity is only valid in reference frames that are rectilinearly accelerating with respect to inertial frames. A non-relativistic\,\footnote{Since every physical theory can be called relativistic, be it with respect to Galilean, Bargmann or Lorentzian symmetries, we will in this review also often use the more apt word non-Lorentzian instead of non-relativistic, when referring to theories that feature symmetries that can be viewed as $c \rightarrow \infty$ limits of Lorentzian ones. Note however that the phrase non-Lorentzian is more generic and is also used e.g. for so-called Carroll symmetries that arise as $c \rightarrow 0$ limits of Lorentzian ones.} (NR) version of general relativity, formulated in arbitrary reference frames, was constructed by Cartan 8 years after Einstein's discovery of general relativity \cite{Cartan1}. This theory is called Newton-Cartan (NC) gravity and is based on a NR version of the Riemannian geometry underlying general relativity. This is a degenerate metric geometry with a co-dimension one foliation in space-like leaves and an absolute time direction connecting them. This geometry is called NC geometry and the corresponding gravity theory is called NC gravity. It is a reformulation of Newtonian gravity in the sense that one can show that it is always possible to choose reference frames that are rectilinearly accelerated with respect to Newtonian inertial frames and in which one recovers Newtonian gravity.

NC gravity and geometry have attracted great attention in the recent literature due to a variety of reasons. A suitable generalization of it plays a crucial role in the original formulation of a NR version of string theory - a candidate theory of quantum gravity \cite{Gomis:2000bd,Danielsson:2000gi}. The possible formulation of such a NR string theory is tied up with the fundamental question whether combining gravity with quantum mechanics requires special relativity yes or no. The formulation of a consistent NR string theory moreover leads to the exciting question whether the holographic description of quantum gravity has an extension with NR gravity in the bulk and a NR conformal field theory at the boundary. Independent of this, starting with \cite{Son:2005rv,Son:2013rqa}, NC gravity also has found several applications to condensed matter physics.
Using an effective field theory description in which NR symmetries are implemented geometrically, the gravitational fields are used in this context as response functions to obtain new insights into the non-perturbative properties of the model under study.

Symmetries often help to resolve a problem in a given model or lead to useful restrictions on a model. A special kind of symmetry is supersymmetry. It is  a spacetime symmetry in the sense that the action of two supersymmetries leads to a spacetime translation. Supersymmetry has been used as a possible symmetry of extensions of the Standard Model of particle physics.  Finding evidence for supersymmetry is one of the research targets of the running LHC experiment. Supersymmetry is also an essential ingredient of superstring theory, where it solves the issue of how to avoid un-physical particles that would otherwise arise in the string spectrum. Assuming a small spacetime curvature, the low-energy limit of superstring theory is described by supersymmetric extensions of general relativity called supergravity theories.

Non-relativistic supersymmetry is central to defining a NR version of superstring theory. The NR supersymmetry that arises in this context is much less well developed than relativistic supersymmetry.\,\footnote{For early work on NR supersymmetry, see \cite{Puzalowski:1978rv,deAzcarraga:1991fa,Bergman:1995zr,Bergman:1996bx}.} For instance, not much is known about NR superspace and the possible NR supermultiplets; these two concepts are central to the formulation of relativistic supersymmetry. Despite a huge amount of research on supergravity, thus far we do not have a supersymmetric extension of NC gravity that includes a supersymmetrization of Newton’s law for gravity in four spacetime dimensions. 
Presently known results about non-Lorentzian supergravity theories fall in two categories. On the one hand, three-dimensional (3$D$) supersymmetric NR supergravity theories have been constructed, whose underlying geometry is NC geometry with a co-dimension one distribution that splits the tangent space in a spatial subspace and a time direction. This class of theories includes a supersymmetric generalization of 3$D$ NC gravity. On the other hand, recent work has also focused on ten-dimensional (10$D$) NR supergravity theories that employ a generalization of NC geometry, called stringy NC geometry, that is characterized by a co-dimension two distribution that splits the tangent space in an eight-dimensional `transversal' subspace and two Minkowskian `longitudinal' directions'. Whereas ordinary NC geometry naturally couples to NR particles, stringy NC geometry forms the natural background geometry for NR strings. Ten-dimensional supergravity theories based on stringy NC geometry can then be viewed as low energy approximations of NR string theory. It is the purpose of this review to give an overview of what is presently known about these 3$D$ and 10$D$ non-Lorentzian supergravity theories and give an outlook on e.g. eleven-dimensional (11$D$) non-Lorentzian supergravity. Below, for the convenience of the reader, we give a short introduction to the recent literature.

For simplicity reasons, the first attempts to construct supersymmetric extensions of non-Lorentzian gravity have taken place in a 3$D$ context. The advantage of working in 3$D$ is not only that the calculations are simpler than in 4$D$ but also that often gravity can be reformulated as a Chern-Simons theory that makes the relation with an underlying symmetry algebra very clear. Roughly speaking one can distinguish between five kinds of efforts:
\begin{enumerate}
\item gauging a supersymmetric extension of the Bargmann algebra, i.e., the centrally extended Galilei algebra
\item taking a non-Lorentzian limit
\item applying a NR superconformal tensor calculus
\item making use of the Chern-Simons formulation based on a non-Lorentzian superalgebra
\item making Lie algebra expansions, either using Maurer-Cartan equations or semigroups
\end{enumerate}

The first attempt to construct a 3$D$ NC supergravity theory, i.e., a supersymmetric version of NC gravity, was undertaken in \cite{Andringa:2013mma}. The construction was based on effort 1., i.e.~the gauging of a $\mathcal{N}=2$ supersymmetric extension of the Bargmann algebra. The reason that one considers $\mathcal{N}=2$ supersymmetry is that any $\mathcal{N}=1$ supersymmetric extension leads to a superalgebra where the time translation or Hamilton operator can not be written as the anti-commutator of two supersymmetries and therefore cannot be used for a physical realization of supersymmetry. The construction of \cite{Andringa:2013mma} has two distinguishing features:
\begin{enumerate}
\item It is based on ordinary NC geometry that is appropriate to particles but not to strings. It can therefore only be identified with the compactification over a spatial longitudinal direction of a low-energy approximation of a non-Lorenzian superstring theory.
\item By hand a geometric constraint is imposed that implies that the underlying NC geometry admits a notion of absolute Newtonian time.
\end{enumerate}
All transformation rules, including supersymmetry, and the expressions for all curvatures follow from the structure constants of the  $\mathcal{N}=2$ super-Bargmann algebra. Next, the curvature tensors are divided into conventional curvatures - these are the curvatures  that are set to zero in order to solve for the spin connections - and the remaining non-conventional curvatures. Setting one of the latter to zero corresponds to the geometric constraint that implies the presence of a notion of absolute time. A finite set of other (bosonic and fermionic) constraints is then obtained as follows. First, the supersymmetric variation of the absolute time constraint leads to two further constraints, one on a projection of the non-conventional gravitino curvature and one on the curvature of spatial rotations.  At the same time the commutator of two supersymmetries is considered and this leads to two further fermionic  constraints that can be considered as  equations of motion for the gravitino field. This shows that we are dealing with an on-shell supergravity multiplet. Next, making use of the Bianchi identity for the boost curvature, one can show that the supersymmetric variation of the two fermionic constraints leads to a singlet component of the boost curvature that, after gauge fixing to rectilinearly accelerating reference frames, gives the the Poisson equation for the Newton potential. There are no further constraints.

The second attempt to construct a $\mathcal{N}=2$ NC supergravity theory is based on effort 2., i.e. taking a non-Lorentzian limit of a 3$D$ on-shell relativistic $\mathcal{N}=2$ supergravity  theory consisting of a Dreibein and gravitino only \cite{Bergshoeff:2015uaa}.\,\footnote{See also references \cite{Zojer:2016aql,Zojer:2016uum}.} It even leads, in a second step,  to an off-shell version of such a non-Lorentzian supergravity theory.  The construction of \cite{Bergshoeff:2015uaa} uses  a second-order formulation where the spin connection fields have been solved by imposing conventional curvature constraints. It makes use of the same $\mathcal{N}=2$ super-Bargmann algebra as in \cite{Andringa:2013mma} but this time it is shown how this non-Lorentzian superalgebra can be viewed as a special In\"on\"u-Wigner contraction of a $\mathcal{N}=2$ Poincar\'e superalgebra with a central extension.  In the non-Lorentzian limit procedure, one introduces a U(1) gauge field for this central extension, even though it is not part of the relativistic supergravity multiplet. In order not to upset the counting of the bosonic and fermionic degrees of freedom its supercovariant curvature is set to zero by hand. The supersymmetric variation of this relativistic constraint leads to two more constraints (one fermionic and one bosonic) that can be considered as equations of motion.

To define a non-Lorentzian limit of the Dreibein, the gravitino and  the extra U(1) gauge field, a contraction is applied that is dual to the In\"on\"u-Wigner contraction of the generators these gauge fields are associated to. The constraint that was imposed on the additional U(1) gauge field serves two purposes. First, substituting this constraint, for finite contraction parameter, into the expression for the dependent spin connection fields, makes it possible to define the non-Lorentzian limit of these spin connections without divergences. The non-Lorentzian limit of the supersymmetry rules is then non-divergent and reproduces the NR rules of \cite{Andringa:2013mma}. Second, the non-Lorentzian limit of this constraint leads to the geometric constraint that endows the resulting NC geometry with an absolute time. Subsequent supersymmetric variations of this non-Lorentzian constraint, analysis of the supersymmetry algebra and making use of the NR Bianchi identities then shows that one obtains the same non-Lorentzian NC supergravity theory as in \cite{Andringa:2013mma} but this time by taking a non-Lorentzian limit.

The non-Lorentzian supergravity theories of \cite{Andringa:2013mma} and \cite{Bergshoeff:2015uaa} are based upon torsionless NC geometries with an absolute time constraint. As we will review in section \ref{sec:NLgeom}, relaxing this constraint requires the introduction of a particular type of torsion, called intrinsic torsion. Newton-Cartan geometries with non-zero intrinsic torsion occur both in applications to condensed matter physics \cite{Gromov:2014vla} as well as in applications to non-Lorentzian string theory \cite{Christensen:2013lma}. The generalization of NC supergravity to include intrinsic torsion was undertaken in \cite{Bergshoeff:2015ija}, by making use of effort 3., i.e. a NR superconformal tensor calculus that is based on a $\mathcal{N}=2$ Schr\"odinger superalgebra that contains dilatations and plays the role of the superconformal algebra in the relativistic case. The gauging of this Schr\"odinger superalgebra leads to an off-shell  Schr\"odinger supergravity multiplet which is the analogue of the relativistic Weyl multiplet. As was shown in \cite{Bergshoeff:2014uea}, this gauging naturally leads to a Schr\"odinger supergravity multiplet with intrinsic torsion. The torsion is provided by the spatial components of the dilatation gauge field, that depend on the other fields of the multiplet. For this reason, the NR superconformal tensor calculus naturally leads to torsionful NC supergravity theories. Following the superconformal tensor calculus, the Schr\"odinger supergravity multiplet is coupled to two different choices of compensator matter multiplets \cite{Bergshoeff:2015ija}: a NR 3$D$ $\mathcal{N}=2$ scalar multiplet and a NR limit of a $\mathcal{N}=2$ vector multiplet. This leads to a so-called old minimal and new minimal off-shell 3$D$ $\mathcal{N}=2$ NC supergravity multiplet.

The fourth approach to 3$D$ non-Lorentzian supergravity consists of constructing Chern-Simons formulations of gravity based on NR superalgebras with a non-degenerate bilinear form. The first example of such a Chern-Simons theory used an extension of the $\mathcal{N}=2$ Bargmann superalgebra \cite{Bergshoeff:2016lwr}, whose bosonic part was considered in \cite{Papageorgiou:2009zc}. This superalgebra leads to an extension of NC supergravity called extended Bargmann supergravity. This is consistent with the fact that it is believed that there is no action for NC supergravity based upon the standard Bargmann algebra.\,\footnote{For a proposal based upon a larger algebra, see \cite{Hansen:2019vqf}.}  The construction of an action for the extended theory is based upon the observation that, in order to write down an action for the single extra vector field that was added to general relativity in \cite{Bergshoeff:2015uaa}, one introduces two vector fields such that one can write down an action containing both vector fields. The underlying algebra is a Bargmann algebra with two central extensions instead of one. The supersymmetric extension of this extended Bargmann algebra was found by trial and error  \cite{Bergshoeff:2016lwr}. Remarkably, it contains beyond the expected two supersymmetry generators an extra fermionic generator.

The paper \cite{Bergshoeff:2016lwr} led to many follow-up papers that make use of the direct connection between a non-Lorentzian superalgebra\,\footnote{Non-Lorentzian superalgebras have been studied in \cite{Puzalowski:1978rv,deAzcarraga:1991fa,Bergman:1995zr} and \cite{Clark:1983ne,Leblanc:1992wu,Lukierski:2006tr}.} with a non-degenerate bilinear form and a
Chern-Simons formulation of a corresponding non-Lorentzian supergravity theory. For instance, in \cite{Ozdemir:2019orp} a 3$D$ supergravity version  was constructed of the 4$D$  non-Lorentzian gravity theory of \cite{Hansen:2019pkl} by using a certain extension of the Bargmann algebra that contains three additional generators. Further extended supergravity theories were constructed   based on different extensions of the $\mathcal{N}=2$ Bargmann superalgebra such as Extended Newton-Hooke supergravity, Exotic Bargmann supergravity, extended Lifshitz supergravity and extended Schr\"odinger supergravity \cite{Ozdemir:2019tby}. More examples on non-Lorentzian Chern-Simons supergravity theories, including extensions with a cosmological constant, were given in \cite{Concha:2019mxx,Concha:2020tqx,Concha:2020eam,Concha:2021jos, Concha:2021llq}

Finally, a fifth method to construct a non-Lorentzian supergravity theory is by applying so-called Lie algebra expansions. These expansions can be performed either using Maurer-Cartan equations or semigroups. This method has been applied to re-derive the 3$D$ extended Bargmann supergravity theory \cite{deAzcarraga:2019mdn}. 

Apart from 3$D$ this review also focuses on recent results in 10$D$. In particular, a non-Lorentzian version of 10$D$ minimal supergravity, whose underlying structure is given by stringy NC geometry, has recently been constructed as a non-Lorentzian limit \cite{Bergshoeff:2021tfn}. In this case, terms in the supersymmetry transformation rules of the fermions that diverge in the limit, can be consistently eliminated by imposing particular geometric constraints. These constraints are invariant under the NR supersymmetry transformation rules that result from the limit. Applying the limit to the equations of motion of relativistic 10$D$ minimal supergravity then leads to a set of NR equations of motion that consistently transform into each other under NR supersymmetry, once the geometric constraints are taken into account. After taking the limit, the theory is characterized by an emerging an-isotropic scale symmetry, along with accompanying fermionic Stueckelberg symmetries. These symmetries imply that the NR supergravity multiplet that the limit leads to is shortened in comparison to its relativistic counterpart.

This review is organized as follows. In section \ref{sec:LorGeom} we first give a review of some basic notions in Lorentzian geometry to contrast them with a discussion of NC and stringy NC geometry in section \ref{sec:NLgeom}. Next, in section 3, we discuss the construction and realization of $\mathcal{N}=2$ NC supergravity in three spacetime dimensions. In particular, we show how, after partially gauge fixing its local symmetries, this gives rise to a supersymmetric version of Newtonian gravity. Subsequently, in section 4 we discuss NR 3$D$ Chern-Simons supergravity theories. We focus on the theory of \cite{Bergshoeff:2016lwr} that is based on an extension of the Bargmann algebra with 3 supercharges in particular. We also review how this algebra can be viewed as a Lie algebra expansion of the 3$D$ $\mathcal{N}=2$ super-Poincar\'e algebra \cite{deAzcarraga:2019mdn}. In section 5, we outline how a NR version of 10$D$ minimal supergravity with an underlying stringy NC geometry arises as a non-Lorentzian limit of relativistic 10$D$ minimal supergravity. Finally, in section 6, we give a brief outlook on future developments.

\section{Lorentzian and non-Lorentzian Geometry}

In both relativistic and non-Lorentzian gravity, the gravitational interaction is modelled as the effect of matter moving in and curving the geometry of space-time. This geometry includes the specification of a metric and affine connection structure on an underlying space-time manifold and is described in a diffeomorphism covariant manner. The metric structure prescribes how to measure space-time distances, while the affine connection is used to define covariant derivatives of tensor fields and the ensuing notion of parallel transport along curves. In this section, we will review how these geometric ingredients are specified in non-Lorentzian geometry, in a manner that is adapted to their use in non-Lorentzian supergravity. To emphasize some important differences with Lorentzian geometry, we will first give a brief overview of aspects of the latter that are relevant for supergravity.

\subsection{Lorentzian geometry} \label{sec:LorGeom}

Relativistic gravity and supergravity use Lorentzian geometry, whose metric structure is given by a symmetric two-tensor field $g_{\mu\nu}$ (with the coordinate indices $\mu$, $\nu$, $\cdots = 0, \cdots, D-1$, where $D$ is the space-time dimension). This metric $g_{\mu\nu}$ is non-degenerate, i.e., constitutes an invertible matrix and one denotes its inverse by $g^{\mu\nu}$. The affine connection is a connection on the tangent bundle of the manifold and corresponds to a three-index field $\Gamma^\rho_{\mu\nu}$. In Lorentzian geometry, it satisfies the following metric compatibility condition:
\begin{equation}
  \label{eq:metriccomp}
  \partial_\mu g_{\nu\rho} - \Gamma^\sigma_{\mu\nu} g_{\sigma\rho} - \Gamma^\sigma_{\mu\rho} g_{\nu\sigma} = 0 \,,
\end{equation}
that ensures that the length of a vector does not change under parallel transport. Splitting $\Gamma^\rho_{\mu\nu}$ in a symmetric part $\Gamma^\rho_{(\mu\nu)}$ and an anti-symmetric part $T^\rho_{\mu\nu} \equiv 2 \Gamma^\rho_{[\mu\nu]}$, the $D^2(D+1)/2$ equations contained in \eqref{eq:metriccomp} then express that the equal number of components of $\Gamma^\rho_{(\mu\nu)}$ are not independent but can instead be written in terms of the metric $g_{\mu\nu}$ and $T^\rho_{\mu\nu}$ as follows:
\begin{equation}
  \label{eq:GammagT}
  \Gamma^\rho_{(\mu\nu)} = \frac12 g^{\rho\sigma} \left(2 \partial_{(\mu} g_{\nu)\sigma} - \partial_\sigma g_{\mu\nu} \right) + g^{\rho\tau} T^\sigma_{\tau(\mu} g_{\nu)\sigma} \,.
\end{equation}
The anti-symmetric part $T^\rho_{\mu\nu}$ is called the torsion tensor, reflecting the fact that (unlike $\Gamma^\rho_{(\mu\nu)}$) it transforms tensorially under general coordinate transformations. Together with $g_{\mu\nu}$, $T^\rho_{\mu\nu}$ forms the independent data that define Lorentzian geometry. In general relativity, the torsion tensor is set to zero and the corresponding affine connection is called the Levi-Civita connection. Relativistic supergravity by contrast contains non-zero torsion, constructed out of gravitino bilinears.

While the affine connection suffices to couple gravity to bosonic matter fields that are sections of tensor products of the tangent bundle and its dual, it can not be used to couple to sections of the spinor bundle, i.e., fermion fields. For this reason, in relativistic supergravity one needs to use a different description of Lorentzian geometry that can accommodate fermions and that is called the Cartan or Vielbein formulation. Its basic variable is the so-called Vielbein one-form $E^{\hat{A}} = E_\mu{}^{\hat{A}} \rmd x^\mu$, where the index $\hat{A}$ can take the values $0,1,\cdots, D-1$. Viewed as a matrix, $E_\mu{}^{\hat{A}}$ is assumed to be invertible and its inverse is denoted by $E_{\hat{A}}{}^\mu$. The Vielbein $E_\mu{}^{\hat{A}}$ corresponds to a square root of the metric:
\begin{equation} \label{eq:metricVielb}
  g_{\mu\nu} = E_\mu{}^{\hat{A}} E_\nu{}^{\hat{B}} \eta_{\hat{A}\hat{B}} \,.
\end{equation}
Vielbeine that are related by the following infinitesimal action of local Lorentz transformations (with parameter $\Lambda^{\hat{A}\hat{B}} = -\Lambda^{\hat{B}\hat{A}}$)
\begin{equation}
  \label{eq:localLorE}
  \delta E_\mu{}^{\hat{A}} =  \Lambda^{\hat{A}}{}_{\hat{B}} E_\mu{}^{\hat{B}} \,,
\end{equation}
give rise to the same metric and should be seen as physically indistinguishable. The local Lorentz transformations \eqref{eq:localLorE} thus ought to be implemented as a gauge symmetry in the description of the geometry. To this end, the Cartan formulation not only introduces an affine connection $\Gamma^\rho_{\mu\nu}$, but also a gauge connection for local Lorentz transformations, i.e, a one-form field $\Omega_\mu{}^{\hat{A}\hat{B}} = -\Omega_\mu{}^{\hat{B}\hat{A}}$, that transforms as follows:
\begin{equation}
  \label{eq:localLorOm}
  \delta \Omega_\mu{}^{\hat{A}\hat{B}} = \partial_\mu \Lambda^{\hat{A}\hat{B}} + 2 \Lambda^{[\hat{A}|\hat{C}|} \Omega_{\mu \hat{C}}{}^{\hat{B}]} \,.
\end{equation}
This field $\Omega_\mu{}^{\hat{A}\hat{B}}$ is referred to as the `spin connection'. Along with $\Gamma^\rho_{\mu\nu}$, it is constrained to obey the so-called Vielbein postulate:
\begin{equation}
  \label{eq:Vielbpost}
  \partial_\mu E_\nu{}^{\hat{A}} - \Omega_\mu{}^{\hat{A}\hat{B}} E_{\nu \hat{B}} - \Gamma^\rho_{\mu\nu} E_{\rho}{}^{\hat{A}} = 0 \,.
\end{equation}
Using \eqref{eq:metricVielb} and the anti-symmetry of $\Omega_\mu{}^{\hat{A}\hat{B}}$ in the $[\hat{A}\hat{B}]$ indices, one sees that this postulate implies the metric compatibility condition \eqref{eq:metriccomp} on the affine connection $\Gamma^\rho_{\mu\nu}$, so that the Cartan formulation reproduces the expression \eqref{eq:GammagT} for $\Gamma^\rho_{(\mu\nu)}$. Anti-symmetrizing \eqref{eq:Vielbpost} in the $[\mu\nu]$ indices leads to the following equation
\begin{equation}
  \label{eq:Lorconvconstr}
 R_{\mu\nu}{}^{\hat{A}} (P) \equiv 2 \partial_{[\mu} E_{\nu]}{}^{\hat{A}} - 2 \Omega_{[\mu}{}^{\hat{A}\hat{B}} E_{\nu] \hat{B}} =  T^\rho_{\mu\nu} E_{\rho}{}^{\hat{A}} \,.
\end{equation}
Viewing this as a set of $D^2(D-1)/2$ linear equations for the equal number of components of $\Omega_\mu{}^{\hat{A}\hat{B}}$, one finds that $\Omega_\mu{}^{\hat{A}\hat{B}}$ is not independent, but is instead fully determined in terms of $E_\mu{}^{\hat{A}}$ and the torsion tensor:
\begin{align}
  \label{eq:exprOmega}
  \Omega_\mu{}^{\hat{A}\hat{B}} &= -2 E^{[\hat{A}|\nu} \partial_{[\mu} E_{\nu]}{}^{\hat{B}]} + E^{\hat{A}\nu} E^{\hat{B}\rho} E_{\mu \hat{C}} \partial_{[\nu} E_{\rho]}{}^{\hat{C}} - \frac12 E^{\hat{A}\nu} E^{\hat{B}\rho} T^\sigma_{\nu\rho} g_{\sigma \mu} \nonumber \\ & \qquad + E^{[\hat{A}|\nu|} E_\rho{}^{\hat{B}]} T_{\mu\nu}^\rho \,.
\end{align}
In this way, the Cartan formulation of Lorentzian geometry defines metric, affine and spin connection structures in terms of an independent Vielbein and torsion tensor. The dependent spin connection \eqref{eq:exprOmega} gives a connection on the spinor bundle that allows one to define covariant derivatives on fermionic fields and couple them to gravity. For instance, on an ordinary spinor field $\psi$, this covariant derivative is given by
\begin{equation}
  \label{eq:spinorcovder}
  D_\mu \psi = \partial_\mu \psi - \frac14 \Omega_\mu{}^{\hat{A}\hat{B}} \gamma_{\hat{A}\hat{B}} \psi \,.
\end{equation}

In the supergravity literature, the Cartan formulation of Lorentzian geometry is often referred to as a `gauging of the Poincar\'e algebra' \cite{Chamseddine:1976bf,MacDowell:1977jt}. This terminology stems from the fact that the transformation rules \eqref{eq:localLorE}, \eqref{eq:localLorOm} are naturally interpreted as gauge transformation rules under Lorentz transformations, if one would naively gauge the Poincar\'e algebra. From this viewpoint, $E_\mu{}^{\hat{A}}$ and $\Omega_\mu{}^{\hat{A}\hat{B}}$ are interpreted as the components of a Poincar\'e algebra-valued gauge field along the translation and Lorentz transformation generators respectively. The two-form fields $R_{\mu\nu}{}^{\hat{A}}(P)$, defined in equation \eqref{eq:Lorconvconstr}, are then similarly interpreted as the gauge-covariant curvature components along the translation generators. In the gauging language equation \eqref{eq:Lorconvconstr} is often referred to as a `conventional constraint'. Conventional constraints are constraints on gauge-covariant curvature components that can be used to express a dependent field (such as the spin connection) in terms of other independent ones. They should be contrasted with what we will call geometric constraints that are constraints on the independent fields. Note that the term `gauging of the Poincar\'e algebra' should not be taken too literally, since $\Omega_\mu{}^{\hat{A}\hat{B}}$ is not an independent field, as would be the case in ordinary gauge theory. Nevertheless, thinking about Lorentzian geometry in gauge theoretic terms is useful for various generalizations, such as to supergravity and non-Lorentzian geometry and we will frequently use this language in this review.

\subsection{Non-Lorentzian Geometry} \label{sec:NLgeom}

Non-Lorentzian geometry refers to differential geometric frameworks for space-times, whose local symmetry group differs from the Lorentz group. In this review, we will restrict ourselves to NR symmetry groups, that arise in or are extensions of $c \rightarrow \infty$ limits of the Lorentz group. The prime example of such a non-Lorentzian geometry is given by NC geometry \cite{Cartan1,Cartan2} that features local Galilean symmetries instead of the local Lorentz symmetries of Lorentzian geometry. Unlike Lorentzian geometry, NC geometry features two degenerate metrics. As a consequence, the structure of metric compatible connections (with or without torsion) also differs from that in Lorentzian geometry in crucial respects. In this section, we will first review (torsionful) NC geometry in subsection \ref{ssec:NCgeometry}. This is the relevant geometry for the 3$D$ supergravity theories that we will discuss in sections \ref{sec:3Dsugra} and \ref{sec:EBGexp}. The 10$D$ supergravity theory of section \ref{sec:10Dsugra} on the other hand uses a recent generalization of NC geometry, called `stringy Newton-Cartan geometry', whose essential features will be reviewed in subsection \ref{ssec:pbraneNCgeometry}. This section is mostly based on \cite{Bergshoeff:2022fzb} to which we refer for further details and references.

\subsubsection{Newton-Cartan geometry} \label{ssec:NCgeometry}

A NC geometry is described by a $D$-dimensional differentiable manifold $\mathcal{M}$ (with local coordinates $x^\mu$, $\mu = 0, \cdots, D-1$) with particular degenerate metric and metric compatible connection structures. Here, we will review NC geometry in a frame field formulation that was developed in \cite{Duval:1984cj}. This formulation can be viewed as a gauging of the Bargmann algebra \cite{Andringa:2010it}, i.e., the centrally extended Galilei algebra, and this is the language we will adopt here.

In $D$ space-time dimensions, the generators of the Bargmann algebra consist of the time translation $H$, spatial translations $P_a$, Galilean boosts $G_a$, spatial rotations $J_{ab}$ and the central charge $M$, where the spatial indices $a$, $b$ assume values from 1 to $D-1$. The non-trivial commutation relations of the Bargmann algebra are given by:
       \begin{alignat}{3} \label{eq:Bargmann}
     [J_{ab}, P_c] &= -2 \delta_{c[a} P_{b]} \,, \qquad & [J_{ab}, G_c] &= -2 \delta_{c[a} G_{b]} \,, \qquad & [G_a, H] &= -P_a \,, \nonumber \\
     [J_{ab},J_{cd}] &= 4 \delta_{[a[c} J_{d]b]} \,, \qquad & [G_a, P_b] &= -\delta_{ab} M \,.
   \end{alignat}
   The central charge $M$ is physically interpretated as the Noether charge that corresponds to mass or particle number conservation in NR theories. Its inclusion in the gauging procedure is warranted if one wants to couple massive particles or fields to NC geometry.

   In the first step of the gauging procedure, one introduces a Bargmann algebra-valued gauge field, whose components along the various algebra generators are denoted by $\tau_\mu$, $e_\mu{}^a$, $m_\mu$, $\omega_\mu{}^{ab}$ and $\omega_\mu{}^a$ as outlined in table \ref{tab:Bargmannfields}.
   \begin{table}[t]
\begin{center}
\begin{tabular}{|c|c|c|}
\hline
symmetry&generators& gauge field\\[.1truecm]
  \hline
  time translation&$H$&$\tau_\mu$\\[.1truecm]
  spatial translations&$P_a$&$e_\mu{}^a$\\[.1truecm]
  central charge&$M$&$m_\mu$\\[.1truecm]
  spatial rotations&$J_{ab}$&$\omega_\mu{}^{ab}$\\[.1truecm]
  Galilean boosts&$G_{a}$&$\omega_\mu{}^{a}$\\[.1truecm]
\hline
\end{tabular}
\caption{Summary of the gauge fields introduced in the gauging of the Bargmann algebra.}
\label{tab:Bargmannfields}
\end{center}
\end{table}
These fields transform as one-forms under general coordinate transformations. Their gauge transformation rules under local spatial rotations (with parameter $\lambda^{ab}$), local Galilean boosts (with parameter $\lambda^a$) and the local central charge transformation (with parameter $\sigma$) are determined by the structure constants of the Bargmann algebra and are given by:
    \begin{align}
     \delta \tau_\mu &= 0\,,  \qquad
     \delta e_\mu{}^a = \lambda^{ab} e_{\mu b} + \lambda^a \tau_\mu \,,  \qquad \delta m_\mu = \partial_\mu \sigma + \lambda^a e_{\mu a} \,, \label{eq:deltatauem} \\
     \delta \omega_\mu{}^{ab} &= \partial_\mu \lambda^{ab} + 2 \lambda^{[a|c|} \omega_{\mu c}{}^{b]} \,, \qquad
     \delta \omega_\mu{}^a = \partial_\mu \lambda^a - \omega_\mu{}^{a b} \lambda_b + \lambda^{ab} \omega_{\mu b} \,. \label{eq:deltaomegas}
    \end{align}
    Here and in the following, we have freely raised and lowered the flat spatial indices $a$, $b=1,\cdots, D-1$ with Kronecker deltas $\delta^{ab}$, $\delta_{ab}$. For future reference, we also note that the field strengths of $\tau_\mu$, $e_\mu{}^a$, $m_\mu$, $\omega_\mu{}^{ab}$ and $\omega_\mu{}^a$ that are covariant with respect to the transformations of \eqref{eq:deltatauem}, \eqref{eq:deltaomegas} are given by:
       \begin{align} \label{eq:bargmanncurvs}
     R_{\mu\nu}(H) &\equiv 2 \partial_{[\mu} \tau_{\nu]} \,, \nonumber \\
     R_{\mu\nu}{}^a(P) &\equiv 2 \partial_{[\mu} e_{\nu]}{}^a - 2 \omega_{[\mu}{}^{ab} e_{\nu]b} - 2 \omega_{[\mu}{}^a \tau_{\nu]} \,, \nonumber \\
         R_{\mu\nu}(M) &\equiv 2 \partial_{[\mu} m_{\nu]} - 2 \omega_{[\mu}{}^a e_{\nu] a} \,, \nonumber \\ R_{\mu\nu}{}^{ab}(J) &\equiv 2 \partial_{[\mu} \omega_{\nu]}{}^{ab} - 2 \omega_{[\mu}{}^{[a|c|} \omega_{\nu] c}{}^{b]} \,, \nonumber \\
         R_{\mu\nu}{}^a(G) &\equiv 2 \partial_{[\mu} \omega_{\nu]}{}^a - 2 \omega_{[\mu}{}^{ab} \omega_{\nu]b} \,.
   \end{align}

  The field $m_\mu$ is the gauge field associated to the central charge $M$. In accordance to the physical interpretation of $M$, mentioned below \eqref{eq:Bargmann}, $m_\mu$ couples to conserved mass or particle number currents in theories of massive particles or fields in an arbitrary NC background. The one-form $\tau_\mu$ is referred to as the time-like Vielbein or clock form, whereas $e_\mu{}^a$ is referred to as the spatial Vielbein. Even though they do not constitute square invertible matrices, one can still define vectors $\tau^\mu$ and $e_a{}^\mu$ that constitute `inverse Vielbein fields' in the sense that the following relations hold:
      \begin{align}
     & \tau^\mu \tau_\mu = 1 \,, \qquad \qquad \tau^\mu e_\mu{}^a = 0 \,, \qquad \qquad \tau_\mu e_a{}^\mu = 0 \,, \nonumber \\
     & e_a{}^\mu e_\mu{}^b = \delta_a^b \,, \qquad \qquad \tau_\mu \tau^\nu + e_\mu{}^a e_a{}^\nu = \delta_\mu^\nu \,.
      \end{align}
      These inverse Vielbein fields transform as follows under local spatial rotations and Galilean boosts:
      \begin{align} \label{eq:invrules}
        \delta \tau^\mu &= - \lambda^a e_a{}^\mu \,, \qquad \qquad \delta e_a{}^\mu =  \lambda_a{}^b e_b{}^\mu \,.
      \end{align}
      The metric structure of NC geometry is then defined in analogy to Lorentzian geometry by Galilean invariants that are quadratic in the Vielbeine or their inverses. Two such invariants can be found:
      \begin{enumerate}
      \item A covariant symmetric 2-tensor of rank 1, called the `time-like metric'
        \begin{align}
          \tau_{\mu\nu} = \tau_\mu \tau_\nu \,.
        \end{align}
      \item A contravariant symmetric 2-tensor of rank $D-1$, called the `spatial (co-)metric'
        \begin{align}
          h^{\mu\nu} = e_a{}^\mu e_b{}^\nu \delta^{ab} \,.
        \end{align}
      \end{enumerate}
      These two metrics are mutually orthogonal in the sense that $h^{\mu\nu} \tau_{\nu \sigma} = 0$. They can be used to measure time-like and spatial distances along particular curves in the space-time. To see this, one first notes that one can use the time-like Vielbein $\tau_\mu$ to distinguish vectors into time-like and spatial ones. In particular, a vector $X^\mu$ is called time-like future (resp. past) directed whenever $\tau_\mu X^\mu > 0$ (resp. $< 0$) and spatial whenever $\tau_\mu X^\mu = 0$. A curve $\gamma \ : \ t \in [0,1] \ \mapsto \ x^\mu(t) \in \mathcal{M}$, whose tangent vectors $\dot{x}^\mu(t) \equiv \rmd x^\mu(t)/\rmd t$ are everywhere time-like future directed, can be regarded as the worldline of a physical observer that moves between two space-time points $x^\mu(0)$ and $x^\mu(1)$. The time interval $\Delta t$, needed by the observer to complete its motion between these two points, is then defined as
      \begin{align} \label{eq:timeint}
        \Delta t \equiv \int_0^1 \rmd t \, \sqrt{\dot{x}^\mu \dot{x}^\nu \tau_{\mu\nu}} = \int_0^1 \rmd t \, \dot{x}^\mu \tau_\mu = \int_\gamma \rmd x^\mu \tau_\mu \,.
      \end{align}
      Likewise, the spatial distance $\ell$ along a curve $\sigma \ : \ s \in [0,1] \ \mapsto \ x^\mu(s) \in \mathcal{M}$, whose tangent vectors $x^{\prime\, \mu} (s) \equiv \rmd x^\mu(s)/\rmd s$ are everywhere spatial, is defined as
      \begin{align} \label{eq:spatialint}
        \ell \equiv \int_0^1 \rmd s \, \sqrt{x^{\prime\, \mu} x^{\prime\, \nu} h_{\mu\nu}} \,, \qquad \qquad \text{with} \quad h_{\mu\nu} = e_\mu{}^a e_\nu{}^b \delta_{ab} \,.
      \end{align}
      Note that $h_{\mu\nu}$ satisfies
      \begin{align}
        h^{\mu \nu} h_{\nu\rho} = \delta^\mu_\rho - \tau^\mu \tau_\rho \,,
      \end{align}
      and thus is a right inverse of the spatial co-metric $h^{\mu\nu}$, when restricting its action to spatial vectors. This right inverse is however not boost invariant:
      \begin{align}
        \delta h_{\mu\nu} = 2 \lambda^a \tau_{(\mu} e_{\nu) a} \,.
      \end{align}
      Nevertheless, since the integral on the right-hand-side of \eqref{eq:spatialint} is along a curve that satisfies $\tau_\mu x^{\prime\, \mu} = 0$, it is boost invariant.

      Having discussed the metric structure on a NC geometry, let us now turn to the definition of metric compatible connections. We have already introduced two one-form fields $\omega_\mu{}^{ab}$ and $\omega_\mu{}^a$, with gauge transformation rules \eqref{eq:deltaomegas}, that can play the role of a NR analogue of the spin connection $\Omega_\mu{}^{\hat{A}\hat{B}}$. We will refer to $\omega_\mu{}^{ab}$ and $\omega_\mu{}^a$ as spin connections for spatial rotations and Galilean boosts respectively. As in the Lorentzian case, these spin connections should not be independent but should instead depend on the other frame fields $\tau_\mu$, $e_\mu{}^a$, $m_\mu$ as well as torsion. This is achieved by requiring that the following conventional constraints, that are NR analogues of \eqref{eq:Lorconvconstr}, hold identically:
      \begin{align} \label{eq:convBargconstraints}
        R_{\mu\nu}(P^a) &\equiv 2 \partial_{[\mu} e_{\nu]}{}^a - 2 \omega_{[\mu}{}^{ab} e_{\nu]b} - 2 \omega_{[\mu}{}^a \tau_{\nu]} = T_{\mu\nu}{}^a \,, \nonumber \\
         R_{\mu\nu}(M) &\equiv 2 \partial_{[\mu} m_{\nu]} - 2 \omega_{[\mu}{}^a e_{\nu] a} = T^{(m)}_{\mu\nu} \,.
      \end{align}
      Here, $T_{\mu\nu}{}^a$ and $T^{(m)}_{\mu\nu}$ are two torsion tensors that we will call the `spatial torsion tensor' and `mass torsion tensor' respectively. One can view \eqref{eq:convBargconstraints} as a set of $D(D-1)^2/2 + D(D-1)/2 = D^2(D-1)/2$ linear algebraic equations for as many components of $\omega_\mu{}^{ab}$ and $\omega_\mu{}^a$. Solving these equations then yields the following expressions for the spin connections in terms of $\tau_\mu$, $e_\mu{}^a$, $m_\mu$, $T_{\mu\nu}{}^a$ and $T^{(m)}_{\mu\nu}$:
      \begin{align}
  \label{eq:spinconnexpr}
  \omega_\mu{}^{a} &= \tau_\mu \tau^\nu e^{a \rho} \partial_{[\nu} m_{\rho]} + e^{a \nu} \partial_{[\mu} m_{\nu]} + e_{\mu b} e^{a \nu} \tau^\rho \partial_{[\nu} e_{\rho]}{}^b + \tau^\nu \partial_{[\mu} e_{\nu]}{}^a \nonumber \\ & \qquad - \tau_\mu \tau^\nu e^{a\rho} T^{(m)}_{\nu\rho} + e_{\mu b} \tau^\nu e^{(a|\rho|} T_{\nu\rho}{}^{b)} - \frac12 e_{\mu b} e^{b \nu} e^{a\rho} T^{(m)}_{\nu\rho} \,, \nonumber \\
  \omega_\mu{}^{ab} &= - 2 e^{[a|\nu|} \partial_{[\mu} e_{\nu]}{}^{b]} + e_{\mu c} e^{a\nu} e^{b\rho} \partial_{[\nu} e_{\rho]}{}^c - \tau_\mu e^{a\nu} e^{b\rho} \partial_{[\nu} m_{\rho]} \nonumber \\ & \qquad + \frac12 \tau_\mu e^{a\nu} e^{b\rho} T^{(m)}_{\nu\rho} + e^{[a|\nu|}  T_{\mu\nu}{}^{b]} - \frac12 e_{\mu c} e^{a\nu} e^{b\rho} T_{\nu\rho}{}^c \,.
      \end{align}
      Two comments are in order. First, note that the inclusion of the central charge of the Bargmann algebra in the gauging procedure is crucial to ensure that all spin connection components can be expressed in terms of other fields. Without the central charge and its associated gauge field $m_\mu$, one would not be able to impose the second conventional constraint of \eqref{eq:convBargconstraints} and the system of equations for the spin connection components would be underdetermined. One thus sees that, while $m_\mu$ does not play a role in defining the metric structure, it has a geometric significance as an ingredient that determines the connection structure of NC geometry.

      As a second comment, we remark that it is convenient to choose the torsion tensors $T_{\mu\nu}{}^a$ and $T^{(m)}_{\mu\nu}$ such that they transform under local spatial rotations and boosts as follows:
      \begin{align} \label{eq:deltaTs}
        \delta T_{\mu\nu}{}^a = \lambda^a{}_b T_{\mu\nu}{}^b + 2 \lambda^a \partial_{[\mu} \tau_{\nu]} \,, \qquad \quad \delta T^{(m)}_{\mu\nu} = \lambda_a T_{\mu\nu}{}^a \,.
      \end{align}
      This ensures that the local spatial rotation and boost transformations (induced by \eqref{eq:deltatauem} and \eqref{eq:deltaTs}) of the expressions \eqref{eq:spinconnexpr} for the spin connections coincide with the rules \eqref{eq:deltaomegas} that are dictated by the Bargmann algebra. This can be checked either by direct calculation or by noting that the set of equations \eqref{eq:convBargconstraints} is invariant under the spatial rotation and boost transformation rules of \eqref{eq:deltatauem}, \eqref{eq:deltaomegas} and \eqref{eq:deltaTs}. In the following, we will assume that \eqref{eq:deltaTs} hold. The fact that $T_{\mu\nu}{}^a$ transforms under boosts to $\partial_{[\mu} \tau_{\nu]}$ indicates that the latter should also be interpreted as torsion. This will be confirmed in the following.

      Having discussed the spin connections $\omega_\mu{}^{ab}$ and $\omega_\mu{}^a$, one can define an affine connection $\Gamma^\rho_{\mu\nu}$ in analogy to the Lorentzian case, by imposing the following Vielbein postulates:
      \begin{align}
  \label{eq:VielbpostB}
  & \partial_\mu \tau_\nu - \Gamma_{\mu\nu}^\rho \tau_\rho = 0 \,, \qquad \qquad
   \partial_\mu e_\nu{}^a - \omega_\mu{}^{ab} e_{\nu b} - \omega_\mu{}^a \tau_\nu - \Gamma_{\mu\nu}^\rho e_{\rho}{}^a = 0 \,,
      \end{align}
      where $\omega_\mu{}^{ab}$ and $\omega_\mu{}^a$ are understood to be given by the expressions \eqref{eq:spinconnexpr}. It follows that $\Gamma^\rho_{\mu\nu}$ is compatible with the two metrics $\tau_{\mu\nu}$ and $h^{\mu\nu}$:
\begin{align}
  \label{eq:metricscomp}
  \nabla_\mu \tau_{\nu\rho } & \equiv \partial_\mu \tau_{\nu\rho} - \Gamma_{\mu\nu}^\sigma \tau_{\sigma \rho} - \Gamma_{\mu\rho}^\sigma \tau_{\nu \sigma} = 0 \,, \nonumber \\ \nabla_\mu h^{\nu\rho} & \equiv \partial_\mu h^{\nu\rho} + \Gamma_{\mu\sigma}^\nu h^{\sigma \rho} + \Gamma_{\mu\sigma}^\rho h^{\nu\sigma}= 0 \,.
\end{align}
From \eqref{eq:VielbpostB} and \eqref{eq:spinconnexpr}, one finds that $\Gamma^\rho_{\mu\nu}$ can be written in terms of the NC metric structure, $m_\mu$ and the torsion tensors $T_{\mu\nu}{}^a$, $T^{(m)}_{\mu\nu}$ as follows:
\begin{align}
  \label{eq:Gammaexpr}
  \Gamma_{\mu\nu}^\rho &= \tau^\rho \partial_{\mu} \tau_{\nu} + \frac12 h^{\rho \sigma} \left( \partial_\mu h_{\sigma \nu} + \partial_\nu h_{\mu\sigma} - \partial_\sigma h_{\mu\nu} \right) + h^{\rho\sigma} \tau_\mu \partial_{[\sigma} m_{\nu]} + h^{\rho\sigma} \tau_\nu \partial_{[\sigma} m_{\mu]} \nonumber \\
  & \qquad + h^{\rho\sigma} \tau_{(\mu} T^m_{\nu)\sigma} - h^{\rho\sigma} e_{(\mu|a|} T_{\nu)\sigma}{}^a + \frac12 e_a{}^\rho T_{\mu\nu}{}^a \,.
\end{align}
One can explicitly check that this formula for $\Gamma^\rho_{\mu\nu}$ is invariant under local spatial rotations and boosts (even though boost invariance is not manifest)\footnote{Note that $\Gamma^\rho_{\mu\nu}$ will only be invariant under local central charge transformations, provided $T_{\mu\nu}{}^a$ and $T^{(m)}_{\mu\nu}$ are. In applications of NC geometry with torsion, e.g. in Lifshitz holography \cite{Christensen:2013lma,Christensen:2013rfa}, one often encounters situations in which it is not possible to choose $T_{\mu\nu}{}^a$ and $T^{(m)}_{\mu\nu}$ such that they are simultaneously invariant under the central charge and transform as in \eqref{eq:deltaTs} under local spatial rotations and boosts. In those cases, it is conventional/convenient to choose $T_{\mu\nu}{}^a$ and $T^{(m)}_{\mu\nu}$ with non-trivial central charge transformations, but such that \eqref{eq:deltaTs} still hold, and to work with an affine connection $\Gamma^\rho_{\mu\nu}$ that is boost but not central charge invariant.}. This is guaranteed by the fact that the dependent spin connections \eqref{eq:spinconnexpr} transform as in \eqref{eq:deltaomegas}.

The formula \eqref{eq:Gammaexpr} gives the affine connection of NC geometry with arbitrary torsion. In particular, we see that its torsion $2\Gamma^\rho_{[\mu\nu]}$ is given by
\begin{align} \label{eq:torsiondecomp}
  & 2 \Gamma^\rho_{[\mu\nu]} = 2 \tau^\rho \partial_{[\mu} \tau_{\nu]} + e_a{}^\rho T_{\mu\nu}{}^a \,.
\end{align}
The time-like component $2 \Gamma^\rho_{[\mu\nu]} \tau_\rho$ is (as anticipated after \eqref{eq:deltaTs}) given by $2 \partial_{[\mu} \tau_{\nu]}$, while $T_{\mu\nu}{}^a$ gives the spatial components $2 \Gamma^\rho_{[\mu\nu]} e_\rho{}^a$. The time-like component $2 \partial_{[\mu} \tau_{\nu]}$ of the affine connection torsion is often called `intrinsic torsion' \cite{Figueroa-OFarrill:2020gpr}, since it does not appear in the dependent spin connections \eqref{eq:spinconnexpr}. Note that the notion of intrinsic torsion is absent in Lorentzian geometry, where the dependent spin connection \eqref{eq:exprOmega} contains all components of $2 \Gamma_{[\mu\nu]}^\rho$. A further difference between Lorentzian and NC geometry is that the latter features an extra mass torsion tensor $T^{(m)}_{\mu\nu}$ that appears in the spin connections \eqref{eq:spinconnexpr} but not in the affine connection.

It is interesting to consider situations in which the torsion is no longer arbitrary, but instead some of its components vanish. For instance, since the intrinsic torsion components $2 \Gamma^\rho_{[\mu\nu]} \tau_\rho$ are boost and rotation invariant, they can be consistently set to zero. This leads to the differential constraint
\begin{align} \label{eq:abstime}
  \partial_{[\mu} \tau_{\nu]} = 0 \,,
\end{align}
on the clock form. This illustrates another important difference with Lorentzian geometry, where truncating torsion tensor components does not give rise to constraints on the metric structure. The constraint \eqref{eq:abstime} has a natural physical interpretation: it covariantly expresses the existence of an absolute time on the space-time. Indeed, if \eqref{eq:abstime} holds, Stokes' theorem implies that two observers that move along different worldlines between the same initial and final space-time points, measure the same time interval \eqref{eq:timeint} needed for their respective journeys. For modern applications of NC geometry, the absolute time constraint \eqref{eq:abstime} is often too stringent and needs to be relaxed. This is for instance the case in Lifshitz holography, where gravity around so-called Lifshitz space-times is conjectured as a holographic dual of particular NR conformal field theories. The boundary of a Lifshitz space-time is described by a NC geometry, whose metric structure is only determined up to an anisotropic local Weyl rescaling with parameter $\Lambda_D$ \cite{Christensen:2013lma,Christensen:2013rfa}:
\begin{align} \label{eq:aniWeyl}
  \delta \tau_\mu = z\, \Lambda_D \tau_\mu \,, \qquad \qquad \delta e_\mu{}^a = \Lambda_D e_\mu{}^a \,,
\end{align}
where $z$ is a real number, called the dynamical exponent. The type of NC geometry that appears in Lifshitz holography can not obey the constraint \eqref{eq:abstime}, as the latter is not invariant under the Weyl rescaling \eqref{eq:aniWeyl}. A weaker constraint that is both boost and Weyl invariant and that can thus feature in Lifshitz holography is the so-called `twistless torsional' constraint in which only the spatial components\footnote{Similar to how curved indices are turned into flat ones in Lorentzian geometry, we will call the spatial components $X_a$/$X^a$ of a one-form $X_\mu$/vector $X^\mu$ (and by extension of tensors) those that are obtained by contraction with $e_a{}^\mu$/$e_\mu{}^a$, according to the rule $X_a \equiv e_a{}^\mu X_\mu$/$X^a \equiv e_\mu{}^a X^\mu$. Likewise, the time-like components $X_0$/$X^0$ are defined as $X_0 \equiv \tau^\mu X_\mu$/$X^0 \equiv \tau_\mu X^\mu$. Since we allow the $a$-index to be raised and lowered with Kronecker deltas, the notation $X^a$/$X_a$ can also stand for $e^{\mu a} X_\mu$/$e_{\mu a} X^\mu$, depending on the context. It is also convenient to raise and lower the $0$-index at the expense of a minus sign, so that $X^0$/$X_0$ can also be used to denote $-X_0 = - \tau^\mu X_\mu$/$-X^0 = -\tau_\mu X^\mu$, depending on the context.} of the intrinsic torsion are set to zero:
\begin{align}
  \label{eq:TTNC}
  \tau_{[\mu} \partial_\nu \tau_{\rho]} = 0 \qquad \qquad \Leftrightarrow \qquad \qquad \tau_{ab} \equiv e_a{}^\mu e_b{}^\nu \partial_{[\mu} \tau_{\nu]} = 0 \,.
\end{align}
A $D$-dimensional NC geometry with this twistless torsional constraint no longer exhibits a notion of absolute time. According to Frobenius' theorem, it can however still be foliated into $(D-1)$-dimensional leaves that can be identified as spatial hypersurfaces of constant time.

\subsubsection{Stringy Newton-Cartan geometry} \label{ssec:pbraneNCgeometry}

As we saw above, NC geometry is characterized by the existence of local frames (defined by $\tau_\mu$ and $e_\mu{}^a$) that feature a time/space split and are related by local (homogeneous) Galilean transformations, as well as a one-form field $m_\mu$ that couples to conserved mass currents. As such, it is the geometrical arena in which the classical mechanics of NR point particles takes place. Recent years have witnessed a renewed interest in NR limits of extended objects in which the speed of light is taken to infinity only in the directions transverse to the objects under consideration. Applied to strings, such a limit leads to NR string theory \cite{Gomis:2000bd,Danielsson:2000gi,Danielsson:2000mu}, whose excitations satisfy NR dispersion relations and interact via NR gravity. The two-dimensional worldvolume of such NR strings is still relativistic and can therefore not naturally be embedded in NC geometry with a NR time/space split as described in the previous subsection. Recently, it has been shown that the natural background geometry in which NR strings move is given by a generalization of NC geometry that is called stringy Newton-Cartan geometry (see, e.g., \cite{Andringa:2012uz}, \cite{Bergshoeff:2019pij}, \cite{Bidussi:2021ujm}, \cite{Bergshoeff:2022fzb} and \cite{Gomis:2005pg} for an early example)  and that we will review here.

Instead of a local time/space split, a $D$-dimensional stringy NC geometry features a split between 2 local so-called `longitudinal' directions and the remaining $D-2$ `transversal' ones. The longitudinal directions are equipped with a rank-2 Minkowski metric and can thus be used to embed the worldvolume of NR strings. A Cartan formulation of stringy NC geometry can be given in analogy to the one of NC geometry of the previous subsection. It is based on a `longitudinal Vielbein' field $\tau_\mu{}^A$, with $A=0,1$, a `transversal Vielbein' $e_\mu{}^a$, with $a = 2,\cdots, D-1$ and a two-form field $b_{\mu\nu}$. In what follows, we will freely raise and lower the $A$-index with a two-dimensional Minkowski metric $\eta_{AB} = \mathrm{diag}(-1,1)$ and the $a$-index with a $(D-2)$-dimensional Euclidean metric $\delta_{ab}$.

The fields $\tau_\mu{}^A$, $e_\mu{}^a$ and $b_{\mu\nu}$ transform under local $(\mathrm{SO}(1,1) \times \mathrm{SO}(D-2)) \rtimes \mathbb{R}^{2(D-2)}$ transformations as follows:
\begin{align}
  \label{eq:SNCtrafos}
    \delta \tau_\mu{}^A &= \lambda_M \epsilon^A{}_B \tau_\mu{}^B \,, \qquad \qquad \qquad \delta e_\mu{}^{a} = \lambda^{a}{}_{b} e_\mu{}^{b} - \lambda_A{}^{a} \tau_\mu{}^A \,, \nonumber \\
  \delta b_{\mu\nu} &= -2 \epsilon_{AB} \lambda^A{}_{a} \tau_{[\mu}{}^B e_{\nu]}{}^{a} \,.
\end{align}
Here, $\lambda_M$, $\lambda^{ab} = - \lambda^{ba}$ and $\lambda^{A a}$ are the parameters of SO$(1,1)$, SO$(8)$ and $\mathbb{R}^{2(D-2)}$ respectively. The SO$(1,1)$ and SO$(8)$ parts of this local symmetry group will be referred to as `longitudinal Lorentz transformations' and `transversal spatial rotations'. The $\mathbb{R}^{2(D-2)}$ part represents a type of boosts that transform transversal into longitudinal directions (but not vice versa) and that are called `string Galilean boosts'. In addition to the string Galilean boost transformation \eqref{eq:SNCtrafos}, the two-form field $b_{\mu\nu}$ is also subjected to a one-form gauge symmetry, acting with parameter $\theta_\mu$ as follows:
\begin{align}
  \label{eq:oneformgauge}
  \delta b_{\mu\nu} = 2 \partial_{[\mu} \theta_{\nu]} \,.
\end{align}
As a consequence, $b_{\mu\nu}$ naturally couples to conserved string tension currents, in analogy to how the central charge gauge field $m_\mu$ of NC geometry couples to conserved mass currents \cite{Bidussi:2021ujm}.

The longitudinal and transversal Vielbeine can be used to define the metric structure of stringy NC geometry. To do this, one first introduces `inverse' longitudinal and spatial Vielbeine $\tau_A{}^\mu$ and $e_a{}^\mu$ that obey:
\begin{alignat}{3}
  \label{eq:invVielbeineSNC}
  & \tau_A{}^\mu \tau_\mu{}^B = \delta_A^B \,, \qquad \qquad & & \tau_A{}^\mu e_\mu{}^{a} = 0 \,, \qquad \qquad \qquad & & e_{a}{}^\mu \tau_\mu{}^A = 0 \,, \nonumber \\
  & e_\mu{}^{a} e_{b}{}^\mu = \delta_{b}^{a} \,, \qquad \qquad  & & \tau_\mu{}^A \tau_A{}^\nu + e_\mu{}^{a} e_{a}{}^\nu = \delta_\mu^\nu \,.
\end{alignat}
One can then construct the following two symmetric two-tensors that are quadratic in the (inverse) Vielbeine and that are invariant under the local $(\mathrm{SO}(1,1) \times \mathrm{SO}(D-2)) \rtimes \mathbb{R}^{2(D-2)}$ transformations \eqref{eq:SNCtrafos}:
\begin{align} \label{eq:SNCmetrics}
  \tau_{\mu\nu} = \tau_\mu{}^A \tau_\nu{}^B \eta_{AB} \,, \qquad \qquad h^{\mu\nu} = e_a{}^\mu e_b{}^\nu \delta^{ab} \,.
\end{align}
The first of these has rank-2 and is called the longitudinal metric, whereas $h^{\mu\nu}$ has rank-$(D-2)$ and is called the transversal (co-)metric. Similar to NC geometry, the longitudinal metric can be used to determine the proper area of string worldsheets, while the transversal metric allows one to measure transversal distances to such worldsheets. We refer to \cite{Bergshoeff:2022fzb} for more details on this.

The metric compatible connection of stringy NC geometry is determined by three spin connections $\omega_\mu$, $\omega_\mu{}^{ab}$ and $\omega_\mu{}^{Aa}$ that transform under longitudinal Lorentz transformations, transversal spatial rotations and string Galilean boosts as follows:
\begin{align}
  \label{eq:localtrafosdepSNC}
    \delta \omega_\mu &= \partial_\mu \lambda_M \,, \qquad \qquad
    \delta \omega_\mu{}^{ab} = \partial_\mu \lambda^{ab} + 2 \lambda^{[a|c|} \omega_{\mu c}{}^{b]} \,, \nonumber \\
  \delta \omega_\mu{}^{A a} &= \partial_\mu \lambda^{A a} + \lambda_M \epsilon^A{}_B \omega_\mu{}^{B a}+ \lambda^{a}{}_{b} \omega_\mu{}^{A b} - \epsilon^A{}_B \lambda^{B a} \omega_\mu + \lambda^{A b} \omega_{\mu b}{}^{a} \,.
\end{align}
These spin connections satisfy the following constraints in analogy to \eqref{eq:convBargconstraints}:
\begin{align}
  \label{eq:SNCtorsionconstraints}
    & 2 \partial_{[\mu} \tau_{\nu]}{}^A - 2 \epsilon^A{}_B \omega_{[\mu} \tau_{\nu]}{}^B= T_{\mu\nu}{}^A \,, \nonumber \\
  &  2 \partial_{[\mu} e_{\nu]}{}^{a} - 2 \omega_{[\mu}{}^{ab} e_{\nu] b} + 2 \omega_{[\mu}{}^{A a} \tau_{\nu] A} = T_{\mu\nu}{}^a \,, \nonumber \\
  & 3 \partial_{[\mu} b_{\nu\rho]} + 6 \epsilon_{AB} \omega_{[\mu}{}^{A b} \tau_\nu{}^B e_{\rho] b} = T^{(b)}_{\mu\nu\rho} \,,
\end{align}
where the left-hand-sides are covariant with respect to \eqref{eq:localtrafosdepSNC} and $T_{\mu\nu}{}^A$, $T_{\mu\nu}{}^a$, $T_{\mu\nu\rho}^{(b)}$ are suitable torsion tensors. Not all of the constraints \eqref{eq:SNCtorsionconstraints} are conventional, since not all of them contain the spin connections $\omega_\mu$, $\omega_\mu{}^{ab}$ and $\omega_\mu{}^{Aa}$. In particular, contracting \eqref{eq:SNCtorsionconstraints} with the inverse Vielbeine $\tau_A{}^\mu$, $e_a{}^\mu$ and taking suitable (anti-)symmetrizations, one finds that \eqref{eq:SNCtorsionconstraints} contains the following components
\begin{align}
  \label{eq:geomconstraints}
  &  2 \tau_{(A|}{}^\mu e_{a}{}^\nu \partial_{[\mu} \tau_{\nu]|B)} = \tau_{(A|}{}^\mu e_{a}{}^\nu T_{\mu\nu |B)} \,, \qquad \quad  2 e_{a}{}^\mu e_{b}{}^\nu \partial_{[\mu} \tau_{\nu]}{}^A = e_{a}{}^\mu e_{b}{}^\nu T_{\mu\nu}{}^A\,, \nonumber \\
  &  3 e_{a}{}^\mu e_{b}{}^\nu e_{c}{}^\rho \partial_{[\mu} b_{\nu\rho]} = e_{a}{}^\mu e_{b}{}^\nu e_{c}{}^\rho T^{(b)}_{\mu\nu\rho} \,,
\end{align}
that are independent of the spin connections. This leaves $D + D(D-1)(D-2)/2 + (D-2)^2$ equations of \eqref{eq:SNCtorsionconstraints} that can be used to solve and express the spin connections in terms of $\tau_\mu{}^A$, $e_\mu{}^a$, $b_{\mu\nu}$, $T_{\mu\nu}{}^A$, $T_{\mu\nu}{}^a$ and $T_{\mu\nu\rho}^{(b)}$. Since $\omega_\mu$, $\omega_\mu{}^{ab}$ and $\omega_\mu{}^{Aa}$ have a total of $D + D(D-2)(D-3)/2 + 2 D (D-2)$ components, one sees that not all spin connection components can be expressed in this way. In particular, the $2(D-2)$ components
\begin{align}
    \label{eq:indepcomp}
  \tau_{\{A|}{}^\mu \omega_{\mu|B\}}{}^{a} \equiv \tau_{(A|}{}^\mu \omega_{\mu|B)}{}^{a} - \frac12 \eta_{AB} \tau_C{}^\mu \omega_\mu{}^{C a} \,,
\end{align}
remain as independent fields in the connection structure of stringy NC geometry. All other spin connection components can be expressed in terms of the longitudinal and transversal Vielbeine, $b_{\mu\nu}$ and the torsion tensors appearing in \eqref{eq:SNCtorsionconstraints}. We refer to \cite{Bergshoeff:2022fzb} for their explicit expressions and for details on how these can be obtained.

Once the spin connections $\omega_\mu$, $\omega_\mu{}^{ab}$ and $\omega_\mu{}^{Aa}$ have been expressed in this way, an affine connection $\Gamma^\rho_{\mu\nu}$ can be introduced via the following Vielbein postulates:
\begin{align}
  \label{eq:VielbpostSNC}
  & \partial_\mu \tau_\nu{}^A - \epsilon^A{}_B \omega_\mu \tau_\nu{}^B -  \Gamma_{\mu\nu}^\rho \tau_\rho{}^A = 0 \,, \nonumber \\
  &  \partial_\mu e_\nu{}^{a} - \omega_\mu{}^{ab} e_{\nu b} + \omega_\mu{}^{A a} \tau_{\nu A} - \Gamma_{\mu\nu}^\rho e_{\rho}{}^{a} = 0 \,.
\end{align}
This connection $\Gamma_{\mu\nu}^\rho$ is by construction compatible with the metrics $\tau_{\mu\nu}$ and $h^{\mu\nu}$ of \eqref{eq:SNCmetrics}
\begin{align}
  \label{eq:metricscompSNC}
  \nabla_\mu \tau_{\nu\rho } & \equiv \partial_\mu \tau_{\nu\rho} - \Gamma_{\mu\nu}^\sigma \tau_{\sigma \rho} - \Gamma_{\mu\rho}^\sigma \tau_{\nu \sigma} = 0 \,, \nonumber \\ \nabla_\mu h^{\nu\rho} & \equiv \partial_\mu h^{\nu\rho} + \Gamma_{\mu\sigma}^\nu h^{\sigma \rho} + \Gamma_{\mu\sigma}^\rho h^{\nu\sigma}= 0 \,.
\end{align}
In analogy to the NC case, one can ensure that $\Gamma^\rho_{\mu\nu}$ is invariant under local longitudinal Lorentz transformations, transversal spatial rotations and Galilean boosts, by choosing the transformation rules of the tensors $T_{\mu\nu}{}^A$, $T_{\mu\nu}{}^a$ and $T_{\mu\nu\rho}^{(b)}$ as follows:
\begin{align}
  \label{eq:trafoTATAp}
  & \delta T_{\mu\nu}{}^A = \lambda_M \epsilon^A{}_B T_{\mu\nu}{}^B \,, \qquad \qquad \qquad \delta T_{\mu\nu}{}^{a} = \lambda^{a}{}_{b} T_{\mu\nu}^{b} - \lambda_A{}^{a} T_{\mu\nu}{}^A \,, \nonumber \\
  & \delta T^{(b)}_{\mu\nu\rho} = -3 \epsilon_{AB} \lambda^A{}_{a} T_{[\mu\nu}{}^B e_{\rho]}{}^{a} + 3 \epsilon_{AB} \lambda^A{}_{a} T_{[\mu\nu}{}^{a} \tau_{\rho]}{}^B \,.
\end{align}
From \eqref{eq:VielbpostSNC} one can solve and express $\Gamma^\rho_{\mu\nu}$ in terms of the stringy NC metrics \eqref{eq:SNCmetrics}, $b_{\mu\nu}$ and the tensors $T_{\mu\nu}{}^A$, $T_{\mu\nu}{}^a$ and $T_{\mu\nu\rho}^{(b)}$ (see \cite{Bergshoeff:2022fzb} for an explicit expression).

The metric compatible affine connection $\Gamma^\rho_{\mu\nu}$ introduced above features generic torsion, given by
\begin{align} \label{eq:SNCtorsiondecomp}
    & 2 \Gamma^\rho_{[\mu\nu]} = \tau_A{}^\rho T_{\mu\nu}{}^A + e_{a}{}^\rho T_{\mu\nu}{}^{a} \,.
\end{align}
As in the NC case, part of this torsion is intrinsic. In particular, the torsion components $T_{(A|a|B)} \equiv \tau_{(A|}{}^\mu e_{a}{}^\nu T_{\mu\nu |B)}$ and $T_{ab}{}^A \equiv e_{a}{}^\mu e_{b}{}^\nu T_{\mu\nu}{}^A$ \footnote{We follow a similar convention as in footnote 8 to turn curved indices into longitudinal or transversal ones. In particular, we will call the longitudinal components $X_A$/$X^A$ of a one-form $X_\mu$/vector $X^\mu$ (and by extension of tensors) those that are obtained by contraction with $\tau_A{}^\mu$/$\tau_\mu{}^A$, according to the rule $X_A \equiv \tau_A{}^\mu X_\mu$/$X^A \equiv \tau_\mu{}^A X^\mu$. Likewise, the transversal components $X_a$/$X^a$ are defined as $X_a \equiv e_a{}^\mu X_\mu$/$X^a \equiv e_\mu{}^a X^\mu$.} do not appear in the dependent spin connections and constitute the intrinsic torsion of the stringy NC affine connection. From \eqref{eq:geomconstraints} one sees that setting this intrinsic torsion equal to zero leads to differential constraints on the longitudinal Vielbein $\tau_\mu{}^A$, similar to what happens in the NC case. As in the NC case, one can consistently set intrinsic torsion tensor components equal to zero in various ways. We refer to \cite{Bergshoeff:2022fzb} for an in-depth analysis of the different possibilities and their geometrical interpretation.

Having reviewed the basics of Lorentzian and non-Lorentzian geometry, we are now ready to discuss different non-Lorentzian supergravity theories.  We start by discussing in the next section 3$D$ Newton-Cartan supergravity.

\section{3$D$ Newton-Cartan supergravity} \label{sec:3Dsugra}

In this section, we will review the 3$D$ supersymmetric version of NC gravity of \cite{Andringa:2013mma}, \cite{Bergshoeff:2015uaa}. Originally, this theory was constructed as a gauging of a $\mathcal{N}=2$ super-Bargmann algebra \cite{Andringa:2013mma}. In \cite{Bergshoeff:2015uaa}, it was rederived by applying a non-Lorentzian limit procedure to relativistic 3$D$, $\mathcal{N}=2$ supergravity. Since recent efforts to construct non-Lorentzian supergravity theories in higher dimensions are mainly based on taking a non-Lorentzian limit, we will here focus in detail on the construction of \cite{Bergshoeff:2015uaa}, in order to allow for comparison with the 10$D$ non-Lorentzian supergravity of section \ref{sec:10Dsugra}. In subsection \ref{ssec:3DN2NCsugra}, we will first review the construction of 3$D$ NC supergravity via a non-Lorentzian limit. In subsection \ref{ssec:gaugefix}, we will then show how a supersymmetric version of Newtonian gravity can be obtained as a gauge fixing of NC supergravity.

\subsection{3$D$ $\mathcal{N}=2$ on-shell Newton-Cartan Supergravity} \label{ssec:3DN2NCsugra}

Our starting point is the
relativistic 3$D$ $\mathcal{N}=2$ supergravity multiplet with field content $\{E_\mu{}^{\hat{A}}\,,\Psi_{\mu i}\}\, (\hat{A} = 0,1,2; i=1,2)$.
Under diffeomorphisms (with parameter $\xi^\mu$) and Lorentz
rotations (with parameter $\Lambda^{\hat{A}\hat{B}}$) these fields transform as
\begin{align}
 \delta E_\mu{}^{\hat{A}} &= \xi^\nu\partial_\nu E_\mu{}^{\hat{A}} +\partial_\mu\xi^\nu E_\nu{}^{\hat{A}}
                     +\Lambda^{\hat{A}}{}_{\hat{B}} E_\mu{}^{\hat{B}} \,, \label{reletrafo}\\[.1truecm]
                     \delta \Psi_{\mu i} &=  \xi^\nu\partial_\nu \Psi_{\mu i} + \partial_\mu\xi^\nu \Psi_{\nu i}
                     + \frac{1}{4} \Lambda^{\hat{A}\hat{B}}\gamma_{\hat{A}\hat{B}} \Psi_{\mu i}\,.
\end{align}
Furthermore, the supersymmetry
transformation rules (with parameter $\eta_i$) are given by
\begin{align}
 \delta E_\mu{}^{\hat{A}} &= \frac12\,\delta^{ij}\,\bar\eta_i\,\gamma^{\hat{A}} \Psi_{\mu j} \,, \label{susyE}\\[.1truecm]
 \delta \Psi_{\mu i} &= D_\mu\eta_i= \partial_\mu\eta_i -\frac14\,\Omega_\mu{}^{\hat{A}\hat{B}}(E,\Psi_i)\gamma_{\hat{A}\hat{B}}\eta_i \label{susyPsi}\,,
\end{align}
where $D_\mu$ is the Lorentz-covariant derivative and the dependent spin connection $\Omega_\mu{}^{\hat{A}\hat{B}}(E,\Psi_i)$ is given by
 \begin{align}\begin{split}\label{relsusyomega}
 \Omega_\mu{}^{\hat{A}\hat{B}}(E,\Psi_i) &= -2\,E^{\nu[\hat{A}}\Big(\partial_{[\mu} E_{\nu]}{}^{\hat{B}]}
                                   -\frac14\,\delta^{ij}\,\bar\Psi_{[\mu i}\gamma^{\hat{B}]}\Psi_{\nu]j}\Big) \\[.1truecm]
   &\quad    +E_{\mu \hat{C}} E^{\nu \hat{A}} E^{\rho \hat{B}} \Big(\partial_{[\nu} E_{\rho]}{}^{\hat{C}}
                                   -\frac14\,\delta^{ij}\,\bar\Psi_{[\nu i}\gamma^{\hat{C}}\Psi_{\rho]j}\Big)\,.
\end{split}\end{align}
From this expression one derives that the supersymmetry transformation of the (dependent)
spin connection is given by
\begin{align}\label{susyomegatrafo}
 \delta \Omega_\mu{}^{\hat{A}\hat{B}}(E,\Psi_i) = -\frac12\,\delta^{ij}\,E^{\nu[\hat{A}}\,\bar\eta_i\,\gamma^{\hat{B}]}\hat\Psi_{\mu\nu j}
         +\frac14\,\delta^{ij}\,E_{\mu \hat{C}}E^{\nu \hat{A}}E^{\rho \hat{B}}\,\bar\eta_i\,\gamma^{\hat{C}}\,\hat\Psi_{\nu\rho j}\,,
\end{align}
where $\hat{\Psi}_{\mu\nu i} \equiv 2 D_{[\mu} \Psi_{\nu] i}$.
Note that this transformation rule is zero on-shell, i.e.~it vanishes upon using the fermionic equations of motion
\begin{align}\label{relsusyeom}
 \hat\Psi_{\mu\nu i}=0 \,.
\end{align}
One may verify that with the transformation rules \eqref{susyE} and \eqref{susyPsi} the supersymmetry algebra closes on-shell on the fields.

Besides the relativistic $\mathcal{N}=2$ supergravity multiplet, we will introduce an additional  field $M_\mu$, that can be associated
to the central charge generator $\cal{Z}$ of the $\mathcal{N}=2$ super-Poincar\'e algebra (see eq.~\eqref{3dpoincare} below).
This gauge field transforms under diffeomorphisms and abelian gauge transformations (with
parameter $\Lambda$) as follows:
\begin{align}\label{Mtransf}
 \delta M_\mu = \xi^\nu \partial_\nu M_\mu + \partial_\mu \xi^\nu M_\nu +
\partial_\mu \Lambda \,.
\end{align}
Its transformation rule under supersymmetry
is determined by the Poincar\'e superalgebra (see eq.~\eqref{3dpoincare} below)
\begin{align}\label{susyM}
 \delta M_\mu = \frac12\,\varepsilon^{ij}\,\bar\eta_i\Psi_{\mu j} \,.
\end{align}
This field is ordinarily not introduced in the supergravity
multiplet. In order not to upset the on-shell counting of bosonic and
fermionic degrees of freedom, we are thus obliged to set the
supercovariant curvature of $M_\mu$ to zero, i.e.
\begin{align}\label{susyMcurv}
 \hat F_{\mu\nu}(M) \equiv 2\,\partial_{[\mu}M_{\nu]} -\frac12\,\varepsilon^{ij}\,\bar\Psi_{[\mu i}\Psi_{\nu]j} =0 \,,
\end{align}
so that this field corresponds to a pure gauge degree of freedom. Note that this constraint also implies that the commutator of two supersymmetry transformations acting on $M_\mu$ closes to a general coordinate transformation and a central charge transformation. Moreover,  this constraint will be important to obtain finite expressions for the
NR spin connections by taking the limit of the relativistic connection.
Starting from expression (\ref{susyMcurv}), the full set of relativistic equations of motion is obtained by the following chain of supersymmetry transformations
\begin{align}\label{relchain}
 \hat F_{\mu\nu}(M)=0 \quad\to\quad \hat \Psi_{\mu\nu i}=0  \quad\to\quad \hat R_{\mu\nu}{}^{\hat{A}\hat{B}}(\Omega)=0 \,.
\end{align}
This concludes the summary of our relativistic starting point.

The non-Lorentzian limit procedure employs a redefinition of the fields that mimics the In\"on\"u-Wigner contraction of the $\mathcal{N}=2$ Poincar\'e superalgebra to the $\mathcal{N}=2$ Bargmann superalgebra. To motivate this redefinition, we will first consider this algebra contraction. The starting relativistic superalgebra is given by
the following $\mathcal{N}=2$ Poincar\'e superalgebra with translation $P_{\hat{A}}$, Lorentz transformations $M_{\hat{A}\hat{B}}$, central extension $\mathcal{Z}$ and supercharges $Q^i$ ($i=1,2$):
\begin{align}\begin{split}\label{3dpoincare}
 \big[M_{\hat{A}\hat{B}} , P_{\hat{C}} \big] &= -2\,\eta_{\hat{C}[\hat{A}} P_{\hat{B}]} \,, \hskip.62cm
 \big[M_{\hat{A}\hat{B}} , M_{\hat{C}\hat{D}} \big] = 4\,\eta_{[\hat{A}[\hat{C}} M_{\hat{D}]\hat{B}]} \,, \\[.1truecm]
 \big[M_{\hat{A}\hat{B}} , Q^i \big] &= -\frac12\,\gamma_{\hat{A}\hat{B}}Q^i \,, \hskip1,2cm
 \big\{ Q^i , Q^j \big\} = -\gamma^{\hat{A}}C^{-1}\,P_{\hat{A}}\,\delta^{ij} +C^{-1}\,\mathcal{Z}\,\epsilon^{ij} \,.
\end{split}\end{align}
Here, the supercharges $Q^i$ are two-component
Majorana spinors.
For the gamma-matrices we choose a real basis,
i.e.~$\gamma^{\hat{A}}=(i\sigma_2,\sigma_1,\sigma_3)$ and the charge
conjugation matrix is taken to be $C=i\gamma^0$.

In order to define the In\"on\"u--Wigner contraction, we first define
the projections
\begin{align}\label{Q+-}
 Q_\pm =\frac{1}{\sqrt{2}}\,\big(Q^1 \pm \gamma_0Q^2\big) \,,
\end{align}
and split the three-dimensional flat indices $\hat{A}$, $\hat{B}$ into time-like
and space-like indices $\{0,a\}$. We set $M_{ab}=J_{ab}$
for the purely spatial rotations.  Next, we perform the following invertible redefinition of the generators:
\begin{align}\begin{split}\label{susyalgcontr}
 Q_- &\to \sqrt{\omega}\,Q_- \,, \hskip2.45cm Q_+\to \frac{1}{\sqrt\omega}\,Q_+ \,, \hskip1.6cm  M_{a0} \to \omega \,G_a \,, \\[.1truecm]
 \mathcal{Z} &\to -\omega M+\frac{1}{2\omega}\,H \,, \hskip1.5cm P_0\to \omega M +\frac{1}{2\omega}\,H \,,
\end{split}
\end{align}
where $\omega$ is a finite dimensionless contraction parameter and we leave the generators $P_a$ and $J_{ab}$ untouched.
Using these redefinitions, the $\mathcal{N}=2$ supersymmetric extension
of the Bargmann algebra is then obtained in the limit $\omega\to\infty$. In particular, we find the
following non-vanishing commutation relations:
\begin{align}
 \big[J_{ab} , P_c \big] &= -2\,\delta_{c[a} P_{b]} \,, \hskip2.48cm
 \big[J_{ab} , G_c \big] = -2\,\delta_{c[a} G_{b]} \,, \nonumber\\[.12truecm]
 \big[G_a , H \big] &= -P_a \,, \hskip3.45cm \big[ G_a, P_b \big] = -\delta_{ab}\,M \,, \nonumber\\[.12truecm]
 \big[J_{ab} , Q_\pm \big] &= -\tfrac12\,\gamma_{ab}Q_\pm \,, \hskip2.32cm
 \big[G_a , Q_+ \big] = -\tfrac12\,\gamma_{a0}Q_- \,,                                           \label{3dsuperbargmann} \\[.12truecm]
 \big\{ Q_+ , Q_+ \big\} &= -\gamma^0C^{-1}\,H \,, \hskip2cm
 \big\{ Q_+ , Q_- \big\} = -\gamma^aC^{-1}\,P_a \,, \nonumber\\[.12truecm]
 \big\{ Q_- , Q_- \big\} &= -2\,\gamma^0C^{-1}\,M \,. \nonumber
\end{align}
The bosonic part of the algebra corresponds to the Bargmann algebra \eqref{eq:Bargmann}. Note that, since we are working in three dimensions, the
spatial rotations are Abelian.

We now extend the above algebra contraction to the fields of the on-shell $\mathcal{N}=2$
supergravity multiplet. For the bosonic fields, we employ the following redefinitions
\begin{eqnarray}
E_\mu{}^0 &=& \omega \tau_\mu + \frac{1}{2\omega} m_\mu\,, \hskip 1truecm E_\mu{}^a = e_\mu{}^a\,,\\[.1truecm]
M_\mu &=& \omega\tau_\mu -\frac{1}{2\omega} m_\mu\,.
\end{eqnarray}
In the limit $\omega \rightarrow \infty$, the fields $\tau_\mu$, $e_\mu{}^a$ and $m_\mu$ become the clock form, spatial Vielbein and central charge gauge field of NC geometry. To derive their bosonic NR transformation rules \eqref{eq:deltatauem},
we first express the new fields in
terms of the old ones, i.e.
\begin{align}\label{tauandm}
 \tau_\mu=\frac{1}{2\omega}\,\big(E_\mu{}^0 +M_\mu\big) \,, \hskip2cm  m_\mu = \omega\,\big(E_\mu{}^0 -M_\mu\big) \,.
\end{align}
By also redefining the symmetry parameters $\Lambda^{ab}$, $\Lambda^{a0}$ and $\Lambda$ as
\begin{align}
\lambda^{ab} = \Lambda^{ab} \,, \hskip2cm \lambda^a=\omega\,\Lambda^a{}_0 \,, \hskip2cm \sigma=-\omega \Lambda \,,
\end{align}
it is straightforward to obtain the bosonic transformation rules \eqref{eq:deltatauem}.
All fields transform under diffeomorphisms in the usual way.

The redefinitions of the gravitini follow from the way we
contract the fermionic generators of the 3$D$ $\mathcal{N}=2$
Poincar\'e superalgebra to get the Bargmann superalgebra, i.e.~we first
define  projected spinors
\begin{align}\label{psi+-}
 \Psi_\pm =\frac{1}{\sqrt2}\,\Big(\Psi_1 \pm \gamma_0\Psi_2\Big) \,,
\end{align}
and we similarly define parameters $\eta_\pm$ from the $\eta_{1,2}$. We then introduce the scalings:
\begin{align}\begin{split}\label{fermscaling}
 \Psi_{\mu+} &= \sqrt{\omega}\,\psi_{\mu+} \,, \hskip2cm
      \eta_+ = \sqrt{\omega}\,\epsilon_+ \,, \\[.1truecm]
 \Psi_{\mu-} &= \frac{1}{\sqrt{\omega}}\,\psi_{\mu-} \,, \hskip1,9cm
      \eta_- = \frac{1}{\sqrt{\omega}}\,\epsilon_- \,.
\end{split}\end{align}
The following NR supersymmetry transformation rules then follow
\begin{align}\begin{split}\label{bosfermtrafo}
 \delta \tau_\mu &= \frac12\,\bar\epsilon_+\gamma^0\psi_{\mu+} \,, \\[.1truecm]
 \delta e_\mu{}^a &= \frac12\,\bar\epsilon_+\gamma^a\psi_{\mu-} +\frac12\,\bar\epsilon_-\gamma^a\psi_{\mu+}\,,\\[.1truecm]
 \delta m_\mu &= \bar\epsilon_-\gamma^0\psi_{\mu-} \,,
\end{split}\end{align}
as well as
\begin{align}\begin{split}\label{fermfermtrafo}
 \delta \psi_{\mu+} &= \partial_\mu\epsilon_+  -\frac14\,\omega_\mu{}^{ab}\gamma_{ab}\epsilon_+ \,, \\
 \delta \psi_{\mu-} &= \partial_\mu\epsilon_-  -\frac14\,\omega_\mu{}^{ab}\gamma_{ab}\epsilon_-
                       +\frac12\,\omega_\mu{}^a\gamma_{a0}\epsilon_+ \,.
\end{split}\end{align}
The transformation rules of the spinors with respect to the NR bosonic symmetries are found to be
\begin{align}\begin{split}\label{fermbostrafo}
  \delta \psi_{\mu+} &= \frac14\,\lambda^{ab}\gamma_{ab}\psi_{\mu+} \,, \\
  \delta \psi_{\mu-} &= \frac14\,\lambda^{ab}\gamma_{ab}\psi_{\mu-}
-\frac12\,\lambda^a\gamma_{a0}\psi_{\mu+} \,.
\end{split}\end{align}

It is understood that the spin connections $\omega_\mu{}^{a},\omega_\mu{}^{ab}$ in \eqref{fermfermtrafo} are dependent,
i.e.~$\omega_\mu{}^{a}=\omega_\mu{}^{a}(e,\tau,m,\psi_\pm)$ and $\omega_\mu{}^{ab}=\omega_\mu{}^{ab}(e,\tau,m,\psi_\pm)$. Their explicit expressions are given by
\begin{align}\nonumber \label{susyomegaab}
 \omega_\mu{}^{ab}(e,\tau,m,\psi_\pm) &= -2\,e^{[a|\nu|}\big(\partial_{[\mu}e_{\nu]}{}^{b]}
                               -\frac12\,\bar\psi_{[\mu+}\gamma^{b]}\psi_{\nu]-}\big)\nonumber\\[.1truecm]
       &+e_\mu{}^ce^{a\nu}e^{b\rho}\big(\partial_{[\nu}e_{\rho]}{}^c
                               -\frac12\,\bar\psi_{[\nu+}\gamma^c\psi_{\rho]-}\big) \nonumber\\[.1truecm]
    &-\tau_\mu e^{a \nu}e^{b \rho}\big(\partial_{[\nu}m_{\rho]} -\frac12\,\bar\psi_{[\nu-}\gamma^0\psi_{\rho]-} \big)
            \,, \\[.2truecm]
 \omega_\mu{}^a(e,\tau,m,\psi_\pm) &= \tau^\nu \big(\partial_{[\mu}e_{\nu]}{}^a
                               -\frac12\,\bar\psi_{[\mu+}\gamma^a\psi_{\nu]-} \big)\nonumber\\[.1truecm]
      &+e_{\mu b} e^{a \nu}\tau^\rho \big(\partial_{[\nu}e_{\rho]}{}^b -\frac12\,\bar\psi_{[\nu+}\gamma^b\psi_{\rho]-} \big) \nonumber\\[.1truecm]
    &+e^{a \nu}\big(\partial_{[\mu}m_{\nu]} -\frac12\,\bar\psi_{[\mu-}\gamma^0\psi_{\nu]-} \big)\nonumber\\[.1truecm]
           &-\tau_\mu e^{a \nu}\tau^\rho \big(\partial_{[\nu}m_{\rho]}
                                      -\frac12\,\bar\psi_{[\nu-}\gamma^0\psi_{\rho]-} \big) \,.\label{susyomegaa}
\end{align}
The expressions for these NR spin connections can be obtained from the relativistic expression given in eq.~\eqref{relsusyomega}, by also using eq.~\eqref{susyMcurv}. In particular, in order to obtain these expressions, we have used
eq.~\eqref{susyMcurv}, for finite $\omega$, to replace terms that diverge in the $\omega \to \infty$
limit, by terms with the expected $\omega$-order. Note that the expressions \eqref{susyomegaab}, \eqref{susyomegaa} are of the form \eqref{eq:spinconnexpr} with
\begin{align}
  T_{\mu\nu}{}^a = \bar\psi_{[\mu+}\gamma^a\psi_{\nu]-} \qquad \text{and} \qquad T_{\mu\nu}^{(m)} = \bar\psi_{[\mu-}\gamma^0\psi_{\nu]-} \,.
\end{align}
According to the discussion that led to eqs.~\eqref{eq:spinconnexpr}, the dependent spin connections \eqref{susyomegaab}, \eqref{susyomegaa} thus identically fulfil the following conventional constraints:
\begin{align}\begin{split}\label{RPRZ}
 \hat R_{\mu\nu}{}^a(P) & \equiv R_{\mu\nu}{}^a(P) -\bar\psi_{[\mu+}\gamma^a\psi_{\nu]-}=0 \,, \\[.1truecm]
 \hat R_{\mu\nu}(M) & \equiv R_{\mu\nu}(M) -\bar\psi_{[\mu-}\gamma^0\psi_{\nu]-}=0\,,
             \end{split}\end{align}
where $R_{\mu\nu}{}^a(P)$, $R_{\mu\nu}(M)$ are defined in \eqref{eq:convBargconstraints} and where $\hat R_{\mu\nu}{}^a(P)$, $\hat R_{\mu\nu}(M)$ are the supercovariant curvatures of $e_\mu{}^a$ and $m_\mu$ respectively. The other supercovariant curvatures that will be used in what follows are given by
\begin{align}\label{curvatures}
{\hat R}_{\mu\nu}{}^a(G) & \equiv 2\partial_{[\mu}\omega_{\nu]}{}^{a} - 2
\omega_{[\mu}{}^{ab} \omega_{\nu]b}\,,
\nonumber\\[.1truecm]
{\hat R}_{\mu\nu}{}^{ab} (J) & \equiv
                               2\partial_{[\mu}\omega_{\nu]}{}^{ab}  \,, \nonumber\\[.1truecm]
 \hat{R}_{\mu\nu}(H) & \equiv 2 \partial_{[\mu} \tau_{\nu]}
-\frac12\,\bar\psi_{[\mu+}\gamma^0\psi_{\nu]+} \,, \nonumber \\[.1truecm]
\hat{\psi}_{\mu \nu +} & \equiv  2
\partial_{[\mu} \psi_{\nu]+} - \frac{1}{2} \omega_{[\mu}{}^{ab}
\gamma_{ab} \psi_{\nu]+} \,, \nonumber \\[.1truecm]
\hat{\psi}_{\mu \nu -} & \equiv  2 \partial_{[\mu} \psi_{\nu]-} -
\frac{1}{2} \omega_{[\mu}{}^{ab} \gamma_{ab} \psi_{\nu]-} +
\omega_{[\mu}{}^a \gamma_{a0} \psi_{\nu]+} \,.
\end{align}

The conventional constraints \eqref{RPRZ} can be used to determine that the supersymmetry transformations of the dependent spin and boost connections
\eqref{susyomegaab} and \eqref{susyomegaa} are given by
\begin{align}
\begin{split}\label{susyomegaabtrafo}
 \delta_Q \omega_\mu{}^{ab}(e,\tau,m,\psi_\pm) &= \frac12\,\bar\epsilon_+\gamma^{[b}\hat\psi^{a]}{}_{\mu-}
         +\frac14\,e_{\mu c} \,\bar\epsilon_+\gamma^c\hat\psi^{ab}{}_-
         -\frac12\,\tau_\mu \,\bar\epsilon_-\gamma^0\hat\psi^{ab}{}_- \\[.1truecm]
   &\quad +\frac12\,\bar\epsilon_-\gamma^{[b}\hat\psi^{a]}{}_{\mu+}
         +\frac14\,e_{\mu c} \,\bar\epsilon_-\gamma^c\hat\psi^{ab}{}_+\,,
\end{split} \\[.1truecm]
\begin{split}\label{susyomegaatrafo}
  \delta_Q \omega_\mu{}^a(e,\tau,m,\psi_\pm) &= \frac12\,\bar\epsilon_-\gamma^0\hat\psi_\mu{}^a{}_-
         +\frac12\,\tau_\mu\,\bar\epsilon_-\gamma^0\hat\psi_0{}^a{}_-
         +\frac14\,e_{\mu b}\,\bar\epsilon_+\gamma^b\hat\psi^a{}_{0-}\\[.1truecm]
        & +\frac14\,\bar\epsilon_+\gamma^a\hat\psi_{\mu0-}
    +\frac14\,e_{\mu b}\,\bar\epsilon_-\gamma^b\hat\psi^a{}_{0+}
         +\frac14\,\bar\epsilon_-\gamma^a\hat\psi_{\mu0+} \,.
\end{split}
\end{align}

The constraint  \eqref{susyMcurv} is not only used to get finite expressions for the NR spin connections. As already alluded to above, its
$\omega \to \infty$ limit also leads to the following constraint:
\begin{align}\label{susyRhcurv}
 \hat R_{\mu\nu}(H)  =0\,.
\end{align}
This constraint is a geometric one (i.e., not a conventional one) and leads to further conditions upon variation under supersymmetry.
One finds  that under supersymmetry transformations, with parameters $\epsilon_+$ and  $\epsilon_-$, the following set of constraints is generated:
\begin{align}
 \hat\psi_{ab-} &=0 \label{moreconstraints}\\
 \hat R_{\mu\nu}(H)=0 \quad \to \quad \hat\psi_{\mu\nu+}=0 \quad \to \quad
 \hat R_{\mu\nu}{}^{ab}(J) &=0 \label{constraints}\\
 \gamma^a\hat\psi_{a0-} &=0 \quad\to\quad \hat R_{0a}{}^a(G)=0\,. \label{onshell}
\end{align}
Note that the  variation of the $\hat\psi_{\mu\nu+}=0$ constraint leads to three different constraints. Of these three constraints only the variation of the constraint
$ \gamma^a\hat\psi_{a0-} =0$ leads to one more constraint.
Using the last constraint  given in eq.~\eqref{constraints} the NR Bianchi identities reduce to
\begin{align}\label{susynrBianchi}
 \hat R_{ab}{}^c(G)=0 \,, \hskip2cm \hat R_{0[a}{}^{b]}(G) =0 \,.
\end{align}
These identities are needed to show that the variation of the constraint given in eq.~\eqref{moreconstraints} does not lead to further constraints and that the variation of the first constraint in \eqref{onshell} leads to the singlet constraint $R_{0a}{}^a(G) = 0$ on the boost curvature. As we will see in the next subsection, this singlet constraint corresponds to a covariant (i.e., valid in arbitrary reference frames) version of the Poisson equation for the Newton potential.

At this point we have  obtained the supersymmetry rules of {\sl all} gauge fields, both the dependent as well as the independent ones. We find that with these supersymmetry transformations the
supersymmetry algebra closes on-shell. To be precise, the commutator
of two supersymmetry transformations  closes and is given by the
following soft algebra:
\begin{align} \label{closbos}
\left[\delta_Q(\epsilon_1),\delta_Q(\epsilon_2)\right] & =
\delta_{\mathrm{g.c.t.}}(\xi^{\mu}) + \delta_{J_{ab}}(\lambda^a{}_b)
+ \delta_{G_a}(\lambda^a{}) + \delta_{Q_+}(\epsilon_+)\ +
 \nonumber \\[.1truecm] & \ \ \ +\ \delta_{Q_-}(\epsilon_-) + \delta_{M}(\sigma) \,,
\end{align}
provided the two fermionic constraints on the gravitino curvature in eqs.~(\ref{moreconstraints}) and (\ref{onshell}) hold.
Here $\mathrm{g.c.t.}$ denotes a general
coordinate transformation and the field-dependent parameters are
given by
\begin{align} \label{parclos}
\xi^\mu & =  \frac{1}{2} \Big(\bar{\epsilon}_{2+} \gamma^0
\epsilon_{1+}  \Big) \tau^\mu + \frac{1}{2} \Big(\bar{\epsilon}_{2+}
\gamma^a \epsilon_{1-} + \bar{\epsilon}_{2-} \gamma^a \epsilon_{1+}
\Big) e_a{}^\mu \,, \nonumber
\\[.1truecm]
\lambda^a{}_b & =  -\xi^\mu \omega_\mu{}^a{}_b  \,, \nonumber
\\[.1truecm]
\lambda^a & =  -\xi^\mu \omega_\mu{}^a  \,, \nonumber \\[.1truecm]
\epsilon_\pm & =  - \xi^\mu \psi_{\mu \pm} \,, \nonumber
\\[.1truecm]
\sigma & =  - \xi^\mu m_\mu + \left( \bar{\epsilon}_{2-} \gamma^0
\epsilon_{1-}  \right) \,.
\end{align}

This concludes our discussion of how by taking a limit one can derive
 the three-dimensional
on-shell $\mathcal{N}=2$ NC supergravity theory first constructed in
\cite{Andringa:2013mma} by gauging a $\mathcal{N}=2$ supersymmetric extension of the Bargmann algebra, i.e., we obtained all constraints, equations of motion and
transformation rules.
To finish the consistency check of our procedure we should check whether the supersymmetry variation of
 the bosonic equation of motion given in \eqref{onshell}, describing the Poisson equation of the Newton potential in arbitrary frames,  does not lead to new constraints and/or equations of motion. Instead of doing this we shall show
in the next subsection that all constraints can be solved after gauge fixing, leading to a consistent system with a closed algebra.

\subsection{Gauge Fixing} \label{ssec:gaugefix}

For clarity, we will first explain how the gauge fixing of local diffeomorphisms, spatial rotations, Galilean boosts and the central charge transformation works in the purely bosonic case with all fermions set to zero. After that we will extend the discussion and include local supersymmetry and the fermions in the gauge fixing procedure.

\subsubsection{The Bosonic Case}

We start by first solving the two bosonic constraints in \eqref{constraints}, with all fermions set to zero, by imposing the gauge fixing conditions\,\footnote{We use a notation where $\emptyset$ indicates a curved $\mu=0$ index.}
\begin{equation}\label{gffirst}
\tau_\mu(x^\nu) = \delta_\mu{}^\emptyset\,,\hskip 1.5truecm \omega_\mu{}^{ab}(x^\nu)=0\,.
\end{equation}
This fixes the local time translations and spatial
rotations to constant ones:
 \begin{equation}
 \xi^\emptyset(x^\nu) = \xi^\emptyset\,,\hskip 2truecm \lambda^{ab}(x^\nu)=\lambda^{ab}\,.
 \end{equation}
 No compensating transformations are induced by these gauge fixings.

Next, we gauge fix the spatial dependence of the spatial translations by imposing the gauge fixing condition
\begin{equation}\label{gffs}
e_i{}^a(x^\nu) = \delta_i{}^a\,.
\end{equation}
Requiring $\delta e_i{}^a=0$ leads to the condition (with $t = x^0$)
\begin{equation}\label{solgf}
\xi^a(x^\nu)=\xi^a(t) -\lambda_{ai}x^i\,.
\end{equation}
 Note that after imposing the gauge fixing \eqref{gffs} the spatial part of space-time corresponds to a flat, Euclidean space. There is thus no longer any need to distinguish the $i$ and $a$ indices and upper and down indices and we will not do so in the following.

 At this stage the independent
 time-like and spatial Dreibein components and their projective inverses are given by
 \begin{alignat}{2}\label{FlatspatialcoordSNC}
\tau_\mu(x^\nu) &= \delta_{\mu}{}^\emptyset \,,&\hskip 2.5truecm  e_\mu{}^{a}(x^\nu) &= \bigl(-\tau^a(x^\nu)\,, \delta_i{}^a\bigr)\,,\nonumber\\[.15truecm]
\tau^\mu(x^\nu) &= \bigl(1\,, \tau^a(x^\nu) \bigr)\,,& e_a{}^{\mu}(x^\nu) &= \bigl(0\,, \delta^i{}_a\bigr)\,,
\end{alignat}
where the $\tau^a(x^\nu)$ are the only non-constant Dreibein components left. The only other independent gauge field left is the central charge gauge field $m_\mu(x^\nu)$. Ta\-king into account the compensating gauge transformation  given in
\eqref{solgf} we find that the remaining independent fields $\tau^a(x^\nu), m_\emptyset(x^\nu)$ and $m_i(x^\nu)$ transform as follows under the left-over transformations:
\begin{align}
\delta \tau^a(x^\nu) &= \lambda^a{}_b \tau^b(x^\nu) - \lambda^c{}_d x^d \partial_c \tau^a(x^\nu) + \xi^{\slashed 0} \partial_{\slashed 0} \tau^a(x^\nu) + \xi^j(t) \partial_j \tau^a(x^\nu)-\nonumber\\[.1truecm]
 &\ \ \ - \dot{\xi}^a(t) - \lambda^a (x^\nu) \,,\label{transftaui}\\[.2truecm]
\delta m_i(x^\nu) & =  \xi^{\slashed 0} \partial_{\slashed 0} m_i(x^\nu) + \xi^j(t) \partial_j m_i(x^\nu) + \lambda_i{}^j m_j(x^\nu) - \lambda^j{}_k x^k \partial_j m_i(x^\nu) +  \nonumber \\[.1truecm] &  \ \ \ +\lambda_i(x^\nu)
 + \partial_i \sigma(x^\nu)\,, \label{transfmi} \\[.2truecm]
\delta m_{\slashed 0}(x^\nu) & =  \xi^{\slashed 0} \partial_{\slashed 0} m_{\slashed 0}(x^\nu) + \dot{\xi}^i(t) m_i(x^\nu) + \xi^i(t) \partial_i m_{\slashed 0}(x^\nu) - \lambda^i{}_j x^j \partial_i m_{\slashed 0}(x^\nu) - \nonumber \\[.1truecm] &  \ \ \ -\lambda^a(x^\nu) \tau_a(x^\nu)+ \dot{\sigma}(x^\nu) \label{transfm0}\,,
\end{align}
where $\dot{}$ denotes a derivative with respect to $t$.

The three fields $\tau^a(x^\nu)$, $m_i(x^\nu)$ and $m_{\slashed 0}(x^\nu)$ are not independent.
Since the gauge field $\omega_\mu{}^{ab}(x^\nu)$ which we gauge fixed to zero, see eq.~\eqref{gffirst}, is dependent we need to investigate its consequences.  Using the other gauge fixing conditions as well, we find that the condition $\omega_\emptyset{}^{ab}(x^\nu)=0$ leads to the following restriction:
\begin{equation}\label{eq}
\partial_{[i}\tau_{j]}(x^\nu) + \partial_{[i}m_{j]}(x^\nu)=0\,.
\end{equation}
This implies that, locally,  one can write
\begin{equation}
\tau_i (x^\nu) + m_i(x^\nu) = \partial_i m(x^\nu) \,.
\end{equation}
Without loss of generality, we can thus eliminate $m_i(x^\nu)$ for $\tau_i(x^\nu)$ and $m(x^\nu)$, which is what we will do in the following.
The transformation rule for $m(x^\nu)$ can be found from $\delta \tau_i(x^\nu)$ and $\delta m_i(x^\nu)$:
\begin{equation} \label{transfm}
\delta m(x^\nu) = \xi^{\slashed 0} \partial_{\slashed 0} m(x^\nu) - \dot{\xi}^k(t) x^k + \xi^j(t) \partial_j m(x^\nu) - \lambda^j{}_k x^k \partial_j m(x^\nu) + \sigma(x^\nu) + Y(t) \,,
\end{equation}
where $Y(t)$ is an arbitrary time-dependent shift.
At this point we are left with three independent fields $\tau^i(x^\nu)$, $m_{\slashed 0}(x^\nu)$ and $m(x^\nu)$ whose transformation laws are given by \eqref{transftaui}, \eqref{transfm0}, \eqref{transfm}, respectively.

From the transformation rule \eqref{transfm}, we see that the central charge transformation, with local parameter $\sigma(x^\nu)$, acts as a St\"uckelberg shift on the field $m(x^\nu)$. We can thus partially fix the central charge transformations by imposing
\begin{equation} \label{centrcond}
m(x^\nu) = 0 \,.
\end{equation}
This fixes the central charge transformations according to
\begin{equation}\label{compcc}
\sigma (x^\mu) = \sigma(t) + \dot{\xi}^a(t)x_a\,,
\end{equation}
where it is understood that we also fix $Y(t)=-\sigma(t)$ in \eqref{transfm}.
After this gauge fixing the transformation rules of the two independent fields $\tau^i(x^\nu)$ and $m_{\slashed 0}(x^\nu)$ are given by:
\begin{align} \label{finallaws}
\delta \tau^i(x^\nu) &=  \lambda^i{}_j \tau^j(x^\nu) - \lambda^j{}_k x^k \partial_j \tau^i(x^\nu) + \xi^{\slashed 0} \partial_{\slashed 0} \tau^i(x^\nu) + \xi^j(t) \partial_j \tau^i(x^\nu) - \dot{\xi}^i(t) \nonumber \\[.1truecm] & \ \ \  - \lambda^i (x^\nu) \,, \nonumber \\[.2truecm]
\delta m_{\slashed 0}(x^\nu) &=  \xi^{\slashed 0} \partial_{\slashed 0} m_{\slashed 0}(x^\nu) - \dot{\xi}^i(t) \tau_i(x^\nu) + \xi^i(t) \partial_i m_{\slashed 0}(x^\nu) + \ddot{\xi}^k (t) x^k \nonumber \\[.1truecm] & \ \ \  - \lambda^i{}_j x^j \partial_i m_{\slashed 0}(x^\nu) - \lambda^i(x^\nu) \tau_i(x^\nu)  + \dot{\sigma}(t) \,.
\end{align}

Next, we note that the local boost transformations, with local parameters $\lambda^i(x^\nu)$, also end up as a Stueckelberg symmetry.
This Stueckelberg symmetry can be fixed by imposing the final gauge condition
\begin{equation}\label{gctau}
\tau^a(x^\nu)=0\,.
\end{equation}
This leads to the following compensating transformations:
\begin{equation}\label{comptau}
\lambda^i(x^\nu) =  - \dot{\xi}^i(t)\,.
\end{equation}
The only independent field left now is
\begin{equation}
 m_{\slashed 0}(x^\nu)\equiv \Phi(x^\nu)\,,
 \end{equation}
 which we will soon identify as the Newton potential.  Using the gauge condition \eqref{gctau} and taking into account the compensating transformations \eqref{comptau} we find that the transformation rule of this field is given by
\begin{equation}
\delta \Phi(x^\nu) = \xi^{\slashed 0} \partial_{\slashed 0} \Phi(x^\nu) + \xi^i(t) \partial_i \Phi (x^\nu)+ \ddot{\xi}^k(t) x^k - \lambda^i{}_j x^j \partial_i \Phi(x^\nu) + \dot{\sigma}(t) \,.
\label{gctPhiNC}
\end{equation}

The fact that we identify the field $m_{\slashed 0}(x^\nu)$ with the Newton potential $\Phi(x^\nu)$ is justified by noting that it satisfies the Poisson equation. To see this, first note that in terms of  $\Phi(x^\nu)$
the expression for the only non-zero dependent boost spin connection components, see eq.~\eqref{susyomegaa}, is given by
\begin{equation}\label{boostcomp}
\omega_\emptyset{}^a(x^\nu) = - \partial^a\Phi(x^\nu)\,.
\end{equation}
If we now plug this expression for the boost spin connection components into the
bosonic equation of motion given in eq.~\eqref{onshell} we find the expected Poisson equation for the Newton potential:
\begin{equation}
\triangle\Phi = \partial_a\partial_a\Phi=0\,.
\end{equation}
This equation is invariant under the so-called acceleration extended Galilei symmetries \eqref{gctPhiNC}.

The transformations \eqref{gctPhiNC}  close an algebra on $\Phi(x^\nu)$. 
One finds the following non-zero commutators:
\begin{align} \label{boscomm}
\left[ \delta_{\xi^{\slashed 0}}, \delta_{\xi^i(t)} \right] \Phi(x^\nu) & = \delta_{\xi^i(t)}\left(-\xi^{\slashed 0} \dot{\xi}^i(t)\right) \Phi(x^\nu) \,, \nonumber \\[.1truecm]
\left[ \delta_{\xi^{\slashed 0}}, \delta_{\sigma(t)} \right] \Phi(x^\nu) & =  \delta_{\sigma(t)} \left(- \xi^{\slashed 0} \dot{\sigma}(t) \right) \Phi(x^\nu) \,, \nonumber \\[.1truecm]
\left[ \delta_{\xi^i_1(t)}, \delta_{\xi^i_2(t)} \right] \Phi(x^\nu) & =  \delta_{\sigma(t)}\left(\dot{\xi}_1^j (t) \xi_2^j (t) - \dot{\xi}_2^j (t) \xi_1^j (t)\right) \Phi(x^\nu)\,, \nonumber \\[.1truecm]
\left[\delta_{\xi^i(t)} ,\delta_{\lambda^{jk}} \right] \Phi(x^\nu) & =  \delta_{\xi^i(t)}\left(\lambda^i{}_j \xi^j(t) \right) \Phi(x^\nu) \,,
\end{align}
where we have indicated the parameters of the transformations on the right-hand-side in the brackets.

This finishes our review of the bosonic case. For the convenience of the reader we have summarized
all gauge conditions and resulting compensating transformations in Table \ref{tabelgfbosonic}.

{\small
\begin{table}[h]
\begin{center}
\begin{tabular}{|c|c|}
\hline\rule[-1mm]{0mm}{6mm}
gauge condition/restriction\ \ \ &compensating transformation\ \ \ \\[.1truecm]
\hline\rule[-1mm]{0mm}{6mm}
$\tau_\mu(x^\nu)=\delta_\mu{}^\emptyset$&$\xi^\emptyset(x^\nu) =\xi^\emptyset$\\[.1truecm]
$\omega_\mu{}^{ab}(x^\nu)=0$&$\lambda^{ab}(x^\nu)=\lambda^{ab}$\\[.1truecm]
$e_i{}^a(x^\nu)=\delta_i{}^a$&$\xi^a(x^\nu) = \xi^a(t)-\lambda_{ai}x^i$\\[.1truecm]
$\tau_i(x^\nu) + m_i(x^\nu) = \partial_i m(x^\nu)$& --\\[.1truecm]
$m(x^\nu)=0$&$\sigma(x^\nu) = \sigma(t)+{\dot{\xi}}^a(t) x_a$\\[.1truecm]
$\tau^a(x^\nu)=0$&$\lambda^i(x^\nu) = -{\dot \xi}^i(t)$\\[.1truecm]
\hline\hline\rule[-1mm]{0mm}{6mm}
$m_\emptyset(x^\nu) = \Phi(x^\nu)$&$\omega_\emptyset{}^a(x^\nu) = -\partial^a\Phi(x^\nu)$\\[.1truecm]
\hline
\end{tabular}
\caption{This table indicates the gauge fixing conditions and corresponding compensating transformations that allow one to obtain Newtonian gravity, in frames that are rectilinearly accelerating with respect to inertial reference frames, from NC gravity. We have also included the restrictions that follow from the fact that the spin connection field $\omega_\mu{}^{ab}$ is dependent. At the bottom of the table we have summarized the expressions of the
non-zero remaining gauge fields in terms of the Newton potential $\Phi(x^\nu)$.\label{tabelgfbosonic}}
\end{center}
 \end{table}
 }

\subsubsection{The Supersymmetric Case}

Our aim is now to include the fermions and local supersymmetry transformations in the discussion of the previous subsection and to perform a partial gauge fixing of the bosonic and fermionic symmetries to
derive the NC supergravity theory from the Galilean observer point of view. We will define a supersymmetric Galilean observer as one for which only a supersymmetric
extension of the acceleration extended Galilei symmetries are retained. We will see below that only half of the supersymmetries are gauge fixed to constant ones, due to the fact that the only translations that are gauge fixed to constants are the time translations.

First, we solve the constraints \eqref{constraints}, but now keeping the fermions,   by imposing the gauge fixing conditions
\begin{equation}\label{gffirstsuper}
\tau_\mu(x^\nu) = \delta_\mu{}^\emptyset\,,\hskip 1truecm \omega_\mu{}^{ab}(x^\nu)=0\,, \hskip 1truecm \psi_{\mu +}(x^\nu)=0\,.
\end{equation}
This fixes the local time translations, spatial rotations and $\epsilon_+$ transformations to constant transformations:
 \begin{equation}
 \xi^\emptyset(x^\nu) = \xi^\emptyset\,,\hskip 1truecm \lambda^{ab}(x^\nu)=\lambda^{ab}\,,\hskip 1truecm \epsilon_+(x^\nu) = \epsilon_+\,.
 \end{equation}
 No compensating transformations are induced by these gauge fixings.

Next, we  partially gauge fix the spatial translations by imposing the gauge choice
\begin{equation}\label{gf11}
e_i{}^a(x^\nu)=\delta_i{}^a\,.
\end{equation}
This gauge choice implies that we may use from now on the expressions \eqref{FlatspatialcoordSNC}
for the time-like and spatial Dreibein components and their projective inverses. We will derive the required compensating transformation below.
First, using the above gauge choices and the fact that the
purely spatial components ${\hat R}_{ij}{}^a(G)$ of the curvature of boost transformations
and the purely spatial components  ${\hat\psi}_{ij -}$ of the curvature of $\epsilon_-$ transformations are zero, we derive from their expressions (see eq.~\eqref{curvatures}) that
\begin{equation}
\partial_{[i}\omega_{j]}{}^a=0\,,\hskip 2truecm \partial_{[i}\psi_{j]-}=0\,.
\end{equation}
The first equation we solve locally by writing
 \begin{equation}
 \omega_i{}^a = \partial_i\omega^a\,,
 \end{equation}
where $\omega^a$ is a dependent field since $\omega_i{}^a$ is dependent. This also explains why we have not added a purely time-dependent
piece to the r.h.s.~of the above solution.

We subsequently partially gauge fix the $\epsilon_-$ transformations by imposing the gauge choice
\begin{equation}\label{gcgravitino}
\psi_{i-}(x^\nu)=0\,.
\end{equation}
This fixes the $\epsilon_-$ transformations according to
\begin{equation}
\epsilon_-(x^\nu) = \epsilon_-(t) - \frac12 \omega^a \gamma_{a0} \epsilon_+ \,.
\end{equation}
Given the gauge choice \eqref{gcgravitino}, the spatial translations are now fixed without
the need for any fermionic compensating transformation. Indeed, from the total variation of the gauge fixing condition \eqref{gf11} we find:
\begin{equation}
\xi^i (x^\nu) = \xi^i(t) - \lambda^i{}_j x^j \,.
\end{equation}

At this point, we are left with the remaining fields $\tau^a$, $m_i$, $m_{\slashed 0}$ and $\psi_{\slashed 0 -}$. These fields are
not independent since the gauge field $\omega_\mu{}^{ab}$ which we gauge fixed to zero is dependent,
 see eq.~\eqref{susyomegaab}. Like in the bosonic case, only the gauge fixing
 $\omega_{\slashed 0}{}^{ab}=0$ leads to a restriction:
\begin{equation}
\partial_{[i} \left(\tau_{j]} + m_{j]} \right)(x^\nu) = 0 \,.
\end{equation}
As in the bosonic case, this implies that we can locally write:
\begin{equation} \label{defm}
\tau_i(x^\nu) + m_i(x^\nu) = \partial_i m (x^\nu)\,.
\end{equation}
Without loss of generality we will use this equation to eliminate $m_i$ in terms of the other two fields. The variation of $m$ is determined by writing the variation of $\tau_i + m_i$ as a $\partial_i$-derivative. This is trivial for most of the terms, except for the $\epsilon_+$ term. Before addressing this issue below, it is convenient to write down the total variation of $\partial_i m$ instead of $m$. From eq.~\eqref{defm} we find
\begin{align}
\delta \partial_i m &= \xi^{\slashed 0} \partial_{\slashed 0} \partial_i m + \xi^j(t) \partial_j \partial_i m
 +\lambda_i{}^j\partial_j m - \lambda^m{}_n x^n \partial_m \partial_i m + \partial_i \sigma(x^\nu) - \dot{\xi}^i(t) \nonumber \\ & \ \ \ - \frac12 {\bar\epsilon}_+\gamma_i  \psi_{\slashed 0-} \,.
\end{align}
Note that the terms proportional to the local boost parameters $\lambda^i(x^\nu)$ have cancelled out.

We  now partially gauge fix the central charge transformations by putting
\begin{equation}
m(x^\nu)=0\,.
\end{equation}
We thus obtain
\begin{equation}
\partial_i\sigma(x^\nu) = {\dot\xi}^i(t) +
\frac12 {\bar\epsilon}_+\gamma_i  \psi_{\slashed 0-}(x^\nu) \,,
\end{equation}
which is sufficient to calculate the transformation rule of $\partial_i m_{\slashed 0}$. After this gauge fixing, taking into account all the compensating transformations, see table \ref{gfsusy} below, and the restriction \eqref{defm} with $m=0$ substituted, we find the following transformation rules for the remaining independent fields:
\begin{align}
\delta \tau_i & =  \xi^{\slashed 0} \partial_{\slashed 0} \tau_i + \xi^j(t) \partial_j \tau_i - \dot{\xi}^i(t) + \lambda_{ij} \tau^j - \lambda^k{}_l x^l \partial_k \tau_i - \lambda_i (x^\nu) - \frac12 \bar{\epsilon}_+ \gamma_i \psi_{\slashed 0 -} \,, \nonumber \\[.2truecm]
\delta \partial_i m_{\slashed 0} & =  \xi^{\slashed 0} \partial_{\slashed 0}\partial_i m_{\slashed 0} + \xi^j(t) \partial_j \partial_i m_{\slashed 0} + \ddot{\xi}^i(t)  - \dot{\xi}^j(t)\partial_i \tau_j +\lambda_i{}^j\partial_j m_{\slashed 0}  - \lambda^m{}_n x^n \partial_m \partial_i m_{\slashed 0} -
\nonumber\\[.1truecm]
& \ \ \ - \partial_i\big(\lambda^j(x^\nu) \tau_j\big)   + \bar{\epsilon}_- (t) \gamma^0\partial_i \psi_{\slashed 0 -} + \frac12\partial_i\big( \omega^a \bar{\epsilon}_+ \gamma_a \psi_{\slashed 0-}\big) + \frac12 {\bar\epsilon}_+ \gamma_i  {\dot \psi}_{\slashed 0-} \,,  \\[0truecm]
\delta \psi_{\slashed 0-} & =  \xi^{\slashed 0} \partial_{\slashed 0} \psi_{\slashed 0-} + \xi^i(t) \partial_i \psi_{\slashed 0 -} - \lambda^i{}_j x^j \partial_i \psi_{\slashed 0-} + \frac14 \lambda^{ab} \gamma_{ab} \psi_{\slashed 0 -}  \nonumber \\[-.1truecm] & \ \ \  + \dot{\epsilon}_-(t) + \frac12 \left(\omega_{\slashed 0}{}^a - \dot{\omega}^a \right) \gamma_{a0} \epsilon_+ \,.\nonumber
\end{align}
Note that $\omega_{\slashed 0}{}^a$ and $\omega^a$  depend on the fields $\tau_i$, $m_{\slashed 0}$.
Using expression \eqref{susyomegaa} for the dependent boost gauge field $\omega_\mu{}^a$ one can calculate that
\begin{align}
\omega_i{}^a &\equiv \partial_i\omega^a = -\partial_i\tau^a \hskip .5truecm \rightarrow\hskip .5truecm \omega^a = -\tau^a\,,\\[.1truecm]
\omega_{\slashed 0}{}^a &= -\dot{\tau}^a - \partial_a \left(m_{\slashed 0} - \frac12 \tau^i \tau^i \right)\,.
\end{align}

As a final step we now  fix the local boost transformations by imposing
\begin{equation}
\tau^i(x^\nu)=0\,,
\end{equation}
which leads to the following compensating transformations:
\begin{equation}
\lambda^i(x^\nu) = -{\dot\xi}^i(t) -\frac12 \bar{\epsilon}_+ \gamma_i \psi_{\slashed 0 -}(x^\nu) \,.
\end{equation}
One now finds that
\begin{equation}
\omega^a=0\,,\hskip 2truecm \omega_{\slashed 0}{}^a = -\partial^a m_{\slashed 0} \equiv - \partial^a\Phi\,,
\end{equation}
where $\Phi$ is the Newton potential. In
terms of the `Newton force' $\Phi_i$ and its supersymmetric partner $\Psi$ defined  by
\begin{equation}
\Phi_i = \partial_i\Phi\,,\hskip 2truecm \Psi = \psi_{\slashed 0 -}\,,
\end{equation}
one thus obtains the following transformation rules:
\begin{align}
\delta  \Phi_i &=  \xi^{\slashed 0} \partial_{\slashed 0} \Phi_i + \xi^j(t) \partial_j \Phi_i + \ddot{\xi}^i(t) + \lambda_i{}^j  \Phi_j- \lambda^m{}_n x^n \partial_m \Phi_i  + \bar{\epsilon}_-(t) \gamma^0 \partial_i \Psi \nonumber \\[.2truecm] & \ \ \  +  \frac12 \bar{\epsilon}_+ \gamma_i \dot{\Psi}\label{transfforce} \,,\\[.2truecm]
\delta \Psi &= \xi^{\slashed 0} \partial_{\slashed 0} \Psi + \xi^i(t) \partial_i \Psi - \lambda^i{}_j x^j \partial_i \Psi + \frac14 \lambda^{ab} \gamma_{ab} \Psi   + \dot{\epsilon}_-(t) - \frac12 \Phi^i  \gamma_{i0} \epsilon_+ \label{transfPsi}\,.
\end{align}
Note that the central charge transformations only act on the Newton potential, not on the Newton force.

 Determining the transformation rule of the Newton potential $\Phi$ is non-trivial, due to the fact that the last term of \eqref{transfforce} cannot be manifestly written as a $\partial_i$-derivative.
The above transformation rules are consistent with the integrability condition
\begin{equation}\label{integrability}
\partial_{[i} \Phi_{j]}(x^\nu) = 0\,,
\end{equation}
 by virtue of the fermionic equations of motion given in eqs.~\eqref{moreconstraints} and \eqref{onshell}  which, after gauge fixing, take on the form
 \begin{equation}\label{feom}
\gamma^i\partial_i\Psi(x^\nu)=0\hskip .5truecm \Leftrightarrow\hskip .5truecm  \partial_{[i}\gamma_{j]}\Psi(x^\nu)=0\,.
 \end{equation}
Under supersymmetry these fermionic equations of motion lead to the following bosonic equation of motion:
\begin{equation}
\partial^i\Phi_i(x^\nu)=0\,.
\end{equation}
The same bosonic equation of motion also follows from the last equation in \eqref{onshell} after gauge fixing.

In order to obtain transformation rules for the Newton potential $\Phi$ and its fermionic superpartner, we need to solve the fermionic equation of motion/constraint \eqref{feom}. The second form of this constraint makes it clear that the equations of motion are solved by a spinor $\chi$, that obeys:
\begin{equation} \label{defchi}
\gamma_i \Psi = \partial_i \chi \,.
\end{equation}
Note that this only determines $\chi$ up to a purely time-dependent shift. From \eqref{defchi}, it follows that $\chi$ obeys the constraint:
\begin{equation} \label{constrchi}
\gamma^1 \partial_1 \chi = \gamma^2 \partial_2 \chi \,.
\end{equation}
$\Psi$ can thus be expressed in terms of $\chi$ in a number of equivalent ways:
\begin{equation}
\Psi = \gamma^1 \partial_1 \chi = \gamma^2 \partial_2 \chi = \frac12 \gamma^i \partial_i \chi \,.
\end{equation}
It is now possible to determine the transformation rule of $\Phi$ by rewriting $\delta \Phi_i$ as a $\partial_i$-derivative:
\begin{equation}
\delta \Phi_i = \partial_i (\delta \Phi) \,.
\end{equation}
The resulting transformation rule for the Newton potential is
\begin{equation}
\delta \Phi = \xi^{\slashed 0} \partial_{\slashed 0} \Phi + \xi^i(t) \partial_i \Phi + \ddot{\xi}^i(t) x^i - \lambda^m{}_n x^n \partial_m \Phi + \frac12 \bar{\epsilon}_-(t) \gamma^{0i} \partial_i \chi + \frac12 \bar{\epsilon}_+ \dot{\chi} + \sigma(t) \,.
\end{equation}
Note that we have allowed for an arbitrary time-dependent shift $\sigma(t)$ in the transformation rule, whose origin stems from the fact that $\Phi_i = \partial_i \Phi$ only determines $\Phi$ up to an arbitrary time-dependent shift.

To determine the transformation rule of $\chi$, we try to rewrite $\gamma_i \delta \Psi$ as a $\partial_i$-derivative:
\begin{equation}
\gamma_i \delta \Psi = \partial_i (\delta \chi) \,.
\end{equation}
Most of the terms in $\gamma_i \delta \Psi$ can be straightforwardly written as a $\partial_i$-derivative. Only for the $\epsilon_+$ transformation, the argument is a bit subtle. We thus focus on the terms in $\gamma_i \delta \Psi$, given by
\begin{equation}
-\frac12 \gamma_i \Phi^j \gamma_{j0} \epsilon_+ = -\frac12 \gamma_i \partial^j \Phi \gamma_{j0}  \epsilon_+ = -\frac12 \partial^j \Phi \gamma_{ij0} \epsilon_+ - \frac12 \partial_i \Phi \gamma_0 \epsilon_+ \,.
\end{equation}
The last term is already in the desired form. To rewrite the first term in the proper form, we note that the Newton potential $\Phi$ can be dualized to a `dual Newton potential' $\Xi$ via
\begin{equation} \label{defxi}
\partial_i \Phi = \varepsilon_{ij} \partial^j \Xi \,, \qquad \partial_i \Xi = - \varepsilon_{ij} \partial^j \Phi\,.
\end{equation}
Using the convention that $\gamma_{ij0} = \epsilon_{0ij} = \epsilon_{ij}$, we then get
\begin{equation}
-\frac12 \gamma_i \Phi^j \gamma_{j0} \epsilon_+ =
\frac12 \partial_i \Xi \epsilon_+ - \frac12 \partial_i \Phi \gamma_0 \epsilon_+ \,.
\end{equation}
One thus obtains the following transformation rule for $\chi$, which includes the dual Newton potential $\Xi$:
\begin{eqnarray} \label{rulechi}
\delta \chi &=& \xi^{\slashed 0} \partial_{\slashed 0} \chi +  \xi^i(t) \partial_i \chi - \lambda^m{}_n x^n \partial_m \chi + \frac14 \lambda^{mn} \gamma_{mn} \chi \nonumber\\[.1truecm]
&&+\,\, x^i \gamma_i\dot{\epsilon}_-(t) + \frac12 \Xi \epsilon_+ - \frac12 \Phi \gamma_0 \epsilon_+ + \eta(t)\,.
\end{eqnarray}
Note that we have again allowed for a purely time-dependent shift $\eta(t)$, whose origin lies in the fact that \eqref{defchi} only determines $\chi$ up to a purely time-dependent shift.

In order to calculate the algebra on $\Phi$ and  $\chi$, we also need the transformation rule of the dual potential $\Xi$. This rule is determined by dualizing the transformation rule of $\Phi$:
\begin{equation}
\partial_i(\delta \Xi) = -\varepsilon_{ij} \partial^j (\delta \Phi) \,.
\end{equation}
By repeatedly using \eqref{defchi} and \eqref{defxi}, we obtain:
\begin{eqnarray}
\delta \Xi &=& \xi^{\slashed 0} \partial_{\slashed 0} \Xi + \xi^i(t) \partial_i \Xi + \ddot{\xi}^i(t) \varepsilon_{ij} x^j - \lambda^m{}_n x^n \partial_m \Xi\nonumber\\[.1truecm]
&& +\,\, \frac12 \bar{\epsilon}_-(t) \gamma^i \partial_i \chi - \frac12 \bar{\epsilon}_+ \gamma_0 \dot{\chi} + \tau(t) \,,
\end{eqnarray}
where we again allowed for a purely time-dependent shift $\tau(t)$.

The algebra then closes on $\Phi$ and $\chi$, using \eqref{defchi}, \eqref{constrchi}, \eqref{defxi} . One finds the following non-zero commutators between the fermionic symmetries:
\begin{align}
\left[\delta_{\epsilon_{1-}(t)},\delta_{\epsilon_{2-}(t)} \right] &= \delta_{\sigma(t)}\left(\frac{\rmd}{\rmd t} \left(\bar{\epsilon}_{2-}(t) \gamma^0 \epsilon_{1-}(t) \right)\right) \,,\nonumber \\
\left[\delta_{\epsilon_{1+}},\delta_{\epsilon_{2+}} \right] &= \delta_{\xi^{\slashed 0}} \left(\frac12 \left(\bar{\epsilon}_{2+} \gamma^0 \epsilon_{1+}  \right) \right) \,, \nonumber \\
\left[\delta_{\epsilon_{+}},\delta_{\epsilon_{-}(t)} \right] &= \delta_{\xi^i(t)} \left( \frac12 \left(\bar{\epsilon}_-(t) \gamma^i \epsilon_+ \right)\right) \,, \nonumber \\
\left[\delta_{\eta(t)}, \delta_{\epsilon_+} \right] &= \delta_{\sigma(t)}\left(\frac12\left( \bar{\epsilon}_+ \dot{\eta}(t) \right) \right) \,.
\end{align}
The non-zero commutators between the bosonic and fermionic symmetries are given by:
\begin{alignat}{2}
\left[\delta_{\xi^i(t)},\delta_{\epsilon_{+}} \right] &= \delta_{\epsilon_{-}(t)} \left(\frac12 \dot{\xi}^i(t) \gamma_{0i} \epsilon_+ \right)\,, &\qquad \left[\delta_{\lambda^{ij}},\delta_{\epsilon_{+}} \right] &= \delta_{\epsilon_+} \left( -\frac14 \lambda^{ij} \gamma_{ij} \epsilon_+ \right) \,, \nonumber \\
\left[\delta_{\xi^{\slashed 0}},\delta_{\epsilon_{-}(t)} \right] &= \delta_{\epsilon_{-}(t)} \left(- \xi^{\slashed 0} \dot{\epsilon}_-(t) \right)\,, & \left[\delta_{\xi^i(t)},\delta_{\epsilon_{-}(t)} \right] &= \delta_{\eta(t)} \left( -\xi^i(t) \gamma_i \dot{\epsilon}_-(t) \right)\,, \nonumber \\
\left[\delta_{\lambda^{ij}},\delta_{\epsilon_{-}(t)} \right] &= \delta_{\epsilon_{-}(t)} \left( -\frac14 \lambda^{ij} \gamma_{ij} \epsilon_-(t) \right)\,, & \left[\delta_{\sigma(t)},\delta_{\epsilon_{+}} \right] &= \delta_{\eta(t)} \left(\frac12\left(\sigma(t) \gamma^0 \epsilon_+ \right) \right) \,, \nonumber \\
\left[\delta_{\xi^{\slashed 0}},\delta_{\eta(t)} \right] &= \delta_{\eta(t)} \left(-\xi^{\slashed 0} \dot{\eta}(t) \right)\,, & \left[\delta_{\lambda^{ij}},\delta_{\eta(t)} \right] &= \delta_{\eta(t)} \left( -\frac14 \lambda^{ij} \gamma_{ij} \eta(t) \right) \,.
\end{alignat}
The bosonic commutators are not changed with respect to the purely bosonic case and are given by \eqref{boscomm}.

This finishes our discussion of the ${\cal N}=2$ Galilean supergravity theory. Like in the bosonic case, we have summarized all gauge fixing conditions and resulting compensating transformations in table 4. In the next section we will discuss another type of 3$D$ non-Lorentzian supergravity theories that are of a Chern-Simons form based on a (super-)Lie algebra.

{\small
\begin{table}[h]
\begin{center}
\begin{tabular}{|c|c|}
\hline\rule[-1mm]{0mm}{6mm}
gauge condition/restriction&compensating transformation\\[.1truecm]
\hline\rule[-1mm]{0mm}{6mm}
$\tau_\mu(x^\nu)=\delta_\mu{}^\emptyset$&$\xi^\emptyset(x^\nu) =\xi^\emptyset$\\[.1truecm]
$\omega_\mu{}^{ab}(x^\nu)=0$&$\lambda^{ab}(x^\nu)=\lambda^{ab}$\\[.1truecm]
$\psi_{\mu +}(x^\nu)=0$&$\epsilon_+(x^\nu)=\epsilon_+$\\[.1truecm]
$e_i{}^a(x^\nu)=\delta_i{}^a$&$\xi^i(x^\nu) = \xi^i(t)-\lambda^i{}_jx^j$\\[.1truecm]
$\psi_{i-}(x^\nu)=0$&\ \ \ $\epsilon_-(x^\nu) = \epsilon_-(t) -\tfrac{1}{2}\omega^a(x^\nu)\gamma_{a0}\epsilon_+$\ \ \ \\[.1truecm]
$\tau_i(x^\nu) + m_i(x^\nu) = \partial_i m(x^\nu)$& --\\[.1truecm]
$m(x^\nu)=0$&$\partial_i\sigma(x^\nu) = {\dot\xi}^i(t) +
\frac12 {\bar\epsilon}_+\gamma_i  \psi_{\slashed 0-}(x^\nu)$\\[.1truecm]
$\tau^a(x^\nu)=0$&$\lambda^i(x^\nu) = -{\dot\xi}^i(t) -\frac12 \bar{\epsilon}_+ \gamma_i \psi_{\slashed 0 -}(x^\nu)$\\[.1truecm]
\hline\hline\rule[-1mm]{0mm}{6mm}
$m_\emptyset(x^\nu) = \Phi(x^\nu)\,,\ \ \omega_\emptyset{}^a(x^\nu) = -\partial^a\Phi(x^\nu) $&$ \psi_{\emptyset -}(x^\nu)=\Psi(x^\nu) $\\[.1truecm]
\hline
\end{tabular}
\caption{This table indicates the gauge fixing conditions and corresponding compensating transformations that lead to  3$D$ NC supergravity from the viewpoint of a Galilean observer. We have also included the restrictions that follow from the fact that the spin connection field $\omega_\mu{}^{ab}$ is dependent. At the bottom of the table we have summarized the expressions of the
non-zero remaining gauge fields in terms of the Newton potential $\Phi(x^\nu)$ and its supersymmetric partner
$\chi(x^\nu)$, which is related to $\Psi(x^\nu)$ via \eqref{defchi}.\label{gfsusy}}
\end{center}
 \end{table}
 }

\section{Non-Relativistic 3$D$ Chern-Simons Supergravity and Lie Algebra Expansions} \label{sec:EBGexp}

Above, we discussed the supergravity version of 3$D$ NC gravity, the diffeomorphism covariant formulation of Newtonian gravity. Although we focused on the construction of this theory via a NR limit, we mentioned that it can also be constructed as a gauging of the supersymmetric extension of the Bargmann algebra, given in \eqref{3dsuperbargmann} \cite{Andringa:2013mma}. This gauging perspective highlights an important structural difference between 3$D$ NC (super)gravity and its relativistic Einstein-Hilbert counterpart: while the latter can be written as a Chern-Simons theory \cite{Achucarro:1986uwr,Achucarro:1989gm,Witten:1988hc}, the former can not. This can be seen from the fact that the Bargmann algebra can not be equipped with an invariant bilinear form that is required to construct a Chern-Simons action.

It is natural to ask whether there also exist NR 3$D$ supergravity theories that admit a first-order Chern-Simons formulation. A hint that this is possible is provided by the existence of a purely bosonic 3$D$ NR gravity theory in Chern-Simons form \cite{Papageorgiou:2009zc}. This relies on the fact that the 3$D$ Bargmann algebra admits an extra central extension. The non-zero commutation relations of this centrally extended Bargmann algebra are given by:
\begin{alignat}{3}
  \label{eq:Bargmann2cc}
  \left[H, G_a\right] &= -\epsilon_{ab} P_b \,, \quad & \left[J, G_a\right] &= -\epsilon_{ab} G_b \,, \quad &  \left[J, P_a\right] &= -\epsilon_{ab} P_b \,, \nonumber \\ \left[G_a, G_b\right] &= \epsilon_{ab} S \,,  \quad &
  \left[G_a, P_b\right] &= \epsilon_{ab} M \,.
\end{alignat}
Here, $S$ denotes the extra central extension and we have replaced $J_{ab}$ by $\epsilon_{ab} J$ and $G_a$ by $\epsilon_{ab} G_b$ in the bosonic part of \eqref{3dsuperbargmann}.
In contrast to the Bargmann algebra, the extended Bargmann algebra \eqref{eq:Bargmann2cc} can be equipped with a trace, i.e., a non-degenerate, invariant bilinear form. Explicitly, this trace takes on the following non-zero values when evaluated on the generators of the extended Bargmann algebra \cite{Papageorgiou:2009zc}:
\begin{equation}
  \label{eq:bilformnonrel}
  \mathrm{tr}(G_a\, P_b) = \delta_{ab} \,, \qquad \quad \mathrm{tr}(H \, S) =  \mathrm{tr}(M\, J) = -1 \,.
\end{equation}
One can then work out the generic form of the Chern-Simons action
\begin{equation}
  \label{eq:CSaction}
  S = \frac{k}{4 \pi} \int \mathrm{tr}\left(A \wedge dA + \frac23 A \wedge A \wedge A \right) \,,
\end{equation}
with $k$ the Chern-Simons coupling constant and $A = A_\mu \rmd x^\mu$ a Lie algebra-valued gauge field, for the algebra \eqref{eq:Bargmann2cc} and trace \eqref{eq:bilformnonrel}. Parametrizing the gauge field $A$ as follows
\begin{equation}
  \label{eq:Aexp}
  A_\mu = \tau_\mu \, H + e_\mu{}^a \, P_a + \omega_\mu\, J + \omega_\mu{}^a \, G_a + m_\mu\, M + s_\mu\, S \,.
\end{equation}
this leads to the following action \cite{Papageorgiou:2009zc}
\begin{align}
  \label{eq:CSactionnonrelbos}
  S &= \frac{k}{4 \pi} \int \, \rmd^3 x \, \Big( \epsilon^{\mu\nu\rho} e_\mu{}^a R_{\nu\rho a}(G) - \epsilon^{\mu\nu\rho} m_\mu R_{\nu\rho}(J) - \epsilon^{\mu\nu\rho} \tau_\mu R_{\nu\rho}(S) \Big) \,,
\end{align}
where
\begin{alignat}{2}
  \label{eq:covcurvsBargmann}
  R_{\mu\nu}{}^a(G) &\equiv 2 \partial_{[\mu} \omega_{\nu]}{}^a + 2 \epsilon^{ab} \omega_{[\mu} \omega_{\nu] b} \,, \qquad \quad & R_{\mu\nu}(J) &\equiv 2 \partial_{[\mu} \omega_{\nu]} \,, \nonumber \\
  R_{\mu\nu}(S) &\equiv 2 \partial_{[\mu} s_{\nu]} + \epsilon^{ab} \omega_{[\mu a} \omega_{\nu] b} \,.
\end{alignat}
Upon examining the gauge transformation rule of $A_\mu$, one finds that the fields $\tau_\mu$, $e_\mu{}^a$ and $m_\mu$ transform under local boosts $G_a$, spatial rotations $J$ and the central charge $M$ as in \eqref{eq:deltatauem}, showing that they can be identified as the time-like and spatial Vielbeine and the central charge gauge field of NC geometry. The fields $\omega_\mu$ and $\omega_\mu{}^a$ can likewise be interpreted as spin connections for local spatial rotations and boosts, respectively. Varying \eqref{eq:CSactionnonrelbos} with respect to $\omega_\mu$ and $\omega_\mu{}^a$ yields:
\begin{align} \label{eq:EBGconv}
  R_{\mu\nu}{}^a(P) &\equiv 2 \partial_{[\mu} e_{\nu]}{}^a + 2 \epsilon^{ab} \omega_{[\mu} e_{\nu]b} - 2 \epsilon^{ab} \omega_{[\mu b} \tau_{\nu]} = 0 \,, \nonumber \\
  R_{\mu\nu}(M) &\equiv 2 \partial_{[\mu} m_{\nu]} + 2 \epsilon^{ab} \omega_{[\mu a} e_{\nu] b} = 0 \,.
\end{align}
These are the conventional constraints \eqref{eq:convBargconstraints} (adapted to the notation and conventions of \eqref{eq:Bargmann2cc}) that can be used to express the spin connections $\omega_\mu$ and $\omega_\mu{}^a$ as dependent fields. Note that the torsion tensors $T_{\mu\nu}{}^a$ and $T_{\mu\nu}^{(m)}$ are zero here, as is the intrinsic torsion, since varying \eqref{eq:CSactionnonrelbos} with respect to $s_\mu$ yields the absolute time constraint \eqref{eq:abstime}. Extremizing the action \eqref{eq:CSactionnonrelbos} thus yields torsionless NC geometries. The other equations of motion that follow from varying \eqref{eq:CSactionnonrelbos} with respect to $\tau_\mu$, $e_\mu{}^a$ and $m_\mu$ are obtained by setting the curvatures \eqref{eq:covcurvsBargmann} equal to zero. The first two of these imply that the NC geometries involved have vanishing Riemann curvature. This is analogous to 3$D$ relativistic gravity, so that the Chern-Simons action \eqref{eq:CSactionnonrelbos} can be viewed as a first-order formulation of a NR gravity theory in 3 dimensions. This theory is called `extended Bargmann gravity'.

A supergravity generalization of extended Bargmann gravity can be constructed, if one can find a superalgebra with a non-degenerate invariant supertrace that contains (\ref{eq:Bargmann2cc}) as a bosonic subalgebra. An example of such a superalgebra was found in \cite{Bergshoeff:2016lwr} by trial and error. It extends \eqref{eq:Bargmann2cc}  with three fermionic generators $Q^+$, $Q^-$ and $R$ (that are Majorana spinors) and its non-zero (anti-)commutation relations are given by \eqref{eq:Bargmann2cc}, as well as the following ones:
\begin{alignat}{2}
  \label{eq:superBargmann2cc}
 & [J, Q^\pm] = -\frac12 \gamma_0 Q^\pm \,,  \ & & \quad [J, R] = -\frac12 \gamma_0 R \,, \quad \qquad
 [G_a, Q^+] = -\frac12 \gamma_a Q^- \,,  \nonumber \\  & [G_a, Q^-] = -\frac12 \gamma_a R \,, \ & & \quad [S, Q^+] = - \frac12 \gamma_0 R \,, \nonumber \\  & \{Q^+_{\alpha},Q^+_{\beta}\} = (\gamma_0 C^{-1})_{\alpha\beta} H \,, \ & & \quad \{Q^+_{\alpha},Q^-_{\beta}\} = -(\gamma^a C^{-1})_{\alpha\beta} P_a \,, \nonumber \\   & \{Q^-_{\alpha},Q^-_{\beta}\} = (\gamma_0 C^{-1})_{\alpha\beta} M \,, \ & & \quad \{Q^+_{\alpha}, R_\beta\} = (\gamma_0 C^{-1})_{\alpha\beta} M \,.
\end{alignat}
This `extended Bargmann superalgebra' can be equipped with an invariant supertrace that is given by (\ref{eq:bilformnonrel}), extended with
\begin{equation}
  \label{eq:supertrace}
  \mathrm{tr}\left(Q^+_\alpha\, R_\beta\right) = 2 (C^{-1})_{\alpha\beta} \,, \quad \mathrm{tr}\left(Q^-_\alpha\, Q^-_\beta\right) = 2 (C^{-1})_{\alpha\beta} \,.
\end{equation}
Introducing the gauge field
\begin{align}
  \label{eq:Asuper}
  A_\mu &= \tau_\mu H + e_{\mu}{}^a P_a + \omega_\mu J + \omega_\mu{}^a G_a + m_\mu M + s_\mu S \nonumber \\ & \qquad  + \bar{\psi}^+_\mu Q^+ + \bar{\psi}^-_\mu Q^- + \bar{\rho}_\mu R \,,
\end{align}
where $\psi_\mu^\pm$, $\rho_\mu$ are Majorana, the Chern-Simons action for the superalgebra \eqref{eq:Bargmann2cc}, \eqref{eq:superBargmann2cc} is found to be given by
\begin{align}
  \label{eq:CSactionnonrelsuper}
  S &= \frac{k}{4 \pi} \int \, \rmd^3 x \, \Big( \epsilon^{\mu\nu\rho} e_\mu{}^a R_{\nu\rho a}(G) - \epsilon^{\mu\nu\rho} m_\mu R_{\nu\rho}(J)  - \epsilon^{\mu\nu\rho} \tau_\mu R_{\nu\rho}(S) \nonumber \\ & \qquad \qquad + \epsilon^{\mu\nu\rho} \bar{\psi}^+_\mu \hat{\rho}_{\nu\rho} + \epsilon^{\mu\nu\rho} \bar{\rho}_\mu \hat{\psi}^+_{\nu\rho}  + \epsilon^{\mu\nu\rho} \bar{\psi}^-_\mu \hat{\psi}^-_{\nu\rho}\Big) \,.
\end{align}
Here, the supercovariant curvatures are given by
\begin{align}
  \label{eq:supcovcurvs}
& \hat{\psi}^+_{\mu\nu} = 2 \partial_{[\mu} \psi^+_{\nu]} + \omega_{[\mu} \gamma_0 \psi_{\nu]}^+ \,, \nonumber \\
&  \hat{\psi}^-_{\mu\nu} = 2 \partial_{[\mu} \psi^-_{\nu]} + \omega_{[\mu} \gamma_0 \psi_{\nu]}^- + \omega_{[\mu}{}^a \gamma_a \psi_{\nu]}^+ \,, \nonumber \\
& \hat{\rho}_{\mu\nu} = 2 \partial_{[\mu} \rho_{\nu]} + \omega_{[\mu} \gamma_0 \rho_{\nu]} + \omega_{[\mu}{}^a \gamma_a \psi_{\nu]}^- + s_{[\mu} \gamma_0 \psi_{\nu]}^+ \,.
\end{align}
Denoting the parameters of the local $Q^\pm$ and $R$ transformations by $\epsilon^\pm$ and $\eta$ respectively, one finds that the action \eqref{eq:CSactionnonrelsuper} is invariant under the following supersymmetry transformation rules
\begin{align}
  \label{eq:susytrafos}
  \delta \tau_\mu &= -\bar{\epsilon}^+ \gamma_0 \psi^+_\mu \,, \nonumber \\
  \delta e_\mu{}^a &= \bar{\epsilon}^+ \gamma^a \psi^-_\mu + \bar{\epsilon}^- \gamma^a \psi^+_\mu \,, \nonumber \\
  \delta m_\mu &= -\bar{\epsilon}^- \gamma_0 \psi_\mu^- - \bar{\epsilon}^+ \gamma_0 \rho_\mu - \bar{\eta} \gamma_0 \psi_\mu^+ \,, \nonumber \\
  \delta \psi_\mu^+ &= \partial_\mu \epsilon^+ + \frac12 \omega_\mu \gamma_0 \epsilon^+ \,, \nonumber \\
  \delta \psi_\mu^- &= \partial_\mu \epsilon^- + \frac12 \omega_\mu \gamma_0 \epsilon^- + \frac12 \omega_\mu{}^a \gamma_a \epsilon^+  \,, \nonumber \\
  \delta \rho_\mu &= \partial_\mu \eta + \frac12 \omega_\mu \gamma_0 \eta + \frac12 \omega_\mu{}^a \gamma_a \epsilon^- + \frac12 s_\mu \gamma_0 \epsilon^+ \,.
\end{align}

Unlike the supersymmetric version of the Bargmann algebra \eqref{3dsuperbargmann}, the extended Bargmann superalgebra \eqref{eq:Bargmann2cc}, \eqref{eq:superBargmann2cc} does not correspond to an In\"on\"u-Wigner contraction of a relativistic superalgebra.\footnote{This statement only holds for the full superalgebra. The bosonic part \eqref{eq:Bargmann2cc} can be obtained as an In\"on\"u-Wigner contraction of a direct product of the Poincar\'e algebra with a two-dimensional abelian algebra \cite{Bergshoeff:2016lwr}.} It can however be obtained from the three-dimensional $\mathcal{N}=2$ super-Poincar\'e algebra \eqref{3dpoincare} by performing a so-called Lie algebra expansion. Two different Lie algebra expansion procedures have been used in the literature, based on expanding Maurer-Cartan equations or using semigroups respectively. In what follows we will show how the extended Bargmann superalgebra corresponds to a Lie algebra expansion, according to the Maurer-Cartan equation method of \cite{deAzcarraga:2002xi,deAzcarraga:2007et}. We refer to the original literature \cite{Izaurieta:2006zz} for an account of semigroup expansions. Let us also note that Lie algebra expansions have not only been used to obtain the extended Bargmann superalgebra, but also various generalizations thereof that include e.g. more generators and/or a non-trivial cosmological constant \cite{Ozdemir:2019orp,Ozdemir:2019tby,Concha:2020tqx,Concha:2020eam,Concha:2021jos}. They have been used to construct examples of four-dimensional NR superalgebras and related supergravity theories \cite{Romano:2019ulw}. A procedure to construct off-shell rigid matter multiplets for NR algebras obtained via a Lie algebra expansion has been given in \cite{Kasikci:2021atn}.

Let us first give a general overview of the Lie algebra expansion procedure of \cite{deAzcarraga:2002xi,deAzcarraga:2007et}. Consider a Lie algebra $\mathfrak{g}$, with generators $T_\alpha$ and structure constants $f^\alpha_{\beta\gamma}$ ($\alpha$, $\beta$, $\gamma = 1, \cdots, \mathrm{dim}(\mathfrak{g})$). The Lie algebra expansion method of \cite{deAzcarraga:2002xi,deAzcarraga:2007et} uses that $\mathfrak{g}$ can be specified in a dual manner in terms of a Maurer-Cartan one-form. This is a Lie algebra valued one-form $A^\alpha = A^\alpha_\mu \rmd x^\mu$, satisfying the so-called Maurer-Cartan equations that state that the covariant curvature two-form of $A^\alpha$ vanishes:
\begin{align} \label{eq:MCeqs}
       F^\alpha \equiv \rmd A^\alpha + \frac12 f^\alpha_{\beta\gamma} A^\beta \wedge A^\gamma = 0 \,.
\end{align}
In this dual formulation, the Jacobi identities for the structure constants $f^\alpha_{\beta\gamma}$ are encoded in the consistency of the Maurer-Cartan equations with $\rmd^2 = 0$. Conversely, given a set of two-form equations of the form \eqref{eq:MCeqs} that are consistent with $\rmd^2 = 0$, one can conclude that the $f^\alpha_{\beta\gamma}$ that appear in them satisfy the Jacobi identities and are thus structure constants of a Lie algebra.

In what follows, we will assume\footnote{This assumption is satisfied in most of the applications of Lie algebra expansions to obtain new NR (super)algebras. Note however that the Lie algebra expansion method of \cite{deAzcarraga:2002xi,deAzcarraga:2007et} is more general and can also be applied to different cases where this assumption does not hold.} that $\mathfrak{g}$ can be decomposed (as a vector space) as $\mathfrak{g} = V_0 \oplus V_1$, such that
     \begin{align} \label{eq:symmdecomp}
       & [V_0, V_0] \subset V_0 \,, \qquad [V_0, V_1] \subset V_1 \,, \qquad [V_1, V_1] \subset V_0 \,.
     \end{align}
We will split the Lie algebra index $\alpha$ into $\alpha_0$ to refer to components along $V_0$ and $\alpha_1$ to refer to components along $V_1$. The Maurer-Cartan equations then take the following form:
     \begin{align} \label{eq:MCsymmdecomp}
       F^{\alpha_0} &= \rmd A^{\alpha_0} + \frac12 f^{\alpha_0}_{\beta_0 \gamma_0} A^{\beta_0} \wedge A^{\gamma_0} + \frac12 f_{\beta_1 \gamma_1}^{\alpha_0} A^{\beta_1} \wedge A^{\gamma_1} = 0 \,, \nonumber \\
       F^{\alpha_1} &= \rmd A^{\alpha_1} + f_{\beta_0 \gamma_1}^{\alpha_1} A^{\beta_0} \wedge A^{\gamma_1} = 0 \,.
     \end{align}
The application of the Lie algebra expansion procedure to $\mathfrak{g}$ proceeds by first considering a formal power series expansion of the Maurer-Cartan one-form $A^\alpha$ in an expansion parameter $\lambda$. For the components of $A^\alpha$ along $V_0$, this expansion starts at order $\lambda^0$, while for those along $V_1$ it starts at order $\lambda$:
     \begin{align} \label{eq:expMCform}
       A^{\alpha_0} &= \sum_{n=0}^\infty \lambda^{2n} A^{\alpha_0}_{(2n)} = A^{\alpha_0}_{(0)} + \lambda^2 A^{\alpha_0}_{(2)} + \lambda^4 A^{\alpha_0}_{(4)} + \cdots \,, \nonumber \\
       A^{\alpha_1} &= \sum_{n=0}^\infty \lambda^{2n+1} A^{\alpha_1}_{(2n+1)} = \lambda A^{\alpha_1}_{(1)} + \lambda^3 A^{\alpha_1}_{(3)} + \lambda^5 A^{\alpha_1}_{(5)} + \cdots \,.
     \end{align}
By plugging these expansions in the Maurer-Cartan equations \eqref{eq:MCsymmdecomp}, collecting like powers of $\lambda$ and setting the coefficients of each power of $\lambda$ equal to zero, one obtains Maurer-Cartan-like equations for an infinite tower of one-forms $A^{\alpha_0}_{(2n)}$, $A^{\alpha_1}_{(2m+1)}$ ($n, m = 0, 1,2,\cdots$). These equations are consistent with $\rmd^2 = 0$ by virtue of the Jacobi identities for $f^\alpha_{\beta\gamma}$, so that they correspond to the Maurer-Cartan equations of an infinite-dimensional Lie algebra. To get a finite-dimensional Lie algebra, one looks for consistent truncations of the expansions \eqref{eq:expMCform}, i.e., one proposes the expansions:
     \begin{align}
       A^{\alpha_0} &= \sum_{n=0}^{N_0/2} \lambda^{2n} A^{\alpha_0}_{(2n)} = A^{\alpha_0}_{(0)} + \lambda^2 A^{\alpha_0}_{(2)}  + \cdots + \lambda^{N_0} A_{(N_0)}^{\alpha_0}  \,, \nonumber \\
       A^{\alpha_1} &= \sum_{n=0}^{(N_1-1)/2} \lambda^{2n+1} A^{\alpha_1}_{(2n+1)} = \lambda A^{\alpha_1}_{(1)} + \lambda^3 A^{\alpha_1}_{(3)} + \cdots + \lambda^{N_1} A_{(N_1)}^{\alpha_1}
     \end{align}
     for even  integers $N_0$ and odd integers  $N_1$.
     These lead to the following Maurer-Cartan-like equations:
     \begin{align} \label{eq:truncMCeqs}
       & \rmd A^{\alpha_0}_{(2n_0)} + \frac12 f^{\alpha_0}_{\beta_0 \gamma_0} \sum_{r=0}^{n_0} A^{\beta_0}_{(2r)} \wedge A^{\gamma_0}_{(2(n_0-r))} + \frac12 f^{\alpha_0}_{\beta_1 \gamma_1} \sum_{r=1}^{n_0} A^{\beta_1}_{(2r-1)} \wedge A^{\gamma_1}_{(2(n_0-r) + 1)} = 0  \,, \nonumber \\
       & \rmd A^{\alpha_1}_{( 2n_1 +1)} + f^{\alpha_1}_{\beta_0 \gamma_1} \sum_{r=0}^{n_1} A^{\beta_0}_{(2r)} \wedge A^{\gamma_1}_{(2 (n_1-r) + 1)} = 0 \,,
     \end{align}
     with $n_0 = 0,1,2, \cdots, N_0/2$ and $n_1 = 0,1,2, \cdots, (N_1 - 1)/2$.
     Consistency with $\rmd^2 = 0$ is again guaranteed by the Jacobi identities for $f^\alpha_{\beta\gamma}$, provided that the above Maurer-Cartan-like equations for $A^{\alpha_0}_{(0)}$, $A^{\alpha_0}_{(2)}$, $\cdots$, $A^{\alpha_0}_{(N_0)}$ do not contain any $A^{\alpha_1}_{(n_1)}$ with $n_1 > N_1$ and that those for $A^{\alpha_1}_{(1)}$, $A^{\alpha_1}_{(3)}$, $\cdots$, $A^{\alpha_1}_{(N_1)}$ do not contain any $A^{\alpha_0}_{(n_0)}$ with $n_0 > N_0$. From \eqref{eq:truncMCeqs} and taking into account that $N_0$ and $N_1$ are even and odd respectively, one finds that this requires that
    \begin{align} \label{eq:trunccond}
      N_1 =  N_0 \pm 1 \,.
    \end{align}
    In case this condition is fulfilled, the truncated Maurer-Cartan-like equations \eqref{eq:truncMCeqs} correspond to the Maurer-Cartan equations of a Lie algebra, whose structure constants can be read off from the left-hand-sides of \eqref{eq:truncMCeqs}. For $N_0 = 0$ and $N_1 = 1$, this Lie algebra is an In\"on\"u-Wigner contraction of $\mathfrak{g}$. For other allowed values of $N_0$ and $N_1$, it has more generators than $\mathfrak{g}$ and is called a Lie algebra expansion of $\mathfrak{g}$.

    To see that the extended Bargmann superalgebra \eqref{eq:Bargmann2cc}, \eqref{eq:superBargmann2cc} is a Lie algebra expansion of the three-dimensional $\mathcal{N}=2$ super-Poincar\'e algebra \eqref{3dpoincare}, one then proceeds as follows \cite{deAzcarraga:2019mdn}. Introducing the $\mathcal{N}=2$ super-Poincar\'e algebra-valued gauge field (with $J_{\hat{A}} = -(1/2) \epsilon_{\hat{A}}{}^{\hat{B}\hat{C}} M_{\hat{B}\hat{C}}$ \footnote{Here, we use the convention that $\epsilon^{012} = - \epsilon_{012} = 1$.})
    \begin{align}
      A_\mu = E_\mu{}^{\hat{A}} P_{\hat{A}} + \Omega_\mu{}^{\hat{A}} J_{\hat{A}} + \bar{\epsilon}^i Q_i \,,
    \end{align}
    the Maurer-Cartan equations corresponding to \eqref{3dpoincare} are obtained by setting the supercovariant curvatures
    \begin{align}
      R_{\mu\nu}{}^{\hat{A}}(P) &\equiv 2 \partial_{[\mu} E_{\nu]}{}^{\hat{A}} + 2 \epsilon^{\hat{A}}{}_{\hat{B}\hat{C}} \Omega_{[\mu}{}^{\hat{B}} E_{\nu]}{}^{\hat{C}} - \bar{\psi}_\mu{}^i \gamma^{\hat{A}} \psi^j_\nu \delta_{ij} \,, \nonumber \\
      R_{\mu\nu}{}^{\hat{A}}(J) & \equiv 2 \partial_{[\mu} \Omega_{\nu]}{}^{\hat{A}} + \epsilon^{\hat{A}}{}_{\hat{B}\hat{C}} \Omega_\mu{}^{\hat{B}} \Omega_\nu{}^{\hat{C}} \,, \nonumber \\
      R_{\mu\nu}(Q^i) & \equiv 2 \partial_{[\mu} \psi_{\nu]}^i + \Omega_{[\mu}{}^{\hat{A}} \gamma_{\hat{A}} \psi^i_{\nu]} \,,
    \end{align}
    equal to zero. Splitting the index $\hat{A}$ as $\hat{A} = \{0, a\}$, with $a = 1,2$ and defining
    \begin{align}
      Q_\pm = \frac{1}{\sqrt{2}} \left(Q^1 \pm \gamma_0 Q^2 \right) \,, \qquad \qquad \psi_{\mu \pm} = \frac{1}{\sqrt{2}} \left(\psi_\mu^1 \pm \gamma_0 \psi_\mu^2 \right) \,,
    \end{align}
    these Maurer-Cartan equations can be rewritten as
    \begin{align} \label{eq:MCpoinc}
      R_{\mu\nu}{}^0(P) & \equiv 2 \partial_{[\mu} E_{\nu]}{}^0 + 2 \epsilon_{ab} \Omega_{[\mu}{}^a E_{\nu]}{}^b - \bar{\psi}_{\mu +} \gamma^0 \psi_{\nu +} - \bar{\psi}_{\mu -} \gamma^0 \psi_{\nu -} = 0\,, \nonumber \\
      R_{\mu\nu}{}^0(J) & \equiv 2 \partial_{[\mu} \Omega_{\nu]}{}^0 + \epsilon_{ab} \Omega_{[\mu}{}^a \Omega_{\nu]}{}^b = 0 \,, \nonumber \\
            R_{\mu\nu}(Q_+) & \equiv 2 \partial_{[\mu} \psi_{\nu]+} + \Omega_{[\mu}{}^0 \gamma_0 \psi_{\nu] +} + \Omega_{[\mu}{}^a \gamma_a \psi_{\nu]-} = 0 \,, \nonumber \\
      R_{\mu\nu}{}^a(P) & \equiv 2 \partial_{[\mu} E_{\nu]}{}^a + 2 \epsilon^a{}_b \Omega_{[\mu}{}^0 E_{\nu]}{}^b - 2 \epsilon^a{}_b \Omega_{[\mu}{}^b E_{\nu]}{}^0 - 2 \bar{\psi}_{[\mu +} \gamma^a \psi_{\nu] -} = 0 \,, \nonumber \\
      R_{\mu\nu}{}^a(J) & \equiv 2 \partial_{[\mu} \Omega_{\nu]}{}^a + 2 \epsilon^a{}_b \Omega_{[\mu}{}^0 \Omega_{\nu]}{}^b = 0 \,, \nonumber \\
      R_{\mu\nu}(Q_-) & \equiv 2 \partial_{[\mu} \psi_{\nu]-} + \Omega_{[\mu}{}^0 \gamma_0 \psi_{\nu] -} + \Omega_{[\mu}{}^a \gamma_a \psi_{\nu]+} = 0 \,.
    \end{align}
    Since the $\mathcal{N}=2$ super-Poincar\'e algebra can be decomposed as $V_0 \oplus V_1$ with
    \begin{align}
      V_0 = \{P^0, J^0, Q_+\} \qquad \qquad \text{and} \qquad \qquad V_1 = \{P^a, J^a, Q_-\} \,,
    \end{align}
    obeying \eqref{eq:symmdecomp}, one can perform the above outlined expansion procedure. Picking $N_0 = 2$ and $N_1 = 1$, one thus expands
    \begin{alignat}{3}
      E_\mu{}^0 &= \tau_{\mu (0)} + \lambda^2 \tau_{\mu (2)} \,, \quad & \Omega_\mu{}^0 &= \omega_{\mu (0)} + \lambda^2 \omega_{\mu (2)} \,, \quad & \psi_{\mu +} &= \psi_{\mu + (0)} + \lambda^2 \psi_{\mu + (2)} \,, \nonumber \\
      E_\mu{}^a &= \lambda e_{\mu (1)}{}^a  \,, & \Omega_\mu{}^a &= \lambda \omega_{\mu (1)}{}^a \,, & \psi_{\mu -} &= \lambda \psi_{\mu - (1)} \,.
    \end{alignat}
    Plugging these in the Maurer-Cartan equations \eqref{eq:MCpoinc} and setting the coefficients of like powers of $\lambda$ equal to zero, one is led to the following equations:
    \begin{align}
      \label{eq:MCextsuperbargmann}
      &  2 \partial_{[\mu} \tau_{\nu](0)} - \bar{\psi}_{\mu + (0)} \gamma^0 \psi_{\nu + (0)} = 0 \,, \nonumber \\
      & 2 \partial_{[\mu} \tau_{\nu](2)} + 2 \epsilon^a{}_b  \omega_{[\mu (1)}{}^a e_{\nu](1)}{}^b - \bar{\psi}_{\mu - (1)} \gamma^0 \psi_{\nu - (1)} - 2 \bar{\psi}_{[\mu + (0)} \gamma^0 \psi_{\nu]+ (2)} = 0 \,, \nonumber \\
      & 2 \partial_{[\mu} \omega_{\nu](0)} = 0 \,, \nonumber \\
      & 2 \partial_{[\mu} \omega_{\nu](2)} + \epsilon_{ab} \omega_{[\mu (1)}{}^a \omega_{\nu](1)}{}^b = 0 \,, \nonumber \\
      & 2 \partial_{[\mu} \psi_{\nu]+(0)} + \omega_{[\mu (0)} \gamma_0 \psi_{\nu]+(0)} = 0 \,, \nonumber \\
      & 2 \partial_{[\mu} \psi_{\nu]+(2)} + \omega_{[\mu (0)} \gamma_0 \psi_{\nu]+(2)} + \omega_{[\mu(2)} \gamma_0 \psi_{\nu] + (0)} + \omega_{[\mu (1)}{}^a \gamma_a \psi_{\nu]- (1)}  = 0 \,, \nonumber \\
      & 2 \partial_{[\mu} e_{\nu](1)}{}^a + 2 \epsilon^a{}_b \omega_{[\mu(0)} e_{\nu](1)}{}^b - 2 \epsilon^a{}_b \omega_{[\mu (1)}{}^b \tau_{\nu](0)} - 2 \bar{\psi}_{\mu + (0)} \gamma^a \psi_{\nu] - (1)} = 0 \,, \nonumber \\
      & 2 \partial_{[\mu} \omega_{\nu](1)}{}^a + 2 \epsilon^a{}_b \omega_{[\mu (0)} \omega_{\nu] (1)}{}^b = 0 \,, \nonumber \\
      & 2 \partial_{[\mu} \psi_{\nu]-(1)} + \omega_{[\mu (0)} \gamma_0 \psi_{\nu] - (1)} + \omega_{[\mu (1)}{}^a \gamma_a \psi_{\nu]+(0)} = 0 \,.
    \end{align}
    Upon identifying
    \begin{align}
     & \tau_{\mu(0)} \rightarrow \tau_\mu \,, \quad \tau_{\mu(2)} \rightarrow m_\mu \,, \quad  e_{\mu(1)}{}^a \rightarrow e_\mu{}^a \,, \nonumber \\ & \omega_{\mu(0)} \rightarrow \omega_\mu \,, \quad \omega_{\mu(2)} \rightarrow s_\mu \,, \quad \omega_{\mu(1)}{}^a \rightarrow \omega_\mu{}^a \,, \nonumber \\ & \psi_{\mu + (0)} \rightarrow \psi_{\mu +} \,, \quad \psi_{\mu + (2)} \rightarrow \rho_\mu \,, \quad \psi_{\mu - (1)} \rightarrow \psi_{\mu -} \,,
    \end{align}
    the left-hand-sides of \eqref{eq:MCextsuperbargmann} turn into the supercovariant curvatures of the gauge field \eqref{eq:Asuper} of the superalgebra \eqref{eq:Bargmann2cc}, \eqref{eq:superBargmann2cc}. This shows that \eqref{eq:MCextsuperbargmann} are the Maurer-Cartan equations of the extended Bargmann superalgebra \eqref{eq:Bargmann2cc}, \eqref{eq:superBargmann2cc} and that the latter is indeed obtained as a Lie algebra expansion of the 3$D$ $\mathcal{N}=2$ super-Poincar\'e algebra.

Above we focused on NR Chern-Simons supergravity theories that are based on extensions of the 3$D$ Bargmann superalgebra. Different types of NR Chern-Simons supergravity theories have been considered in the literature as well. For instance, theories of this kind with extensions of Lifshitz and Schr\"odinger symmetries were considered in \cite{Ozdemir:2019tby}, a teleparallel NR supergravity theory was given (based on earlier work \cite{Caroca:2021njq,Salgado:2005pg}) in a Chern-Simons formulation in \cite{Concha:2021llq}. A NR Chern-Simons supergravity theory that is based on an extension of a $c \rightarrow \infty$ limit of the Maxwell algebra (i.e., the extension of the Poincar\'e algebra with generators $Z_{\hat{A}\hat{B}}$ such that $[P_{\hat{A}}, P_{\hat{B}}] = Z_{\hat{A}\hat{B}}$) was constructed in \cite{Concha:2019mxx}.

This finishes our discussion of non-Lorentzian supergravity theories in 3$D$. In the next section we will focus our attention on higher dimensions and discuss recent results on 10$D$ non-Lorentzian supergravity.

\section{10$D$ Minimal Supergravity} \label{sec:10Dsugra}

The 10$D$ minimal supergravity theory is the first pure NR supergravity theory constructed in a dimension $D>4$. It was obtained by taking  the NR limit of the 10$D$ relativistic $\mathcal{N}=1$ supergravity theory without Yang-Mills matter couplings \cite{Bergshoeff:2021tfn}. Before taking the limit, let us first define the relativistic $\mathcal{N}=1$ supergravity theory \cite{Bergshoeff:1981um,Chamseddine:1980cp}.  The field content is given by
\begin{equation}
\{E_\mu{}^{\hat A}\,, B_{\mu\nu}\,, \Phi; \Psi_\mu\,, \lambda\}\,,
\end{equation}
where $\{E_\mu{}^{\hat A}\,, B_{\mu\nu}\,, \Phi\}$ are the Vielbein field, Kalb-Ramond field and the dilaton field, respectively and $\{\Psi_\mu\,,\uplambda\}$ are the left-handed Majorana-Weyl spinor gravitino field and the right-handed dilatino, respectively. For the discussion to follow we only need the specific form of the bosonic part of the $\mathcal{N}=1$ supergravity action\,\footnote{The fermionic part can be found in
\cite{Bergshoeff:2021tfn} but will not be needed here.}
\begin{align} \label{eq:Nisoneaction}
S = \frac{1}{2\kappa^2}\int \rmd^{10}x\,E\,\rme^{-2\Phi}\bigg\{&\mathcal R+ 4\,\partial_\mu\Phi\,\partial^\mu\Phi - \frac{1}{12}\,\mathcal H_{\mu\nu\rho}\mathcal H^{\mu\nu\rho}\bigg\}\,,
\end{align}
where $\kappa$ denotes the gravitational coupling constant, $E = \mathrm{det}(E_\mu{}^{\hat{A}})$, $\mathcal{R}$ is the Ricci scalar and
\begin{align}
\label{eq:defH}
\mathcal{H}_{\mu\nu\rho} = 3 \partial_{[\mu} B_{\nu\rho]}
\end{align}
is the field strength of the Kalb-Ramond field.

The fields of ten-dimensional $\mathcal{N}=1$ supergravity transform as follows under local Lorentz transformations with parameter $\Lambda^{\hat{A}\hat{B}}$, a one-form symmetry of the Kalb-Ramond field with parameter $\Theta_\mu$ and supersymmetry with a left-handed Majorana-Weyl spinor parameter $\varepsilon$:
\begin{subequations}\label{eq:trafosrel}
\begin{align}
\delta E_\mu{}^{\hat{A}} &= \Lambda^{\hat{A}}{}_{\hat{B}} E_\mu{}^{\hat{B}} + \bar{\varepsilon}\,\Gamma^{\hat{A}}\Psi_\mu\,,\label{extra}  \\[.1truecm]
\delta B_{\mu\nu} &= 2 \partial_{[\mu} \Theta_{\nu]} + 2\,\bar{\varepsilon}\,\Gamma_{[\mu}\Psi_{\nu]}\,,  \qquad
\delta \Phi = \frac{1}{2}\,\bar{\varepsilon}\,\uplambda\,, \label{eq:trafosVielBDil} \\[.1truecm]
\delta \Psi_\mu &= \frac{1}{4} \Lambda^{\hat{A}\hat{B}} \Gamma_{\hat{A}\hat{B}} \Psi_\mu + D_\mu(\Omega^{(+)}) \varepsilon \,, \label{eq:trafosGravo} \\[.1truecm]
\delta \uplambda &= \frac{1}{4} \Lambda^{\hat{A}\hat{B}} \Gamma_{\hat{A}\hat{B}} \uplambda + \Gamma^\mu\varepsilon\,\mathcal \partial_\mu \Phi - \frac{1}{12}\,\Gamma^{\hat{A}\hat{B}\hat{C}}\varepsilon\, \mathcal{H}_{\hat{A}\hat{B}\hat{C}}\,, \label{eq:trafosDilo}
\end{align}
\end{subequations}
where the 10$D$ gamma-matrices are denoted by $\Gamma_{\hat{A}}$ and we have ignored terms quadratic in the fermion fields in the supersymmetry rules of the fermions. We have furthermore defined the following torsionful covariant derivative of $\varepsilon$
\begin{align} \label{eq:DOmegapm}
D_\mu(\Omega^{(+)})\varepsilon &= \partial_\mu\varepsilon - \frac14\,\Omega^{(+)}_\mu{}^{\hat A\hat B}\Gamma_{\hat A\hat B}\varepsilon \hskip .2truecm \text{with} \hskip .2truecm  \Omega^{(+)}_\mu{}^{\hat A\hat B} = \Omega_\mu{}^{\hat A\hat B} + \frac12 \mathcal{H}_\mu{}^{\hat A\hat B} \,.
\end{align}

In order to define the NR limit, we introduce a (dimensionless) parameter $\omega$ and perform the following invertible field redefinition
\begin{alignat}{4}\label{eq:rescale}
\tau_\mu{}^A &= \omega^{-1}\,E_\mu{}^A\,, \hskip .3truecm  \ \ & e_\mu{}^{a} &= E_\mu{}^{a}\,, \qquad \ \ &\hskip .3truecm  b_{\mu\nu} &= B_{\mu\nu}+\epsilon_{AB}\,E_\mu{}^A E_\nu{}^B\,, \notag\\[.1truecm]
\phi &= \Phi - \log \omega\,,&\psi_{\mu\pm} &= \omega^{\mp 1/2}\Pi_{\pm}\Psi_\mu\,,&\hskip .3truecm  \lambda_\pm &= \omega^{\mp 1/2}\Pi_\pm\uplambda\,,
\end{alignat}
where we have split the Lorentz index $\hat{A}$ into a longitudinal index $A=0,1$ and a transversal index $a=2,\cdots,9$. Note that the redefinition of the spinor fields involves the `worldsheet chirality' projections
\begin{align}
\label{eq:Piproj}
\chi_\pm = \Pi_\pm\chi \hskip .5truecm \rm{with} \hskip .5truecm \Pi_\pm = \frac12 \left(\mathds{1} \pm \Gamma_{01}\right) \hskip .5truecm \textrm{for\ any spinor}\ \ \chi\,.
\end{align}
In the $\omega \to \infty$ limit, the fields $\tau_\mu{}^A$, $e_\mu{}^a$ and $b_{\mu\nu}$ become the longitudinal and transversal Vielbeine and two-form field of stringy NC geometry. This can be verified by noting that the limit of the Lorentz and one-form symmetry transformation rules of the first line of \eqref{eq:rescale}, taken in a similar way as in the discussion around eq.~\eqref{tauandm}, leads to the transformation rules \eqref{eq:SNCtrafos} and \eqref{eq:oneformgauge}.

The following  issues complicate taking the $\omega \to \infty$ limit of the action and supersymmetry transformation rules of $\mathcal{N}=1$ supergravity:

\begin{enumerate}
\item the emergence of  both bosonic and fermionic Stueckelberg symmetries,
\vskip .1truecm
\item the occurrence of divergent terms in the action,
\vskip .1truecm
\item the presence of divergent terms in the supersymmetry transformation rules.
\end{enumerate}
On top of this, after taking the limit, the two-form field $b_{\mu\nu}$, in contrast to its relativistic counterpart $B_{\mu\nu}$, transforms under Galilean boosts (see eq.~\eqref{eq:SNCtrafos}) and should therefore be considered on par with the Vielbein fields $\tau_\mu{}^A$ and $e_\mu{}^{a}$.
The above three complications lead to the following new features \cite{Bergshoeff:2021tfn}.
\begin{enumerate}
\item Due to the emergent Stueckelberg symmetries, the NR superalgebra is realized on a smaller set of field components than in the relativistic case.
\vskip .1truecm
\item The limit we take is critical in the sense that it is defined in such a way that the divergent terms in the action cancel amongst each other.
\vskip .1truecm
\item The divergent terms in the supersymmetry rules are tamed by imposing {\sl geometric constraints} that we will identify below.
\end{enumerate}

Restricting to the bosonic case, the emergence of local dilatation Stueckelberg symmetries can already be seen by looking at the symmetries of the NR Polyakov action \cite{Bergshoeff:2019pij,Bergshoeff:2021bmc}. These scaling symmetries act non-isotropically in the sense that the Vielbein fields transform as follows:
\begin{equation}\label{D1}
\delta_D \tau_\mu{}^A = \lambda_D\,\tau_\mu{}^A\,,\hskip 1truecm \delta e_\mu{}^{a} = 0\,.
\end{equation}
The fact that the transversal Vielbeine are scale-invariant is typical for the string. A similar an-isotropic scaling symmetry occurs in the 11$D$ membrane case where the transversal Vielbeine have a non-zero scaling weight \cite{Blair:2021waq}. Besides the Vielbeine the only other bosonic field that is not scale-invariant is the dilaton:
\begin{equation}\label{D2}
\delta \phi = \lambda_D\,.
\end{equation}
From this it follows  that the combination $e^{-\phi}\tau_\mu{}^A$ is scale-invariant. Usually, the (vacuum expectation value of) the dilaton is associated with the string coupling constant $g_s$ and, when the longitudinal spatial direction is compactified (as is the case in NR string theory), $\tau_\mu{}^1$ encodes the radius $R$ of the compactification circle. The above scaling symmetries imply that only a product of the dilaton and $\tau_\mu{}^1$ has an invariant meaning. This means that effectively we only have one invariant modulus instead of two.

In the supersymmetric case there are additional fermionic fields that transform under the scaling symmetries. Since we lack a NR supersymmetric Green-Schwarz superstring sigma model, it is easiest to derive the scaling weights of the fermionic fields by first deriving the fermionic Stueckelberg symmetries. To derive the latter, we can make use of the fact that there are divergent terms in the action that cancel against each other. To discuss these cancellations we consider the  redefinitions \eqref{eq:rescale}
that define the NR limit. Substituting these redefinitions into the relativistic supergravity action leads to an expansion in powers of $\omega^{-2}$ of the form
\begin{align} \label{eq:actionexpansion}
S = \omega^{2} S^{(2)} + S^{(0)} + \omega^{-2} S^{(-2)} + \omega^{-4} S^{(-4)} \,,
\end{align}
where each of the $S^{(i)}$ now depends on the fields $\tau_\mu{}^A$, $e_\mu{}^{a}$, $b_{\mu\nu}$, $\phi$, $\psi_{\mu\pm}$ and $\lambda_\pm$. The fact that we find that
\begin{equation}
S^{(2)} =0
\end{equation}
combined with the fact that the action is supersymmetric leads to the emergence of fermionic Stueckelberg symmetries as follows. Due to the occurrence of divergent terms in the supersymmetry rules we can write the infinitesimal  action $\delta_Q$ of a generic supersymmetry $Q$ as follows
\begin{align}
\delta_Q F = \omega^2 \delta_Q^{(2)} F + \delta_Q^{(0)} F + \omega^{-2} \delta_Q^{(-2)} F \,,
\end{align}
where  $F$ is any of the fields $\tau_\mu{}^A$, $e_\mu{}^{a}$, $b_{\mu\nu}$, $\phi$, $\psi_{\mu\pm}$, $\lambda_\pm$.
The supersymmetry variation $\delta_Q S$ of the action can then be expanded as
\begin{align} \label{eq:susyvarSexp}
\delta_Q S = \omega^2 \delta_Q^{(2)} S^{(0)} + \omega^0 \left( \delta_Q^{(0)} S^{(0)} + \delta_Q^{(2)} S^{(-2)} \right) + \mathcal{O}(\omega^{-2}) \,.
\end{align}
Requiring invariance of $S$  imposes that every order of $\omega^{-2}$ in \eqref{eq:susyvarSexp} is separately zero. This in particular leads to the following two requirements
\begin{eqnarray}
&&(1)\ \ \ \ \  \delta_Q^{(2)} S^{(0)} = 0\,,  \label{eq:deltaQconstraints} \\[.1truecm]
&&(2)\ \ \ \ \   \delta_Q^{(0)} S^{(0)} = - \delta_Q^{(2)} S^{(-2)} \,.\label{eq:deltaQconstraints2}
\end{eqnarray}
From the first condition it follows that, when varying $S^{(0)}$ all divergent terms in the supersymmetry rules must cancel amongst each other. It turns out that these divergent terms only occur in the supersymmetry variation of two fermionic field components and that they occur in the form of a shift transformation. From this it follows that the cancellation of the divergences is equivalent to the statement that the action $S^{(0)}$ is invariant under a set of fermionic Stueckelberg symmetries. A more detailed analysis shows that these Stueckelberg symmetries are given by \cite{Bergshoeff:2021tfn}\footnote{Here and in the following, we will often use a light-cone notation $\pm$ for the index $A$: $A=(+,-)$, defined according to e.g. $\tau_\mu{}^\pm = (1/\sqrt{2})(\tau_\mu{}^0 \pm \tau_\mu{}^1)$.}
\begin{align}
\label{eq:STsymm}
\delta_S \psi_{\mu+} &= \frac12 \tau_\mu{}^+ \Gamma_+ \eta_- \,, \qquad \qquad \delta_S \lambda_- = \eta_- \,, \nonumber \\
\delta_T \psi_{\mu-} &= \tau_\mu{}^+ \rho_- \,,
\end{align}
where $\eta_-$ and $\rho_-$ are the parameters of the shift symmetries that we will refer to as $S-$ and $T-$supersymmetries in what follows.

Whereas the divergent parts $\delta_Q^{(2)}$ of the supersymmetry transformation rules can, according to the above discussion, be identified as (special cases of) the fermionic $S-$ and $T-$Stueckelberg symmetries, the $\delta_Q^{(0)}$ part has the right structure to give NR supersymmetry transformation rules.  Note however that the second condition given in eq.~\eqref{eq:deltaQconstraints2} then implies that $S^{(0)}$, the NR limit of the action, is not invariant under NR supersymmetry. This can be remedied by imposing by hand additional geometric constraints setting certain so-called geometric tensors equal to zero. The occurrence of these geometric tensors is characteristic of NC and stringy NC geometry. In the current context, they are defined by dividing the properly supercovariantized curvature tensors $T_{\mu\nu}{}^A$ and $T_{\mu\nu}{}^a$ of the Vielbeine $\tau_\mu{}^A$ and $e_\mu{}^{a}$, as well as the supercovariantized curvature tensor $h_{\mu\nu\rho}$ of the two-form $b_{\mu\nu}$ into conventional tensors and geometric tensors as follows:\,\footnote{Note that a similar division does not apply to the supercovariantized curvatures of the spin connection fields.}
\vskip .2truecm

\noindent A {\bf geometric} tensor is a  curvature component that does not contain (components of) a stringy NC spin connection (for longitudinal Lorentz transformations, transversal spatial rotations or string Galilean boosts) or the dilatation gauge field. Setting a geometric tensor to zero leads to a geometric constraint on the underlying geometry.
\vskip .2truecm

\noindent A {\bf conventional} tensor is a curvature component that contains a stringy NC spin connection (for longitudinal Lorentz transformations, transversal spatial rotations or string Galilean boosts) and/or the dilatation gauge field, multiplied by an invertible longitudinal or transverse Vierbein field. Setting a conventional tensor to zero leads to a conventional constraint that can be used to solve for a spin connection or dilatation gauge field component.

\vskip .2truecm

Using these definitions we find the following geometric tensors\footnote{Note the similarity with the discussion around eq.~\eqref{eq:geomconstraints} in subsection \ref{ssec:pbraneNCgeometry}. The main difference between subsection \ref{ssec:pbraneNCgeometry} and the current discussion is that the former does not include fermionic symmetries and the dilatation symmetry. In particular, the absence of dilatations is responsible for the fact that in subsection \ref{ssec:pbraneNCgeometry}, $T_{a(AB)}$ appears as a geometric tensor, instead of $T_{a\{AB\}}$ here.}
\begin{equation}
T_{ab}{}^A\,,\ \  T_{a}{}^{\{AB\}}\,, \ \ h_{abc}\,,
\end{equation}
where $\{AB\}$ indicates the symmetric traceless part of $AB$. All other components are conventional tensors and setting them to zero can be used to solve for the spin connection fields and dilatation gauge field except for the components
\begin{equation}
\omega_{\{AB\}a}\ \ {\rm and}\ \  b_A\,.
\end{equation}

It turns out that, in order to ensure that $S^{(0)}$ is invariant under NR supersymmetry we  need to set to zero the following subset of geometric tensors:
\begin{equation}\label{g.c.}
T_{ab}{}^-\ =\   T_{a+}{}^{-}\ =\ 0\,.
\end{equation}
These geometric constraints can equivalently be described by the following foliation constraint:
\begin{equation}
\tau_{[\mu}{}^-\partial_\nu \tau_{\rho]}{}^- = 0\,.
\end{equation}

Once we have established that the action $S^{(0)}$, after taking the limit, is invariant under the regular NR $Q-$supersymmetry and the new emergent $S-$ and $T-$supersymmetries, we are also able to derive the emergent an-isotropic scale or $D-$symmetry. The easiest way to derive the $D$-symmetry is to require that the action $S^{(0)}$ must be invariant under the commutator of a $Q-$ with a $S-$ or $T-$supersymmetry, i.e.
\begin{equation}
[\bar \epsilon Q, \bar\eta S\ \textrm{or}\ \bar \eta T] \ \sim \ \lambda_D D\,.
\end{equation}
This yields precisely the an-isotropic scale transformations given in eqs.~\eqref{D1}, \eqref{D2} together with
\begin{equation}\label{D3}
\delta_D \psi_{\mu\pm} = \pm \frac12 \lambda_D \psi_{\mu\pm} \,, \qquad \qquad  \delta_D \lambda_{\pm} = \pm \frac12 \lambda_D \lambda_{\pm} \,.
\end{equation}

An important simplifying feature of the $\mathcal{N}=1$ supersymmetric case is that the geometric constraints \eqref{g.c.} are invariant under NR supersymmetry and therefore do not lead to further constraints. The constraints should be imposed with care: they should not be substituted in the action $S^{(0)}$ but only in the supersymmetry variation of the action. Actually, the action $S^{(0)}$ serves the purpose of a pseudo-action: it is a convenient  way to derive a subset of the  equations of motion that we denominate the `bulk' equations of motion $B$. Due to the emerging an-isotropic dilatations and Stueckelberg symmetries there are also so-called `missing' equations of motion $M$ that can only be derived by taking the NR limit of the equations of motion instead of the action. The complete set of equations of motion $\{M,B\}$ form a reducible indecomposable representation in the sense that under Galilean boosts $G$ we have
\begin{equation}
\delta_G M \sim B\hskip 1truecm {\rm but\ not}\hskip 1truecm \delta_G B \sim M\,.
\end{equation}
This is not the case for the supersymmetry transformations $Q$ that connects all equations of motion back and forth, i.e.
\begin{equation}
\delta_Q M \sim  B\hskip 1truecm {\rm and} \hskip 1truecm \delta_Q B \sim M\,.
\end{equation}
The reason that the bulk equations of motion  do not form a separate multiplet is that the action $S^{(0)}$ is only invariant under supersymmetry after imposing the geometric constraints in the supersymmetry variation of the action. Such an action does not satisfy the criteria considered in \cite{Vanhecke:2017chr}.

The above shows that taking the non-Lorentzian limit of a relativistic action and next determining the equations of motion of the resulting non-Lorentzian action is not the same as varying the relativistic action and then taking the non-Lorentzian limit of the resulting  relativistic equations of motion. Nevertheless, the construction of a non-Lorentzian pseudo-action is a useful tool to collect all bulk equations of motion with just one action.

At the end of the day, the final result for 10$D$ minimal supergravity is a set of constraint equations (originally called bulk equations of motion, missing equations of motion and geometric constraints) that forms a closed collection under all the symmetries of the model.
In the absence of a true action it is irrelevant to distinguish  between equations of motion and additional constraint equations.
More details about minimal supergravity can be found in  \cite{Bergshoeff:2021tfn}. For later reference we give the full final answer, in a self-explanatory way, in a separate subsection below.

\subsection{The Complete Result}

The purpose of this subsection is to present, up to quartic fermion terms,  the relevant expressions of 10$D$ NR minimal supergravity, including some basic definitions, in a self-contained manner that can be used for later reference.   It is useful to split the NR action $S_{NR}$ into a part $S_B$ that is purely bosonic, a part $S_{\psi\psi}$ that is quadratic in the gravitini $\psi_{\mu\pm}$, a part $S_{\lambda\lambda}$ that is quadratic in the dilatini $\lambda_{\pm}$ and a remaining quadratic fermion part $S_{\lambda\psi}$ that contains both a gravitino and a dilatino:
\begin{align}\label{eq:NRaction}
S_{NR} = S_B + S_{\lambda\lambda} + S_{\lambda\psi} + S_{\psi\psi} + \text{quartic fermion terms} \,.
\end{align}
As mentioned above, we will ignore all quartic fermion terms and only require supersymmetry up to cubic fermion terms.

The bosonic part $S_B$ of the action has been given in \cite{Bergshoeff:2021bmc} and reads:
\begin{subequations} \label{eq:NRaction2}
\begin{align} \label{eq:BosAction}
S_B = \frac{1}{2\,\kappa^2}\int \rmd^{10}x \,e\,\rme^{-2\,\phi}\bigg(&\rmR(J)+4\,\partial_{a}\phi\,\partial^{a}\phi-\frac{1}{12}\,h_{abc}h^{abc} \notag\\
&-4\,e_{a}{}^{\mu}(\partial_{\mu} b^{a}-\omega_{\mu}{}^{ab}b_{b}-\omega_{\mu}{}^{Ab}\tau^{a}{}_{bA}) \notag\\&- 4\,b_{a}b^{a}- 4\,\tau_{a\{AB\}}\tau^{a\{AB\}}  \bigg)\,,
\end{align}
where $e = \mathrm{det}(\tau_\mu{}^A, e_\mu{}^{a}), h_{\mu\nu\rho} = 3\partial_{[\mu}b_{\nu\rho]}$ and  $\rmR(J) = -e_a{}^\mu e_b{}^\nu R_{\mu\nu}(J)^{ab}$
with
\begin{align}
\rmR_{\mu\nu}(J)^{ab}  &=
2\,\partial_{[\mu}\omega_{\nu]}{}^{ab} + 2\,\omega_{[\mu}{}^{ac}\omega_{\nu]}{}^{b}{}_{c}\notag\\
&\quad +2\,e_{[\mu}{}^{c}\big( 2\,\omega_{\nu]}{}^{C[a}\tau^{b]}{}_{cC}{-\omega_{\nu]Cc}\tau^{abC}}\big) \notag\\
&\quad+8\,\tau_{[\mu}{}^A\big(\omega_{\nu]}{}^{B[a}\tau^{b]}{}_{\{AB\}}-\frac18\,\epsilon_{A}{}^B\,\omega_{\nu]Bc}h^{abc}\big)\,.
\end{align}
The different dependent gauge fields occurring in the above expressions are given by
\begin{subequations}\label{eq:galspinconn}
\begin{align}
b_\mu &= e_\mu{}^{a}\,\tau_{aA}{}^A +\tau_\mu{}^A\partial_A\phi \,,\\
\omega_\mu &= \big(\,\tau_\mu{}^{AB}-\frac12\,\tau_\mu{}^C\tau^{AB}{}_C \big)\epsilon_{AB} - \tau_\mu{}^A\,\epsilon_{AB}\partial^B\phi\,,\\
\omega_\mu{}^{Aa} &= -e_\mu{}^{Aa}+e_{\mu b}e^{Aab} + \frac12\,\epsilon^A{}_B\,h_\mu{}^{Ba}  + \tau_{\mu B} W^{BAa} \,,\\
\omega_\mu{}^{ab} &= -2\, e_{\mu}{}^{[ab]}+e_{\mu c}e^{abc} - \frac12\,\tau_\mu{}^A\,\epsilon_{AB}\,h^{Bab} \,,
\end{align}
\end{subequations}
where
\begin{equation}
\tau_{\mu\nu}{}^A = \partial_{[\mu} \tau_{\nu]}{}^A\hskip .5truecm {\rm and}\hskip .5truecm e_{\mu\nu}{}^{a} = \partial_{[\mu} e_{\nu]}{}^{a}\,.
\end{equation}
Note that not all components of the above spin connections can be solved for, which is reflected by the undetermined $W^{ABa}$ which is traceless symmetric in the $(AB)$ indices, but otherwise arbitrary. Since all the relevant expressions---such as action, equations of motion, and symmetry transformation rules---follow from a limit it is clear that nothing depends on $W^{ABa}$.

The part of the action that is quadratic in the dilatini reads
\begin{align} \label{eq:SNR2}
S_{\lambda\lambda} = \frac{1}{2\kappa^2}\int \rmd^{10}x \,e\,\rme^{-2\,\phi}\bigg( &2\,\bar{\lambda}_{\pm}\Gamma^{a}D_{a}\lambda_{\mp}+
2\bar{\lambda}_{+}\Gamma^{A}D_{A}\lambda_{+}\notag\\
&-\frac{1}{6}h_{abc}(\bar{\lambda}_{+}\Gamma^{abc}\lambda_{-})+\tau_{bcA}(\bar{\lambda}_{-}\Gamma^{bcA}\lambda_{-})\bigg)\,,
\end{align}
where the covariant derivatives are covariant with respect to the Galilean symmetries and dilatations. The notation $\bar\lambda_\pm\Gamma \lambda_\mp$ is a shorthand for $\bar\lambda_+\Gamma\lambda_- + \bar\lambda_-\Gamma\lambda_+$, and will be used also below.

Next, the off-diagonal terms in the action read
\begin{align} \label{eq:SNR3}
S_{\lambda\psi} = \frac{1}{2\kappa^2}\int \rmd^{10}x \,e\,\rme^{-2\,\phi}\bigg(&-4\,\bar{\lambda}_{\pm}\Gamma^{ab}e_{a}{}^\mu e_{b}{}^\nu D_{[\mu}\psi_{\nu]\mp}-8\,\bar{\lambda}_{+}\Gamma^{Ab}\tau_A{}^{\mu}e_{b}{}^{\nu}D_{[\mu}\psi_{\nu]+}\nonumber\\[.1truecm]
&-4\,\bar{\lambda}_{\pm}\Gamma^{ab}\psi_{a\mp}\,D_{b}\phi -4\,\bar{\lambda}_{+}\Gamma^{Ab}\psi_{A+}\,D_{b}\phi \nonumber\\[.1truecm]
&+\frac{1}{6}\,h_{abc}(\bar{\lambda}_{\pm}\Gamma^{abcd}\psi_{d\mp})+\frac{1}{2}h_{abc}(\bar{\lambda}_{+}\Gamma^{abcD}\psi_{D+})\nonumber\\[.1truecm]
&-(\eta^{DA}+\epsilon^{DA})\tau_{bcD}(\bar{\lambda}_{-}\Gamma^{bc}\psi_{A+}-\bar{\lambda}_{+}\Gamma^{bc}\psi_{A-})\nonumber\\[.1truecm]
&+2\,\tau_{bc}{}^A\bar{\lambda}_{\pm}\Gamma^{bc}\psi_{A\mp} + 2\,\tau^{c\{AB\}}\bar{\lambda}_{+}\Gamma_{cA}\psi_{B+}\nonumber\\[.1truecm]
&-2\tau_{bcA}\bar{\lambda}_{-}\Gamma^{Abcd}\psi_{d-}\bigg)\,.
\end{align}

Finally, the pure gravitino terms are given by
\begin{align} \label{eq:SNR4}
S_{\psi\psi} = \frac{1}{2\kappa^2}\int \rmd^{10}x \,e\,\rme^{-2\,\phi}\bigg(&
-2\,\bar{\psi}_{A+}\Gamma^{Abc}e_{b}{}^\mu e_{c}{}^\nu D_{[\mu}\psi_{\nu] +}-4\,\bar{\psi}_{a+}\Gamma^{abC}e_{b}{}^\mu \tau_C{}^\nu D_{[\mu}\psi_{\nu] +}\nonumber\\[.1truecm]
&-2\,\bar{\psi}_{a\pm}\Gamma^{abc}e_{b}{}^\mu e_{c}{}^\nu D_{[\mu}\psi_{\nu]
\mp}+\frac{1}{2}\,h^{abc}(\bar{\psi}_{a\pm}\Gamma_{b}\psi_{c\mp})\nonumber\\[.1truecm]
&-\frac{1}{6}h_{abc}\big(\bar{\psi}_{d+}\Gamma^{abcdE}\psi_{E+}+ \frac12\,\bar{\psi}_{d\pm}\Gamma^{abcde}\psi_{e\mp}\big)+\nonumber\\[.1truecm]
&-4\,\big(\bar{\psi}_{a\pm}\Gamma^{a}\psi_{b\pm}+\,\bar{\psi}_{A+}\Gamma^{A}\psi_{b+}\big)D^{b}\phi \nonumber\\[.1truecm]
&-2\,(\eta^{AD}+\epsilon^{AD})\tau^{bc}{}_D\,\bar{\psi}_{c\pm}\Gamma_{b}\psi_{A\mp} +2\,\tau^{bc}{}^A(\bar{\psi}_{b-}\Gamma_{A}\psi_{C'-})\notag\\[.1truecm]
&- 2\big(\eta_{BC} -\epsilon_{BC}\big)\tau^{c\{AB\}}\bar\psi^C{}_+\Gamma_A\psi_{c+}\notag\\[.1truecm]
&+(\eta_{AB}+\epsilon_{AB})\tau_{bc}{}^A\,\bar{\psi}_{d\pm}\Gamma^{BbcdE}\psi_{E\mp}\nonumber\\[.1truecm]
&+\tau_{bc}{}^A\,\bar{\psi}_{d-}\Gamma_A\Gamma^{bcde}\psi_{e-}\bigg)\,.
\end{align}
\end{subequations}

The NR supersymmetry transformation rules that leave the action $S_{NR}$ defined above invariant (up to cubic fermion terms), upon imposition of the geometric constraints \eqref{g.c.}, are given by
\begin{subequations} \label{eq:bossusy}
\begin{align}
\delta \tau_\mu{}^A &=\bar{\epsilon}_{+}\Gamma^{A}\psi_{\mu +}\,,\\
\delta e_{\mu}{}^{a} &= \bar{\epsilon}_{+}\Gamma^{a}\psi_{\mu -}+\bar{\epsilon}_{-}\Gamma^{a}\psi_{\mu +}\,,\\
\delta \phi &=\frac{1}{2}(\bar{\epsilon}_{+}\lambda_{-}+\bar{\epsilon}_{-}\lambda_{+})\,,\\
\delta b_{\mu\nu}&= 4\,\tau_{[\mu}{}^{A}\bar{\epsilon}_{-}\Gamma_{A}\psi_{\nu]-} + 2\Big(e_{[\mu}{}^{a}\bar\epsilon_+\Gamma_{a}\psi_{\nu]-}+e_{[\mu}{}^{a}\bar{\epsilon}_{-}\Gamma_{a}\psi_{\nu]+}\Big)\,.
\end{align}
\end{subequations}
as far as the bosonic fields are concerned.

Decomposing the supersymmetry rules of the gravitino and dilatino as follows
\begin{subequations} \label{eq:deltasusy0}
\begin{align}
\delta \psi_{\mu + } &= \delta_+\psi_{\mu+} + \delta_-\psi_{\mu-}\,,\\
\delta \psi_{\mu - } &= \delta_+\psi_{\mu-} + \delta_-\psi_{\mu-}\,,\\
\delta \lambda_+ &= \delta_+\lambda_+ + \delta_-\lambda_+ \,,\\
\delta \lambda_-  &= \delta_+\lambda_- + \delta_-\lambda_-\,,
\end{align}
\end{subequations}
we find that the supersymmetry rules of the fermionic fields are, up to terms quadratic in $\psi_{\mu\pm}\,,\lambda_\pm$,  given by
\begin{subequations}
\label{eq:NRsupersymmetry}
\begin{align}
&\delta_+\psi_{\mu+} =\mathcal D_\mu\epsilon_+ - \frac18\,e_{\mu c}h^{cab}\Gamma_{ab}\epsilon_+ \,,\\
&\delta_-\psi_{\mu+} = \big(e_{\mu b}\tau^{ba+} + \tau_\mu{}^-\tau^{a++}\big)\Gamma_{a+}\epsilon_- \,,\\
&\delta_+\psi_{\mu-} = -\frac12\,\omega_\mu{}^{-a}\Gamma_{-a}\epsilon_+ \,,\\
&\delta_-\psi_{\mu-} = \mathcal D_\mu\epsilon_- - \frac18\,e_{\mu c}h^{cab}\Gamma_{ab}\epsilon_- \,,\\
&\delta_+\lambda_+ = \big(\mathcal \partial_{a}\phi\,\Gamma^{a} - b_{a}\,\Gamma^{a} - \frac{1}{12}\,h^{abc}\Gamma_{abc}\big)\epsilon_+\,,\\
&\delta_-\lambda_+ = \frac12\,\tau^{ab+}\Gamma_{ab+}\epsilon_-\,,\\
&\delta_+\lambda_- = 0\,,\\
&\delta_-\lambda_- = \big(\mathcal \partial_{a}\phi\,\Gamma^{a} - b_{a}\,\Gamma^{a}- \frac{1}{12}\,h^{abc}\Gamma_{abc}\big)\epsilon_-\,,
\end{align}
\end{subequations}
where the covariant derivative  $\mathcal D_\mu \epsilon_\pm$ is given by
\begin{align}
\mathcal D_\mu\epsilon_\pm = \bigg(\partial_\mu - \frac14\,\omega_\mu{}^{ab}\Gamma_{ab}\pm\frac12\,\omega_\mu\mp\frac12\,b_\mu\bigg)\epsilon_\pm\,.
\end{align}

This finishes our presentation of the action and symmetries of the 10D minimal  supergravity theory.

\section{Conclusions}

In this review we gave an overview of the different non-Lorentzian supergravity theories that have been constructed so far in the literature. In 3$D$ and 10$D$ we explained the construction method based on  taking a non-Lorentzian limit of a relativistic supergravity theory. Moreover, we gave the explicit results for the non-Lorentzian supergravity theory in these two cases. We also illustrated the construction of several 3$D$ Chern-Simons supergravity theories using the Lie algebra and/or semigroup expansion.

Obviously, more work needs to be done. At the time of writing this review efforts are made to extend the work of \cite{Blair:2021waq} and to take the non-Lorentzian limit of 11$D$ supergravity based upon a non-Lorentzian geometry with a membrane distribution of co-dimension 3 \cite{inpreparation}. It is expected that the gauge fixing of this theory leads to a 11$D$ supersymmetric version of Newtonian gravity much in the same spirit of the 3$D$ Newtonian supergravity theory we discussed in subsection  4.2.2. We expect that in the same way non-Lorentzian versions of 10$D$ IIA and IIB supergravity can be constructed. Another issue that needs attention is a heterotic extension of the 10$D$ minimal supergravity theory that we discussed in this review. This theory should contain a non-Lorentzian version of the Yang-Mills and Lorentz Chern-Simons term that has played such an important role in the Green-Schwarz anomaly cancellation mechanism \cite{Green:1984sg}. Such anomaly cancellations are expected to also happen in the non-Lorenzian case.

Finally, in the longer term we hope that knowledge about the web of non-Lorentzian supergravity theories in diverse dimensions, as low-energy limits of non-Lorentzian string theories,  will help to understand the role they might play in a holographic formulation for describing a new class of NR conformal field theories at the boundary along the lines of  \cite{Brugues:2006yd,Gomis:2005pg,Bagchi:2009my}.

\subsection*{Acknowledgements}
This review is partly based on several articles that we wrote with our collaborators and that we refer to in the reference list.
We wish to thank them all for the illuminating discussions we had with them.


\end{document}